\documentclass[aps, pra, twocolumn, superscriptaddress, amsmath, showpacs, tightenlines, footinbib, longbibliography]{revtex4-1}

\usepackage{graphicx,subfigure}
\usepackage{enumerate}
\usepackage{color}
\usepackage{dcolumn}
\usepackage{epstopdf}
\usepackage{multirow}
\usepackage{appendix}
\usepackage{braket}
\usepackage[colorlinks, breaklinks, citecolor=blue, linkcolor=blue,  urlcolor={blue}]{hyperref}
\usepackage{breakurl}
\usepackage{amsfonts}
\usepackage{amsmath}
\usepackage{amssymb}
\usepackage{mathrsfs}
\usepackage{epsfig}
\usepackage{bm}

\usepackage{colortbl}
\definecolor{mygray}{gray}{.9}
\definecolor{darkblue}{rgb}{1,1,.70}
\definecolor{lightblue}{rgb}{1,1,.90}

\begin{document}

\title{Detecting degenerate bands topological invariants in optical lattice }

\author{Jing-Xin Liu}
\affiliation{National Laboratory of Solid State Microstructures and School of Physics, Nanjing University, Nanjing 210093, China}
\affiliation{Key Laboratory of Atomic and Subatomic Structure and Quantum Control (Ministry of Education), Guangdong Basic Research Center of Excellence for Structure and Fundamental Interactions of Matter, School of Physics, South China Normal University, Guangzhou 510006, China}

\author{Jian-Te Wang}
\affiliation{National Laboratory of Solid State Microstructures and School of Physics, Nanjing University, Nanjing 210093, China}
\affiliation{Key Laboratory of Atomic and Subatomic Structure and Quantum Control (Ministry of Education), Guangdong Basic Research Center of Excellence for Structure and Fundamental Interactions of Matter, School of Physics, South China Normal University, Guangzhou 510006, China}

\author{Hai-Tao Ding}
\affiliation{National Laboratory of Solid State Microstructures and School of Physics, Nanjing University, Nanjing 210093, China}
\affiliation{Key Laboratory of Atomic and Subatomic Structure and Quantum Control (Ministry of Education), Guangdong Basic Research Center of Excellence for Structure and Fundamental Interactions of Matter, School of Physics, South China Normal University, Guangzhou 510006, China}
\affiliation{Department of Physics, National University of Singapore, Singapore 117551}

\begin{abstract}
In this paper, we present a novel experimental approach for simulating and detecting topological invariants using ultracold fermions confined in two-dimensional hexagonal optical lattices. We propose achieving degenerate four-band models with non-trivial topologies in both the AII and A classes by introducing additional inertial forces, Raman processes, or periodic driving. By implementing various quench sequences and observing the evolution of the time-of-flight pattern, we can gather comprehensive information about the ground states and determine the topological property of the valence bands. Through the analysis of tomographic results, we are able to extract and calculate the spin Chern number. Additionally, we demonstrate the robustness of the quantized topological invariants and discuss the effects of various experimental parameters.
\end{abstract}

\maketitle


\preprint{APS/123-QED}

\section{Introduction}

The discovery of the quantum spin Hall(QSH) effect in materials with strong spin-orbit coupling(SOC) has led to the emergence of a new class of topological states called $\mathbb{Z}_2$ topological insulators \cite{RevModPhys.82.3045,RevModPhys.83.1057,PhysRevLett.96.106802}. With advancements in experimental techniques, researchers can now investigate quantum phenomena in artificial systems such as Rydberg-excited atoms \cite{Weimer2010}, trapped ions \cite{barreiro2011open,PRL2006.97}, superconductor circuit\cite{nakamura1999coherent} and nanostructured materials \cite{kitagawa2012observation}. These platforms offer a unique opportunity to explore the quantum behaviors of matter and their potential applications in quantum information processing.

Studies of topological effects often require ultrastrong gauge fields or spin-orbit couplings. Cold atoms confined in optical lattices provide an excellent platform for emulating a wide range of systems in condensed matter physics \cite{RevModPhys.80.885,lewenstein2007ultracold,APDWZhang2018}. Synthetic gauge fields and SOC can be realized using various techniques, including trap rotation \cite{RevModPhys.81.647}, microrotation \cite{PhysRevLett.94.086803,PhysRevLett.100.130402,PhysRevA.82.013616,PhysRevB.82.235114,Mei2012}, and Raman laser-induced transitions \cite{RevModPhys.83.1523,lin2009synthetic,lin2011spin,SLZhu2011,PhysRevLett.102.130401,PhysRevLett.107.255301,PhysRevLett.95.010403,JZLi2022,jaksch2003creation,gerbier2010gauge,PhysRevLett.103.035301,PhysRevLett.97.240401,PhysRevResearch.5.L012006}. By combining laser-induced tunneling and superlattice techniques, strong Abelian \cite{jaksch2003creation} and non-Abelian \cite{PhysRevLett.95.010403} gauge fields can be achieved, enabling the simulation of topological insulators and other models. These technologies have been proposed for realizing quantum Hall and quantum spin Hall states \cite{gerbier2010gauge,PhysRevLett.103.035301,PhysRevLett.97.240401,goldman2007quantum,PhysRevLett.101.246810,PhysRevLett.101.186807,PhysRevResearch.5.L012006,PhysRevA.79.053639,PhysRevLett.107.235301,PhysRevLett.107.145301,PhysRevLett.105.255302}.

The discovery of topological matters has opened up new avenues for quantum control and measurement. The probing techniques for different topological phase are also important for quantum simulation. For these techniques, it \textbf{}can be divided into two main types. The first type involves directly extracting the topological properties locally. Such as extract  Berry curvature of the entire Brillouin zone(BZ), which can be achieved using interferometers \cite{atala2013direct,duca2015aharonov,grusdt2016interferometric,PhysRevLett.110.165304,PhysRevA.89.043621} or studying the semiclassical dynamics of wavepackets \cite{PhysRevA.85.033620,PhysRevLett.111.135302,jotzu2014experimental,aidelsburger2015measuring}. Quantum state tomography method which measure the entire BZ can also be viewed as measure interference pattern from the time-of-flight(ToF) image\cite{PhysRevLett.107.235301,PhysRevA.90.041601}. Several experiments about tomography of two-level models in cold atom systems has realized based on the dynamics after different quench sequences\cite{PhysRevLett.113.045303,doi:10.1126/science.aad4568,PhysRevResearch.5.L032016}, projection on different momenta by fast acceleration\cite{doi:10.1126/science.aad5812}, and off-resonant coupling to higher bands\cite{PhysRevLett.118.240403}. As for the second type, these involves directly detecting physical observables that act as a response to the topological phase or topological invariants, such as density profile plateaus\cite{Streda_1982,PhysRevLett.100.070402,PhysRevLett.101.246810,SLZhu2013,PhysRevA.84.063629,PhysRevA.78.053617,PhysRevLett.108.220401,SCDWZhang2020} or edge states detected using Bragg spectroscopy\cite{PhysRevA.81.033622,PhysRevLett.108.255303,PhysRevA.82.013608,PhysRevA.85.063614,PhysRevA.85.061606,PhysRevA.89.063628,Zhu2007}. All of these techniques are being used to study topological phases of matter in artificial systems and to explore their potential applications in quantum technology.

Without considering the time-reversal symmetry(TRS), the simple non-trivial two band model can be realized easily. If we consider the regime with TRS and other symmetry, the lattice model is complicated for realization. Optical lattice systems have been used to realize models with both TRS and non-trivial topology. This has been achieved through the application of gradient magnetic fields and Raman laser fields \cite{PhysRevLett.111.225301,PhysRevLett.111.185301,PhysRevLett.105.255302}. However, realization of non-trivial topology and breaking of spin conservation simultaneously remains a challenge. In this paper, we propose the realization of a hexagonal geometric structure that exhibits non-trivial topology in two-fold degenerate four-band models. In our first model, we introduce additional inertial forces and Raman fields to achieve a non-trivial phase belonging to class AII with additional $s_x$ spin symmetry. In the second model, we combine two Haldane models with opposite Chern numbers and introduce a mass term that preserves degeneracy while breaking spin conservation. We can define the spin-Chern number(SCN) for such a system when $s_z$ spin conservation is broken, and we present an experimental scheme to simulate and probe this topological invariants. Our proposal involves trapping cold atoms in a 2D spin-dependent optical lattice subjected to periodic driving and modulation. We aim to measure the topology of the degenerate bands through the evolution of ToF images with quench dynamics.

The paper is organized as follows. Section~\ref{sec:2} shows the conditions when a four-band model holds global degeneracy with Clifford matrices. Section~\ref{sec:3} provides a definition of SCN. In Section~\ref{sec:4}, we discuss the realization of the topological phase of a four-band model in class AII and class A with two-fold degeneracy. In Section~\ref{sec:5}, we introduce a method for band tomography by analyzing ToF images with the aid of quench dynamics. This method allows for the direct calculation of the spin Hamiltonian and Berry curvature of the entire BZ in the spin non-conserving case, thereby obtaining the topological invariants. Finally, Section~\ref{sec:6} provides a discussion about experimental parameters and conclusion of the paper.

\section{Conditions of four-band model with degeneracy}
\label{sec:2}

Firstly, we discuss the conditions when the 4-band Hamiltonian shows degeneracy. An arbitrary four-band model can be represented by 15 Clifford matrices where Clifford matrices are defined as $\Gamma_{(1,2,3,4,5)} = (I \otimes \sigma_x, I \otimes \sigma_z, s_x \otimes \sigma_y, s_y \otimes \sigma_y, s_z \otimes \sigma_y)$, and their ten commutators are $\Gamma_{ij} = [\Gamma_i,\Gamma_j]/(2i)$. If there exists global degeneracy, the Hamiltonian satisfies $H^2 \propto I $. Suppose $H = \sum_{i} g_i \Gamma_i + \sum_{i<j} g_{ij} \Gamma_{ij}$ and its square is
\begin{equation}
\begin{split}
H^2 &= \left(\sum_{i} g_i^2 + \sum_{i<j} g_{ij}^2 \right) I_4 + \sum_{i < j} g_i g_j \left\{\Gamma_i , \Gamma_j \right\} \cr
&+ \sum_{i,j<k} g_i g_{jk} \left\{\Gamma_i , \Gamma_{jk} \right\} + \sum_{i<j,k<l} g_{ij} g_{kl} \left\{\Gamma_{ij} , \Gamma_{kl} \right\},
\end{split}
\end{equation}
and this implies all anticommutators are zero. Firstly, $\{\Gamma_i,\Gamma_j\} = 2\delta_{ij}$. For the last two terms with $\Gamma_{ij}$, $\{\Gamma_{i},\Gamma_{jk}\} = 0$ if and only if $i=j$ or $i=k$. Similarly, if one index in $\{i,j\}$ is equal to $\{k,l\}$, $\{\Gamma_{ij},\Gamma_{kl}\}$ also takes zero. Above all, we get two types of 2-fold global degenerate cases:
\begin{equation}
\begin{split}
H &= \sum_{i} g_i \Gamma_i, \cr
H &= g_i \Gamma_i + \sum_{j} g_{ij} \Gamma_{ij}, \cr
\end{split}
\end{equation}
and these two types of Hamiltonians are equivalent by performing one unitary transformation. We could also observe that one four band model with global degeneracy can have at most five different Clifford matrices.

\section{Quantum spin hall effect model and spin Chern number}

\label{sec:3}

In the study of topological insulators within the AII class, the system's behavior is typically characterized by a $\mathbb{Z}_2$ topological invariants\cite{PhysRevLett.95.146802}. When additional symmetries, like spin conservation ($s_z$), are introduced, the classification shifts from $\mathbb{Z}_2$ to $\mathbb{Z}$. In such cases, the SCN can be defined based on the differences in Chern numbers between distinct spin components. Even without spin conservation, as explored in \cite{PhysRevLett.97.036808,PhysRevB.82.165104}, the SCN remains applicable. In these situations, the expected spin values for the two valence bands do not simply fall at $-1$ and $+1$, but form distinct regions. As long as these differences remain discernible, the edge states can be viewed as clusters with different spin orientations.

To determine the SCN, researchers employ a method that involves decomposing the occupied valence bands into two sectors by diagonalizing the expression $\hat{P} \hat{s}_z \hat{P}$, where $\hat{P}$ represents the valence band projection operator. This diagonalization allows us to represent $\hat{P} \hat{s}_z \hat{P}$ as a $2\times 2$ matrix $\bra{u_{\alpha}(k)}\hat{s}_z\ket{u_{\beta}(k)}$ within the valence band. To compute the SCN, we calculate the spin Berry curvature $\mathcal{F}_{\pm}(k) = i \varepsilon_{\mu\nu} \braket{\partial_{\mu} \psi_{\pm}(k)|\partial_{\nu} \psi_{\pm}(k)}$ where $\pm$ represent two orthogonal degenerate sectors. Using this, we define $C_\pm = 1/(2\pi) \int \mathrm{d}^2 k \mathcal{F}_{\pm}(k)$ and define $C_s = C_{+} - C_{-}$\cite{PhysRevLett.93.206602}. It has been proven that this method remains robust against continuous deformations of the system Hamiltonian, including symmetry-breaking perturbations, as shown in \cite{PhysRevB.80.125327,PhysRevLett.127.136802}. Using this method, the SCN can also be defined when TRS is broken\cite{PhysRevLett.107.066602}, and there exists QSHE but without Kramer pairs. In the next section, we will introduce two models that can be realized in an optical lattice with ultracold Fermi gas, following the idea of SCN described above.

\section{Realization in optical lattice}
\label{sec:4}

We present two possible approaches for realizing and detecting non-trivial topology with two-fold degeneracy. The first model presents a method to realize a system that maintains TRS($\mathcal{T}^2 = -1$) in the AII class with $s_x$ conservation, without introducing additional SOC. The second model is based on two Haldane models with opposite Chern numbers. We also introduce additional mass terms to break spin conservation while maintaining the condition of the double global degeneracy. All of these models could exhibit non-trivial topological invariants.

\begin{figure}[tbph]
	\centering\includegraphics[width=8cm]{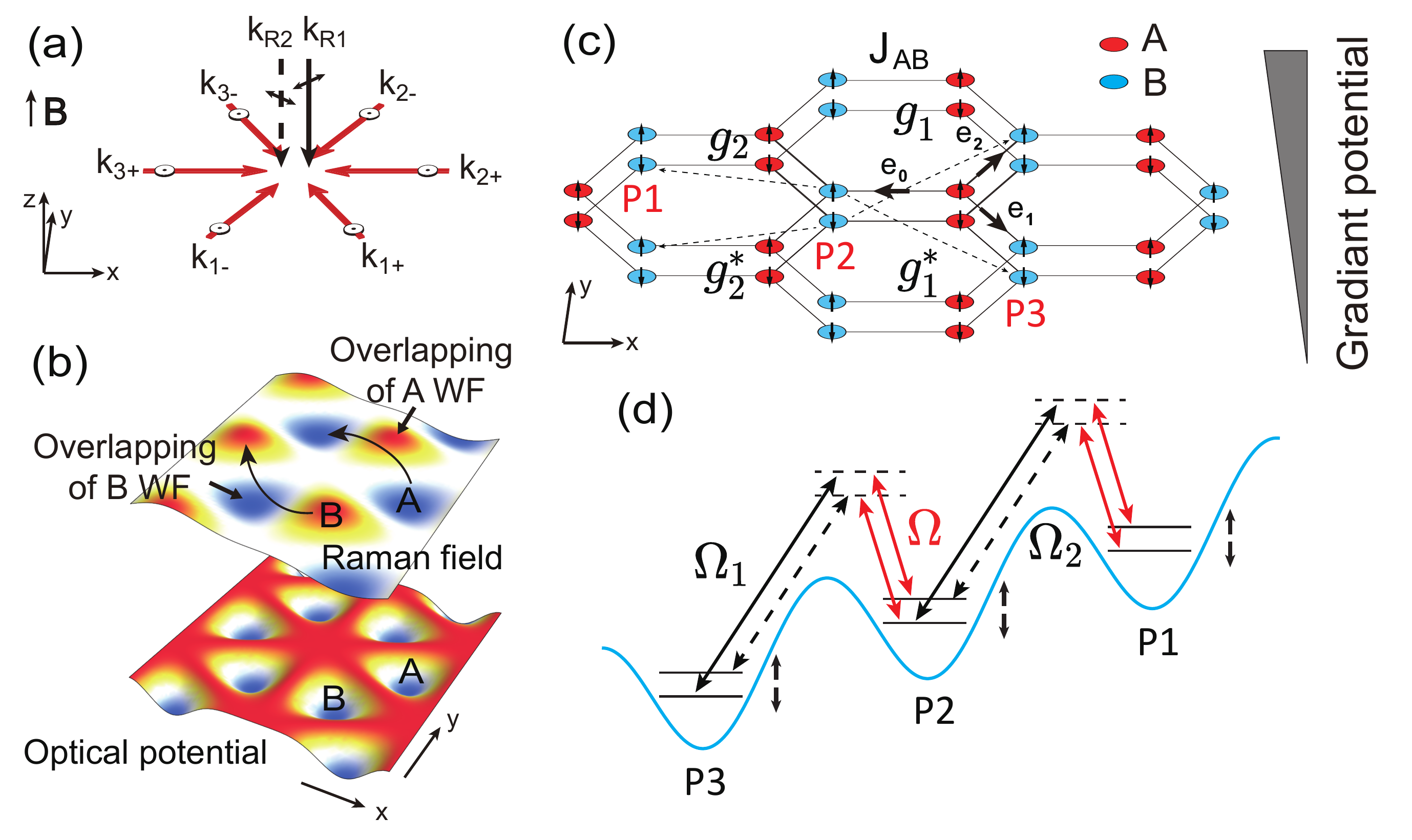}
	\caption{(a) The laser configuration for the two-dimensional hexagonal optical lattice. (b) The hexagonal optical potential and Raman field $\mathbf{M}(\mathbf{r})$. The overlapping region of Wannier functions indicates that the Raman process takes opposite values for different sublattices. (c) The tight-binding model of this system, with some spin-flipping hopping indicated by black dashed arrows. (d) Spin-flipping hopping in the optical lattice tilted by an additional gradient potential. Two laser fields (black solid and dashed arrows) match the frequency difference of spin-flipping hopping along different directions. }
	\label{fig:model 1}
\end{figure}

\subsection{Model 1: AII class with $s_x$ conservation }

In this section, we discuss the realization of a model with TRS ($\mathcal{T}^2 = -1$) and conserved $s_x$. The model's topological invariant is SCN which falling under $\mathbb{Z}$ classification. The implementation of a $\mathcal{T}$-symmetric model has been achieved in an optical lattice by incorporating gradient magnetic and Raman laser fields\cite{PhysRevLett.111.225301,PhysRevLett.111.185301,PhysRevLett.105.255302,DWZhang2016}. A similar strategy could be applied on honeycomb lattice. However, the difference in our scheme is that in order to realize the topological phase, an additional external force should be introduced.

An optical potential can be generated using six laser fields with wavevectors $\mathbf{k}_{i\pm}$ and polarization along $\hat{z}$, as illustrated in Fig.\ref{fig:model 1}(a) with red arrows. By carefully choosing frequencies for extra laser fields propagating along the $\hat{z}$ direction, we establish additional two-photon Raman processes. This mechanism facilitates the coupling of different spins at specific positions with two different Raman fields.

The complete Hamiltonian encompasses the optical potential $V(\mathbf{r})$ and Raman fields $\mathbf{M}_1(\mathbf{r}), \mathbf{M}_2(\mathbf{r})$, taking into account the influence of the Zeeman field and external force in the Raman field:
\begin{equation}
\begin{split}
H = \frac{\mathbf{p}^2}{2m} + V(\mathbf{r}) + \left( \mathbf{M}_{1}(\mathbf{r}) + \mathbf{M}_{1}(\mathbf{r}) \right) \cdot \boldsymbol{S}.
\end{split}
\end{equation}
The optical potentials are formed by laser fields with wavevectors $\mathbf{k}_{i\pm} = \sqrt{3}R_z(\pm\pi/6) \mathbf{k}_i$ where $\mathbf{k}_1 = (0,1,0)$, $\mathbf{k}_2 = (-\sqrt{3}/2, -1/2, 0 )$, $\mathbf{k}_3 = (\sqrt{3}/2, -1/2, 0 )$, and $R_z(\theta)$ is the rotation operator along the $z$ axis. The combined electric field of all laser fields propagating in the $x-y$ plane is
\begin{equation}
\begin{split}
\mathcal{E}(\mathbf{r}) & = E_0 \sum_{j} \left[ \exp{(\mathbf{k}_{j+} \cdot \mathbf{r})} - \exp{(\mathbf{k}_{j-} \cdot \mathbf{r})} \right] \mathbf{e}_z, \cr
\end{split}
\end{equation}
where lasers with opposite wavevectors exhibit an additional $\pi$ phase difference. The associated optical potential is given by $V(\mathbf{r}) = - \alpha_s |\mathcal{E}(\mathbf{r})|^2$, as depicted in Fig.\ref{fig:model 1}(b), where $\alpha_s$ represents the atomic polarizability and $E_0$ represents the strength of the electric field.

With the help of Raman fields, the inertial force generated by accelerating the lattice in a specific direction facilitates the coupling of spins at particular positions. The Zeeman field introduces an energy difference, allowing for the independent coupling of different spins. By carefully choosing the frequency of laser fields $\mathcal{E}_{R1}(\mathbf{r})$ and $\mathcal{E}_{R2}(\mathbf{r})$ in Fig.\ref{fig:model 1}(a) with solid and dashed black lines propagating along the $\hat{z}$ direction, we can control the next-nearest-neighbor(NNN) spin-flipping hopping independently. The laser fields $\mathcal{E}_{R1}(\mathbf{r})$ and $\mathcal{E}_{R2}(\mathbf{r})$ are considered constant at the $z = 0$ plane. By adjusting the polarization of these laser fields, we can introduce $\sigma_+$ or $\sigma_-$ processes independently. For instance, in Fig.\ref{fig:model 1}(d), the red arrows represent the $\pi$ transition process, while the black solid and dashed lines depict the $\sigma_+$ and $\sigma_-$ transitions, discernible through the changes in spin.

The determination of tight-binding parameters relies on the analysis of Wannier-Stark functions in the presence of gradient field casued by inertial force. The terms governing NNN spin-flipping hopping are associated with the overlap of two distinct Wannier-Stark functions with different spin on the same sublattice. Therefore, an analysis of the Raman field properties is crucial. The Rabi frequencies $\Omega$, $\Omega_1$, and $\Omega_2$ in Fig.\ref{fig:model 1}(d) are proportional to $\mathcal{E}(\mathbf{r})$, $\mathcal{E}_{R1}(\mathbf{r})$, and $\mathcal{E}_{R2}(\mathbf{r})$. Consequently, the Raman field term can be expressed as $\mathbf{M}_1(\mathbf{r}) = \Omega_1^\ast(\mathbf{r})\Omega(\mathbf{r})/\Delta_1 \hat{S}^{-} + \text{h.c.}$, as shown in Fig.\ref{fig:model 1}(b), representing the coupling between $\ket{\uparrow}$ at P2 and $\ket{\downarrow}$ at P3 in Fig.\ref{fig:model 1}(d). Similarly, coupling in the opposite direction is achieved by the Raman field $\mathbf{M}_2(\mathbf{r}) = \Omega_2(\mathbf{r})\Omega^\ast(\mathbf{r})/\Delta_2 \hat{S}^{-} + \text{h.c.}$.

Observing the Raman field values in Fig.\ref{fig:model 1}(b), it becomes apparent that NNN spin-flipping hopping takes opposite values on different sublattices. The NNN spin-flipping terms along the $y$ direction, induced by the Raman process, can be disregarded due to the anti-symmetric structure of the Raman field $\mathbf{M}(\mathbf{r})$ relative to the middle of the A-B sublattice. Consequently, we can formulate the tight-binding Hamiltonian as depicted in Fig.\ref{fig:model 1}(c)
\begin{equation}
\begin{split}
\hat{H}_{tb} &= \sum_{\braket{\mathbf{r},\mathbf{s}}} J_{AB} a_{\mathbf{r},\sigma}^\dag b_{\mathbf{s},\sigma} + \sum_{\braket{\braket{\mathbf{r},\mathbf{s}}}} \left( g_1 a_{\mathbf{r},\downarrow}^\dag a_{\mathbf{s},\uparrow} \right. \\
& \left. + g_2 a_{\mathbf{r},\uparrow}^\dag a_{\mathbf{s},\downarrow} -  g_1 b_{\mathbf{r},\downarrow}^\dag b_{\mathbf{s},\uparrow} - g_2 b_{\mathbf{r},\uparrow}^\dag b_{\mathbf{s},\downarrow} \right) + \text{h.c.},
\end{split}
\end{equation}
using annihilation operators $\hat{a}_{\mathbf{r},\sigma}/\hat{b}_{\mathbf{r},\sigma}$ for particles at sublattice A/B with spin $\sigma$ and position $\mathbf{r}$. Here, $\braket{\mathbf{r},\mathbf{s}}$ calculates all nearest-neighbor(NN) terms, and $\braket{\braket{\mathbf{r},\mathbf{s}}}$ calculates the four possible NNN hopping terms in Fig.\ref{fig:model 1}(c). The SOC terms $g_j \propto \Omega_j \Omega^\ast$ illustrate that these terms can be controlled independently by adjusting the amplitude and phase difference of $\mathcal{E}_{R1}$ and $\mathcal{E}_{R2}$. By setting $g_1 = g_2 = i g$, the Bloch Hamiltonian is
\begin{equation}
H(\mathbf{k}) = B_x(\mathbf{k}) \sigma_x + B_y(\mathbf{k}) \sigma_y + B_{xz}(\mathbf{k}) s_x \sigma_z,
\end{equation}
where
\begin{equation}
\begin{split}
B_x &= \mathrm{Re}\left[\sum_{j} J_{AB,j} \exp{(-i\mathbf{k} \cdot \mathbf{e}_j)} \right], \cr
B_y &= \mathrm{Im}\left[\sum_{j} J_{AB,j} \exp{(-i\mathbf{k} \cdot \mathbf{e}_j)} \right], \cr
B_{xz} &= g\sin{[\mathbf{k}\cdot(\mathbf{e}_0 - \mathbf{e}_2)]} + g\sin{[\mathbf{k}\cdot(\mathbf{e}_1 - \mathbf{e}_0)]},
\end{split}
\end{equation}
and $\mathbf{e}_{0} = -a \hat{\mathbf{x}},~\mathbf{e}_{1} = a \left( \hat{\mathbf{x}}/2 - \sqrt{3}/2 \hat{\mathbf{y}} \right),~\mathbf{e}_{2} = a \left( \hat{\mathbf{x}}/2 + \sqrt{3} \hat{\mathbf{y}}/2 \right) $. This model exhibits $\mathcal{P} = s_x \sigma_x$, $\mathcal{T} = i s_y K$, and $s_x$ symmetries, falling into a non-trivial $\mathbb{Z}$ SCN. Utilizing the $s_x$ symmetry, the Hamiltonian can be block-diagonalized, and the SCN can be determined from the Chern number of one of these blocks. However, a limitation of this model is the inability to experimentally introduce additional terms to break the conserved spin symmetry or close the gap. This prevents the observation of a phase transition to topological insulator.

\subsection{Model 2: A class }

\begin{figure}[tbph]
	\centering\includegraphics[width=8cm]{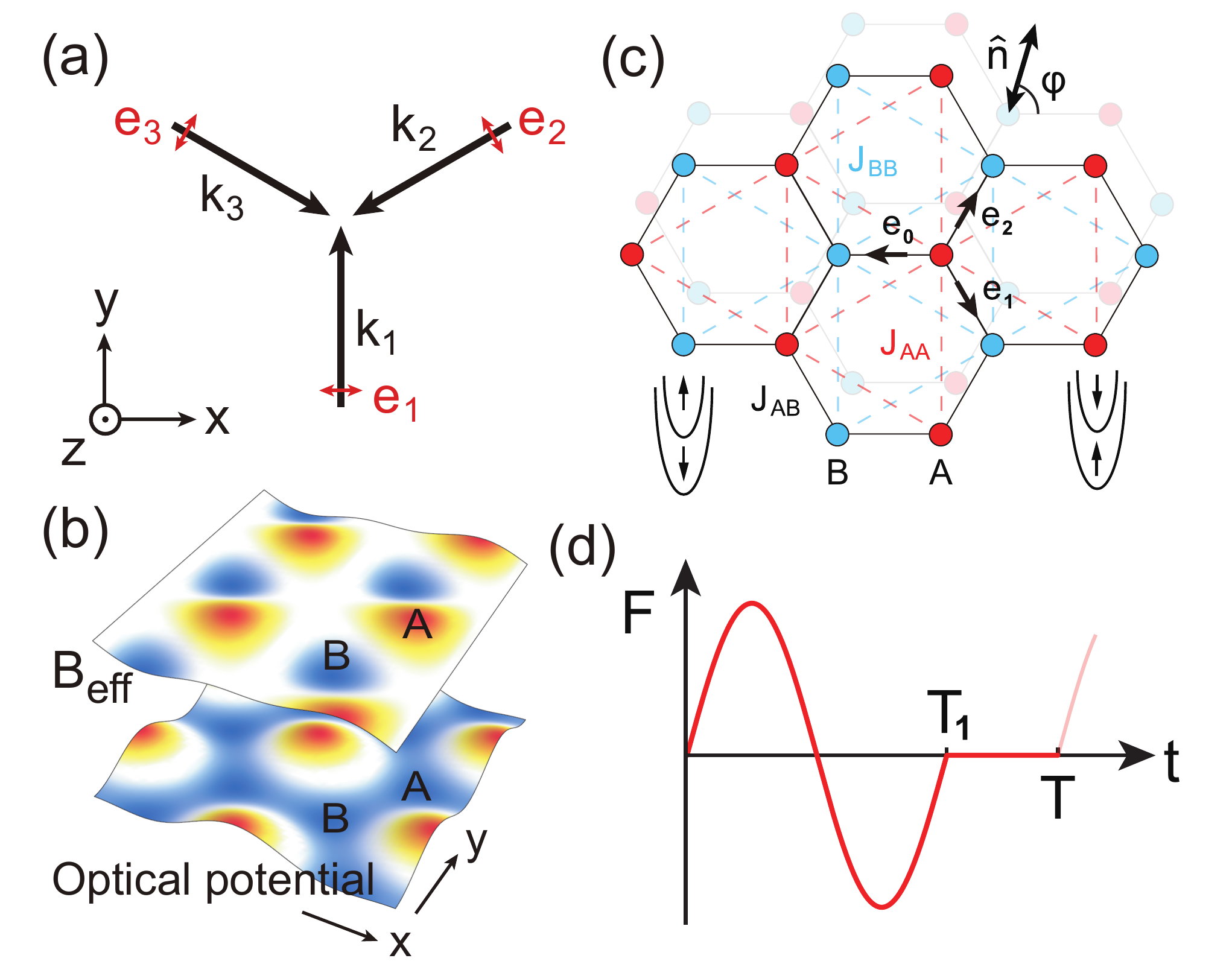}
	\caption{(a) Laser setup of the two-dimensional hexagonal optical lattice. A spin-dependent sublattice potential $\Delta$ is introduced by controlling the local polarization. The trap's depth is not extremely strong, such that the next-nearest-neighbor (NNN) hopping terms $J_{AA}$ and $J_{BB}$ cannot be ignored. (b) The strength of optical potential and effective magnetic field. (c) Tight-binding discription of model. (d) Periodic modulation function.}
	\label{fig:model 2}
\end{figure}

Because of SCN also can be defined when TRS is broken\cite{PhysRevLett.107.066602}, here we consider a construction of SCN without considering TRS. In this section, we can realize two Haldane models with opposite Chern numbers for different spins. We also consider additional mass terms that break spin conservation while maintaining the condition of not disrupting the double global degeneracy.

We start with cold fermions in a 2D setup, confined in a spin-dependent hexagonal optical lattice. This confinement is achieved by adjusting the light polarization between $\sigma^+$ and $\sigma^-$, as successfully demonstrated in experiments\cite{soltan2011multi}. The hexagonal lattice can be regarded as a triangular lattice with a sublattice structure. In this arrangement, atoms occupy $\sigma^+$ and $\sigma^-$ sites, experiencing a spin-dependent a.c. Stark shift within the light field. The overall potential is given by $V(\mathbf{r}) = V_{\mathrm{hex}}(\mathbf{r}) + m_z V_{\mathrm{tri}}(\mathbf{r})$. The fermion dynamics follow the Hamiltonian $H = \mathbf{p}^2/2m + V(\mathbf{r})$.

The formation of a spin-dependent hexagonal optical lattice \cite{soltan2011multi,PhysRevA.80.043411} is achieved through the intersection of three laser beams at a $120^\circ$ angle, each linearly polarized in the $x-y$ plane. The configuration of the three laser fields is as follows:
\begin{equation}
\begin{split}
\mathcal{E}_1(\mathbf{r},t) &= E_0 e^{i(\mathbf{k}_1 \cdot \mathbf{r} - \phi_1) - i \omega_L t} \mathbf{e}_x \cr
\mathcal{E}_2(\mathbf{r},t) &= -E_0 e^{i(\mathbf{k}_2 \cdot \mathbf{r} - \phi_2) - i \omega_L t} \left( \frac{1}{2} \mathbf{e}_x - \frac{\sqrt{3}}{2}\mathbf{e}_y \right) \cr
\mathcal{E}_3(\mathbf{r},t) &= -E_0 e^{i(\mathbf{k}_3 \cdot \mathbf{r} - \phi_3) - i \omega_L t} \left( \frac{1}{2} \mathbf{e}_x + \frac{\sqrt{3}}{2}\mathbf{e}_y \right) \cr
\end{split}
\end{equation}
The spin-independent potential of 2D-lattice $V_s = -\alpha_s |\mathcal{E}|^2$, given by $ V_s (\mathbf{r}) = - \alpha_s E_0^2 [ 3 - \cos{(\mathbf{b}_1 \cdot \mathbf{r})} - \cos{(\mathbf{b}_2 \cdot \mathbf{r})} - \cos{( (\mathbf{b}_1 - \mathbf{b}_2) \cdot \mathbf{r})} ] $ and $\mathbf{b}_1 = \mathbf{k}_2 - \mathbf{k}_1,~\mathbf{b}_2 = \mathbf{k}_3 - \mathbf{k}_2,~\mathbf{b}_3 = \mathbf{k}_1 - \mathbf{k}_3$. The spin-dependent potential of 2D-lattice $V_v = i\alpha_v (\mathcal{E}^\ast \times \mathcal{E})\cdot \mathbf{F}$, given by $V_v (\mathbf{r}) = \sqrt{3} \alpha_v E_0^2 \mathbf{F}_z \left[ \sin{(\mathbf{b}_1 \cdot \mathbf{r})} + \sin{(\mathbf{b}_2 \cdot \mathbf{r})} - \sin{( \mathbf{b}_3 \cdot \mathbf{r})} \right] $ and shown in Fig.~\ref{fig:model 2}(b) as effective magnetic field $\mathbf{B}_{eff}$.

The tight-binding model, with a weak potential corresponding to the recoil energy $E_R$, enables the inclusion of a significant NNN term and can be expressed as:
\begin{equation}
\begin{split}
\hat{H}_{tb} &= \sum_{\mathbf{r}}  m_z \Delta ( a_{\mathbf{r},\sigma}^\dag a_{\mathbf{r},\sigma} - b_{\mathbf{r},\sigma}^\dag b_{\mathbf{r},\sigma}) + \left( \sum_{\mathbf{r},\mathbf{s}} J_{AB}^{(\sigma)} a_{\mathbf{r},\sigma}^\dag b_{\mathbf{s},\sigma} + \right. \cr
& \left. \sum_{\braket{\braket{\mathbf{r},\mathbf{s}}},\sigma} J_{AA}^{(\sigma)} a_{\mathbf{r},\sigma}^\dag a_{\mathbf{s},\sigma} + J_{BB}^{(\sigma)} b_{\mathbf{r},\sigma}^\dag b_{\mathbf{s},\sigma} + \text{h.c.} \right)
\end{split}
\end{equation}
The tunneling amplitudes can be calculated from the overlap of the Wannier function. Since the potential for a spin-up atom at sublattice A is identical to a spin-down atom at sublattice B, the Wannier packet with different spins should satisfy $w_{\uparrow,A}(\mathbf{r}) = w_{\downarrow,B}(\mathbf{r})$ and $w_{\downarrow,A}(\mathbf{r}) = w_{\uparrow,B}(\mathbf{r})$. This property implies that the NN hopping is spin-independent, and the NNN hopping satisfies $J_{AA}^{\uparrow} = J_{BB}^{\downarrow},~J_{AA}^{\downarrow} = J_{BB}^{\uparrow}$.

Here we consider the periodic driving optical lattice shown in Fig.~\ref{fig:model 2}(c), described by the free fermion Hamiltonian
\begin{equation}
\hat{H}(t) = \hat{H}_{tb} + \sum_{\mathbf{r}} \boldsymbol{F}(t)\cdot \mathbf{r} (\hat{n}_{A,\mathbf{r}} + \hat{n}_{B,\mathbf{r}} ).
\end{equation}
The term $\hat{H}_{os}$ collects on-site terms describing many-body interactions or a weak static potential. In our model, the modulation involves moving the lattice along a periodic trajectory $\mathbf{r}(t)$, which introduces an inertial force $\boldsymbol{F}(t) = - m \ddot{\mathbf{r}}_{\mathrm{lat}}(t)$ acting on the atoms. This additional time-dependent term can be canceled by a unitary transformation $U(t) = \mathrm{exp} \left[ -i \sum_{\mathbf{r}} m \dot{\mathbf{q}}_{\mathrm{lat}}(t) \cdot \mathbf{r} (\hat{n}_{A,\mathbf{r}} + \hat{n}_{B,\mathbf{r}} )  \right]$, where $\mathbf{q}_{\mathrm{lat}} = m [\dot{\mathbf{r}}_{\mathrm{lat}}(t) - \dot{\mathbf{r}}_{\mathrm{lat}}(0)]$ introduces a complex phase factor to the tunneling amplitudes. If there is no spin-flipping term, $s_z$ is conserved and the Hamiltonian can be divided into two individual parts.

For a simple case in our floquet gauge, if we only consider the zero-order term\cite{PhysRevLett.108.225304,PhysRevLett.109.145301,PhysRevA.89.061603}, the Floquet Hamiltonian is the time average $H_{\mathrm{eff}} = \braket{H(t)}_T$ as long as the recoil energy $E_R$ is significantly larger than both the nearest-neighbor hopping $J$ and the energy scales of Hamiltonian. In this treatment, the effective tunneling after one period is given by
\begin{equation}
|J_{ij}^{(\mathrm{eff})}|e^{i\theta_{ij}} = \left\langle J_{ij} e^{-i \mathbf{q}_{\mathrm{lat}} \cdot \mathbf{r}_{ij} } \right\rangle_{T}.
\label{eq:10}
\end{equation}

\begin{figure}[tbph]
	\centering\includegraphics[width=8cm]{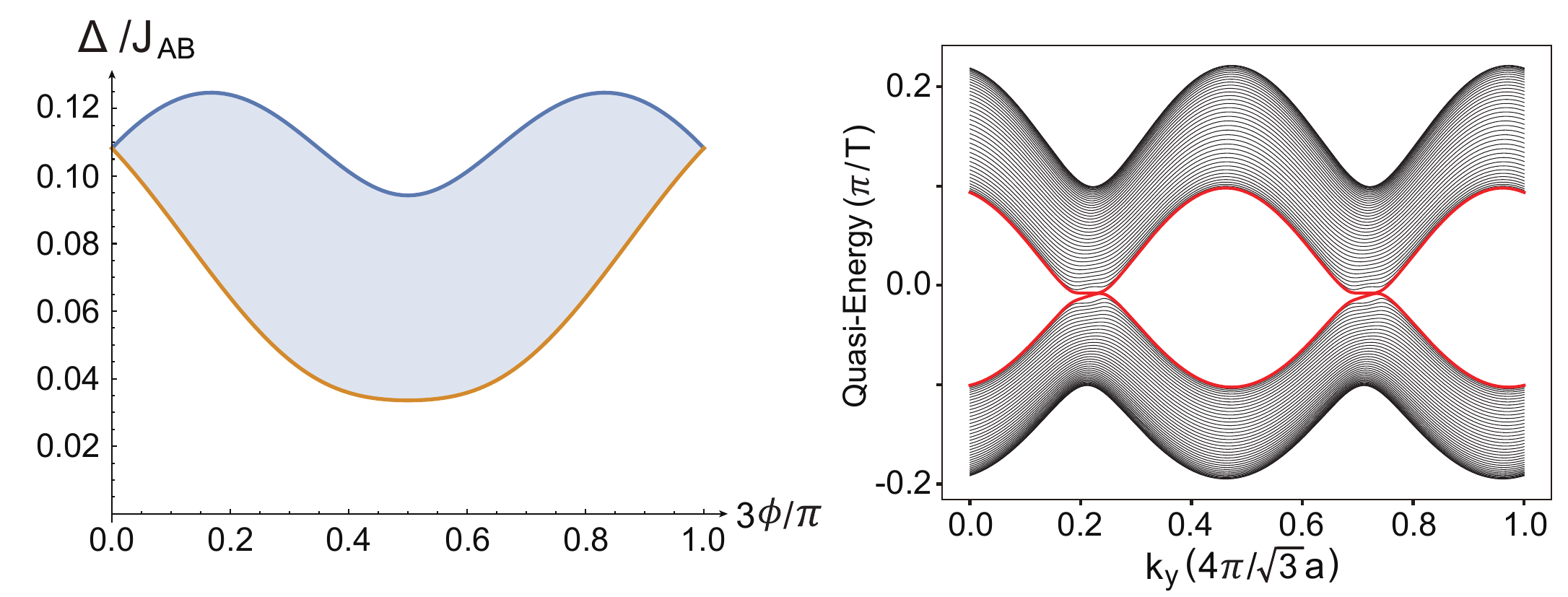}
	\caption{ The phase diagram of the topological phase, and the corresponding floquet band structure. The calculated parameters include $J_{AA} = 0.1 J_{AB}$, $J_{BB} = 0$, $T_1 = T/2$, $\omega/J_{AB} = 20$, and $F_0 a/\omega_1 \approx 1.89$. Additional parameters relevant to the experiment are provided in Sec.~\ref{sec:5}.	}
	\label{fig:model 2 result}
\end{figure}

Moreover, for sinusoidal forcing, such dynamics modification of tunneling is restricted to $\theta_{ij} = 0$ or $\pi$\cite{Gro_mann_1992,PhysRevB.34.3625,PhysRevLett.69.351,PhysRevLett.95.260404}. This phenomenon has been observed in several experiments\cite{PhysRevLett.100.190405,PhysRevLett.102.100403,science.1207239}. The Peierls phase $\theta_{ij}$ also can be smoothly tuned to any value through appropriate driving\cite{PhysRevLett.109.145301,PhysRevA.85.053613}. Here, we consider a linear shaking scheme: the inertial force comprises a series of sinusoidal pulses separated by periods of rest with a periodicity of $T = T_1 + T_2$, as illustrated in Fig.~\ref{fig:model 2}(d).
\begin{equation}
\begin{split}
\boldsymbol{F}(t) &= \left\{
\begin{aligned}
 F_0 \sin{\omega_1 (t \mod T)} \hat{\mathbf{n}} &, \quad 0 < (t \mod T) < T_1, \\
 0 &, \quad T_1 < (t \mod T) < T.
\end{aligned} \right.
\end{split}
\end{equation}
where $\omega_1 = 2\pi/T_1$ and the shaking axis is given by $\hat{\mathbf{n}} = (\cos{\varphi} \mathbf{e}_x + \sin{\varphi} \mathbf{e}_y)$. The effective hopping will be renormalized and multiplied by a phase factor
\begin{equation}
\begin{split}
\left\langle e^{-i \mathbf{q}_{\mathrm{lat}} \cdot \mathbf{r}_{ij}} \right\rangle_{T} &= \frac{T_1}{T} e^{i \alpha_{ij}} \mathcal{J}_0(\alpha_{ij}) + \frac{T_2}{T}.
\end{split}
\end{equation}
where $\alpha_{ij} = F_0 \hat{\mathbf{n}} \cdot \mathbf{r}_{ij}/\omega_1$ and $\mathcal{J}_n$ represents n-order Bessel function. If $T_2 \neq 0$, the Peierls phase $\theta_{ij}$ is not a linear function of $\mathbf{r}_{ij}$, but
\begin{equation}
\begin{split}
\tan{\theta_{ij}} &= \frac{\sin{\alpha_{ij}}}{\cos{\alpha_{ij}} + \frac{T_2}{T_1 \mathcal{J}_0(\alpha_{ij})}},
\end{split}
\end{equation}
and this gives rise to an inhomogeneous finite artificial magnetic flux through the elementary triangular plaquettes of 2D lattices. The effective Floquet Hamiltonian in momentum space can always be expressed as two distinct parts (details in Appendix~\ref{app:1})
\begin{equation}
\hat{H}_{\mathrm{eff}}(\mathbf{k}) = \lambda_+ I_4 + B_x(\mathbf{k}) \sigma_x + B_y(\mathbf{k}) \sigma_y + B_z(\mathbf{k}) s_z\sigma_z.
\label{eq: model 2}
\end{equation}
and the ground state is always two-fold degenerate. This model exhibits a topological phase transition when the gap closes, achieved by controlling the shaking amplitude $F$, shaking direction $\varphi$, and sublattice potential $\Delta$. The phase diagram and Floquet bands are shown in Fig. \ref{fig:model 2 result}, illustrating the possibilities for realizing a system with non-trivial topological invariants.

To break $s_z$ symmetry but still keep the degeneracy of ground states, we consider a position-dependent onsite spin-flipping term that breaks the conservation of each spin component
\begin{equation}
\begin{split}
H_{\mathrm{SF}} &= \sum_{\mathbf{r}} \left( g_A a_{\mathbf{r},\uparrow}^\dag a_{\mathbf{r},\downarrow} + g_B b_{\mathbf{r},\uparrow}^\dag b_{\mathbf{r},\downarrow} + h.c.\right). \cr
\end{split}
\label{eq: model 2 mass}
\end{equation}
If $g_{A}$ and $g_{B}$ are controllable with a $\pi$ phase difference, we could set $g_{A} = - g_{B} = g_R - i g_I$. The Hamiltonian in Eq.\ref{eq: model 2} with spin-flipping terms can be expressed using Clifford $\Gamma$ matrices: $H(\mathbf{k}) = B_x(\mathbf{k}) \Gamma_{1} + B_y(\mathbf{k}) \Gamma_{12} + B_z(\mathbf{k}) \Gamma_{15} + g_R \Gamma_{13} - g_I \Gamma_{14}$.

The wavefunctions of the valence bands can be written as a tensor product: $\ket{E_i(\mathbf{k})} = \ket{s_i(\mathbf{k})} \otimes \ket{\tau_i(\mathbf{k})}$, where $\ket{s_i}$ and $\ket{\tau_i}$ represent wavefunctions for spin and sublattice, respectively. For the Hamiltonian given by Eq.\ref{eq: model 2} with the spin-flipping term, we can introduce two rotations along the $\sigma_x$ and $\sigma_y$ directions for each spin component simultaneously using $U_x = e^{-i \eta s_x/2} \otimes I_2$ and $U_y = e^{-i \xi s_y/2} \otimes I_2$. By performing the transformation $U^\dag_x U^\dag_y H U_y U_x = \mathcal{H}$, the spin-flipping term can be eliminated, with $\tan{\xi} = g_R/B_z$ and $\tan{\eta} = g_I/\sqrt{B_z^2 + g_R^2}$ (The introduction of $\eta$ is unnecessary when $g_I=0$). The $\Gamma_{15}$ component of the block diagonalized Hamiltonian changes from $B_z$ to $ \widetilde{B}z = g_I \sin{\eta} + \cos{\eta}(B_z \cos{\xi} + g_R \sin{\xi})$. These terms associated with the SOC introduce an additional onsite energy offset and modify the $\Gamma_{15}$ term. The corresponding wavefunctions for spin and sublattice components are:
\begin{equation}
\begin{split}
\ket{s_1} &= \left(\cos{\frac{\eta}{2}}\cos{\frac{\xi}{2}} + i \sin{\frac{\eta}{2}}\sin{\frac{\xi}{2}}\right)\ket{\uparrow} \cr
& + \left(\cos{\frac{\eta}{2}}\sin{\frac{\xi}{2}} - i \sin{\frac{\eta}{2}}\cos{\frac{\xi}{2}}\right) \ket{\downarrow} ,\cr
\ket{s_2} &= -\left(\cos{\frac{\eta}{2}}\sin{\frac{\xi}{2}} + i \sin{\frac{\eta}{2}}\cos{\frac{\xi}{2}}\right) \ket{\uparrow} \cr
&+ \left(\cos{\frac{\eta}{2}}\cos{\frac{\xi}{2}} - i \sin{\frac{\eta}{2}}\sin{\frac{\xi}{2}}\right) \ket{\downarrow} ,\cr
\ket{\tau_1} &= \sin{\frac{\theta}{2}} \ket{A} - e^{i\varphi} \cos{\frac{\theta}{2}}\ket{B},\cr
\ket{\tau_2} &= e^{-i\varphi} \cos{\frac{\theta}{2}} \ket{A} - \sin{\frac{\theta}{2}}\ket{B},\cr
\end{split}
\end{equation}
where $\tan{\theta} = \sqrt{B^2_x + B^2_y}/\widetilde{B}_z$ and $\tan{\varphi} = B_y/B_x$. It should be noted that $\braket{s_i|s_j} = \delta_{ij}$ and $\braket{\tau_i|\tau_j} \neq \delta_{ij}$.


By introducing additional mass terms to break the $\hat{s}_z$ symmetry, we can follow the definition in Section \ref{sec:3} and express $\hat{P} \hat{s}_z \hat{P}$ to obtain a new reduced spin Hamiltonian denoted as $\mathscr{H}(\mathbf{k})$, that
\begin{equation}
\begin{split}
\mathscr{H}(\mathbf{k}) & = \left(\begin{matrix}
M(\mathbf{k}) & t(\mathbf{k})\\
t^\ast(\mathbf{k}) & -M(\mathbf{k})\\
\end{matrix}\right),
\end{split}
\end{equation}
where $M(\mathbf{k}) = \bra{E_{-,1}} \hat{s}_z \ket{E_{-,1}} = -\bra{E_{-,2}} \hat{s}_z \ket{E_{-,2}}$ and $t(\mathbf{k}) = \bra{E_{-,1}} \hat{s}_z \ket{E_{-,2}}$. These parameters can be represented by $\theta$, $\phi$, $\xi$, and $\eta$, with $M = \cos{\eta} \cos{\xi}$ and $t = -e^{-i\varphi}\sin{\theta}(\sin{\xi} + i \cos{\xi}\sin{\eta})$. The gap of $\mathscr{H}$ will only close if both $M$ and $t$ are equal to 0, which corresponds to $\theta = 0$, $\xi = \eta = \pi/2$. The phase diagram takes the same form as the result shown in Fig.~\ref{fig:model 2 result}, and the magnitude of $g$ doesn't relate to the phase transition. Up to this point, we can apply the same approach to calculate the SCN using the reduced spin Hamiltonian in such gapped models.

\section{Detect method in cold atom system}
\label{sec:5}

Cold atom systems with optical lattices provide a clean platform for simulating and studying lattice models in condensed matter physics. Information about the interference pattern of Bloch states in this system can be extracted by measuring the ToF image. In general, the tomography method can provide complete information about the occupied bands. The tomography of a system with many bands is complicated. Therefore, we limit our consideration to a four-band model with two-fold global degeneracy, which simplifies the tomography procedure. In cold atom systems, various quench sequences can be employed to achieve tomography and investigate Bloch states in optical lattices through the measurement of ToF images. In the preceding section, we give the method to realize tomography of two-fold degenercy space.

Firstly, we will outline the theory behind tomography measurements. In the absence of a Rashba SOC term that couples different spin components, information about each spin component can be independently extracted by measuring the ToF image and spin \cite{PhysRevLett.113.045303,doi:10.1126/science.aad4568,APDWZhang2018}. In a finite-size system, the density distribution of the ToF image with spin $\sigma$ in momentum space can be calculated using
\begin{equation}
\begin{split}
n_{\sigma}(\mathbf{k}) &= f(\mathbf{k})\sum_{i,j} e^{-i \mathbf{k} \cdot (\mathbf{r}_i - \mathbf{r}_j)} e^{i(\mu_{i,\sigma} - \mu_{j,\sigma})t} \cr
&\times \bra{G} \left[ a_{i,\sigma}^\dag, b_{i,\sigma}^\dag\right]
\left[ \begin{matrix}
a_{j,\sigma} \\
b_{j,\sigma}
\end{matrix} \right] \ket{G}, \cr
\end{split}
\end{equation}
where $f(\mathbf{k})$ represents a broad envelope function determined by the momentum distribution of the Wannier function, and $\mu_{i,\sigma}$ denotes the global trap potential strength at site $i$. In the absence of a global trapping difference, one has $\mu_{i,\sigma} = \mu_s$.

The density distribution of the ToF image is given by $n_{\sigma}(\mathbf{k}) = f(\mathbf{k})\bra{G}\sigma_0 + \sigma_x\ket{G}$, where $\ket{G}$ represents the many-particle states with Fermi energy $E_f$ in an ideal periodic system without trapping differences. A quench involving the Pauli matrix $\sigma_z$ generates the evolution operator $\exp{(-i \sigma_z t/2)}$, which transforms the ToF image from $\sigma_x$ to $\sigma_x \cos{t} - \sigma_y \sin{t}$. By choosing $t = 0$ and $t = \pi/2$, the expectation values $\braket{\sigma_x}$ and $\braket{\sigma_y}$ can be extracted. Extracting $\braket{\sigma_z}$ can follow a similar approach to that used for the $\sigma_y$ quench. However, in the case of isolated band tomography, one $\sigma_z$ quench is sufficient to recover $\sigma_z$ by analyzing the evolution curve of quench dynamics\cite{PhysRevLett.113.045303,doi:10.1126/science.aad4568}.

\begin{figure*}[tbph]
	\centering\includegraphics[width=16cm]{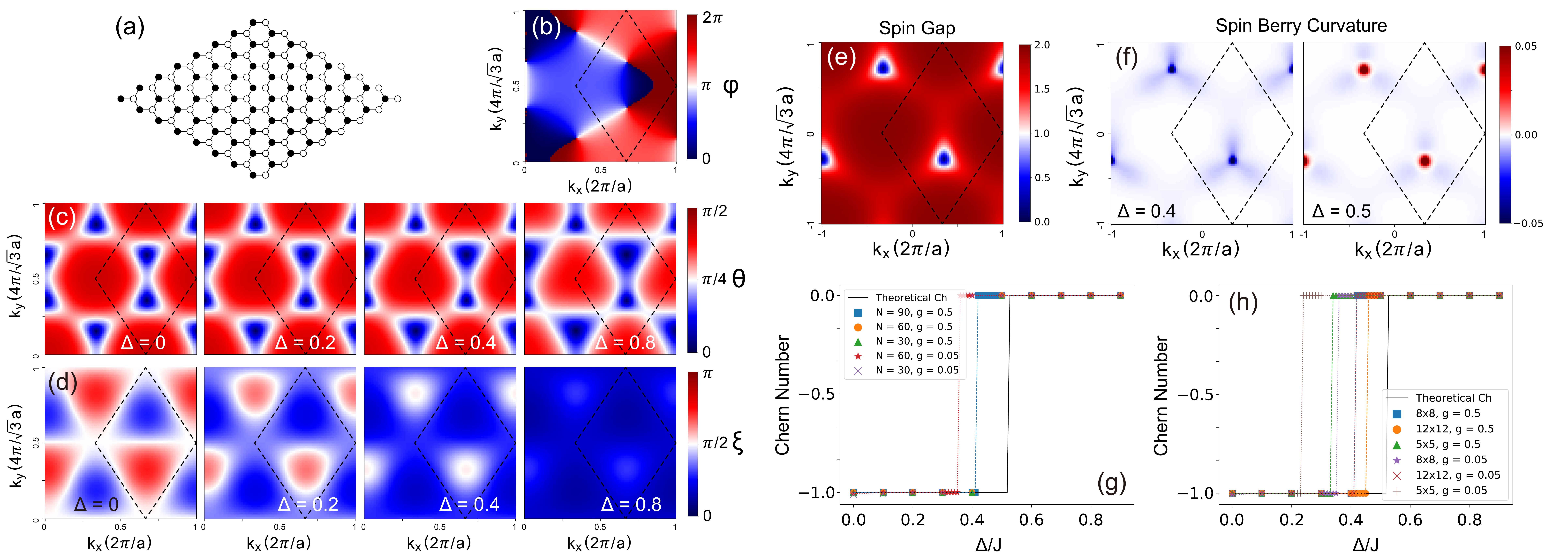}
	\caption{ (a) Calculations are performed on a honeycomb lattice with $N = 8 \times 8 \times 2$ sites and two spin degrees of freedom. Parameters are extracted from the ToF image: (b) $\varphi$, (c) $\theta$, and (d) $\xi$. The black dashed line represents the boundary of BZ. (e) Spin gap near the phase transition point. (f) Spin Berry curvature of the topological and trivial phases. (g) Extracted spin Chern number versus different resolution ratios with strong/weak mass term. (h) Extracted spin Chern number versus different lattice sizes with strong/weak mass term. The calculated parameters are as follows: $J_{NNN}^{\pm} = \pm 0.1 i J_{NN}$, $J_{BB} = 0$, and $g = 0.5 J_{NN}$. }
	\label{fig:tomography}
\end{figure*}

For the four-band model with degeneracy, it is necessary to measure all five components of the Dirac matrices from tomography. In the case of a finite-size system, the overall density distribution of the ToF image in momentum space can be calculated using
\begin{equation}
\begin{split}
n(\mathbf{k}) &= f(\mathbf{k})\sum_{i,j} e^{-i \mathbf{k} \cdot (\mathbf{r}_i - \mathbf{r}_j)} e^{i(\mu_{i,\sigma} - \mu_{j,\sigma})t} \bra{G} C_j^\dag C_i \ket{G}, \cr
\end{split}
\label{eq: 4-TOF}
\end{equation}
where $C_j = [a_{j,\uparrow},b_{j,\uparrow},a_{j,\downarrow},b_{j,\downarrow}]^T$. For an ideal periodic system without trapping differences, the density distribution of the ToF image in Eq.\ref{eq: 4-TOF} can be expressed by $n(\mathbf{k}) = f(\mathbf{k})\bra{G}\Gamma_0 + \Gamma_1 - \Gamma_{23} + \Gamma_{45}\ket{G}$. For example, since the system in above section only has $\Gamma_1$, $\Gamma_{12}$, $\Gamma_{13}$, $\Gamma_{14}$, and $\Gamma_{15}$ components, the expectation value of other components should be zero. In this context, we consider a simplified case where $\eta = 0$ or $g_I = 0$. Three possible quench sequences can simplify the tomography of such a system in an optical lattice: $\Gamma_2$, $\Gamma_{12}$, and $\Gamma_{14}$ (Appendix \ref{app:5}). The parameters $\theta$, $\phi$, and $\xi$ can all be extracted from the ToF images obtained with these different quench sequences.

For the first model generated by Raman lattice, the topological invariants could be extracted directly by measure ToF image of each spin part. For the second model, additional mass terms gap the system but we could still reconstruct the spin Hamiltonian using these parameters from tomography and get a non-trivial SCN.

The SCN can be expressed as momentum-space integrals of the spin Berry curvature $F_{xy}^{(s)}(\mathbf{k}) = \partial_{k_x} A_y^{(s)}(\mathbf{k}) - \partial_{k_y} A_x^{(s)}(\mathbf{k}) $ and connection $A_{\mu}^{(s)}(\mathbf{k}) = i\bra{u_s(\mathbf{k})} \partial_{k_\mu}\ket{u_s(\mathbf{k})}$ associated with the Bloch state $\ket{u_s(\mathbf{k})}$. The integration is over the whole BZ, forming a compact manifold. If the gauge potential $A^{(s)}_{\mu}(\mathbf{k})$ is globally well defined over the BZ, the SCN vanishes because the torus has no boundary. If some topological obstacle exists, the curvature can be solved by Fukui's $U(1)$ link method\cite{JPSJ.74.1674}: $U_{\mu}(\mathbf{k}) = \braket{u(\mathbf{k})|u(\mathbf{k} + \delta \mathbf{k}_{\mu})}/|\braket{u(\mathbf{k})|u(\mathbf{k} + \delta \mathbf{k}_{\mu})}|$ and $\mathcal{F}_{\mu\nu}(\mathbf{k}) = i \ln\left[ U_{\mu}(\mathbf{k}) U_{\nu}(\mathbf{k}+\delta \mathbf{k}_{\mu}) U_{\nu}^{-1}(\mathbf{k}) U_{\mu}^{-1}(\mathbf{k}+\delta \mathbf{k}_{\nu}) \right]$, where $\mathcal{F}_{\mu\nu}(\mathbf{k}) \in \left(-\pi,\pi \right]$ as a discrete version. It can be obtained from the ToF image evolution associated with the pixel $\mathbf{k}$. The topological invariants of the reduced spin Hamiltonian $\mathcal{H}$ can be calculated from $\mathcal{F}^{(s)}_{\mu\nu}(\mathbf{k})$ by directly summing over all pixels of the BZ.

To demonstrate the efficiency of tomography method, we utilize a finite-size system to compute its ToF images. For a quadratic Hamiltonian in real space that is characterized by free fermions, the Hamiltonian is given by
\begin{equation}
H = \sum_{\textbf{r},\textbf{s},\alpha,\beta} \left( a_{\textbf{s},\beta}^\dag , b_{\textbf{s},\beta}^\dag \right) \mathcal{H}_{\textbf{r}\alpha,\textbf{s}\beta}
\left(
\begin{matrix}
a_{\textbf{r},\alpha} \\
b_{\textbf{r},\alpha} \\
\end{matrix}
\right),
\end{equation}
where $\mathbf{r}$ and $\mathbf{s}$ represent position indices, and $\alpha$ and $\beta$ represent spin indices. A unitary operation exists to diagonalize this Hamiltonian, resulting in $\mathcal{H} = U^\dag \Lambda U$. The unitary transformation and the diagonalized Hamiltonian can be expressed as follows:
\begin{equation}
\begin{split}
c_{\textbf{q},\delta} &= \sum_{\textbf{s},\beta} U^{(a \to c)}_{\textbf{q}\delta,\textbf{s}\beta} a_{\textbf{s},\beta} + U^{(b \to c)}_{\textbf{q}\delta,\textbf{s}\beta} b_{\textbf{s},\beta}, \cr
d_{\textbf{q},\delta} &= \sum_{\textbf{s},\beta} U^{(a \to d)}_{\textbf{q}\delta,\textbf{s}\beta} a_{\textbf{s},\beta} + U^{(b \to d)}_{\textbf{q}\delta,\textbf{s}\beta} b_{\textbf{s},\beta}, \cr
\end{split}
\end{equation}
and
\begin{equation}
H = \sum_{\textbf{q},\delta}\Lambda^{(c)}_{\textbf{q},\delta} c_{\textbf{q},\delta}^\dag c_{\textbf{q},\delta} + \Lambda^{(d)}_{\textbf{q},\delta} d_{\textbf{q},\delta}^\dag d_{\textbf{q},\delta},
\end{equation}
where the quasi-particles $c_{\mathbf{q},\delta}$ and $d_{\mathbf{q},\delta}$ correspond to lower and higher energy levels, respectively, with $\delta$ representing the pseudo-spin index. The particle number $\mathcal{N}$ is conserved and defined as $\mathcal{N} = \sum_{\mathbf{r},\alpha} \left( a_{\mathbf{r},\alpha}^\dag a_{\mathbf{r},\alpha} + b_{\mathbf{r},\alpha}^\dag b_{\mathbf{r},\alpha} \right)$. In such a system, particles occupy the top $\mathcal{N}$ eigenmodes with the lowest eigenenergy at absolute zero temperature. Consequently, by manipulating the Fermi energy $E_F$ of this system—effectively controlling $\mathcal{N}$—the system's ground state can be described as $\ket{G} = \prod_{j=1}^{\mathcal{N}_d} d_j^\dag \prod_{i=1}^{\mathcal{N}_c} c_i^\dag \ket{0}$, where indices $i$ and $j$ are arranged based on the energy of the quasi-particles $c_{\mathbf{q},\delta}$ and $d_{\mathbf{q},\delta}$. By substituting the terms $a_{\mathbf{r},\sigma}$ and $b_{\mathbf{r},\sigma}$ into Eq.\ref{eq: 4-TOF}, the momentum density distribution of a finite-size system can be computed (Appendix~\ref{app:3}).

Here, we present the numerical simulations of the model we constructed in Eq.\ref{eq: model 2}, augmented with mass terms Eq.\ref{eq: model 2 mass}. Despite the complex parameters generated by periodic shaking, we focus solely on the model described in Eq.\ref{eq: model 2} with a simplified parameter configuration. Specifically, the NNN hopping is spatially uniform and set at $0.1 J_{NN}$ with a phase of $\pm\pi/2$ from different spin components. We consider a finite lattice with dimensions $8 \times 8 \times 2$ sites, investigate quench dynamics and obtained results displayed in Fig.~\ref{fig:tomography}. The reconstructed spin Hamiltonian exhibits a gapless point if and only if $\theta = 0$ and $\xi = \pi/2$. Analyzing the data in Fig.~\ref{fig:tomography}(c), we observe that $\theta$ consistently features two zero points across the entire BZ. On the other hand, for the parameter $\xi$, as $\Delta$ increases, the $\xi = \pi/2$ regions contract to discrete points. Remarkably, these points also adhere to $\theta = 0$, establishing them as gapless points in the reduced spin Hamiltonian. Consequently, we can find a change in the SCN through calculations of spin-Berry curvature.

The extracted spin gap and spin Berry curvature are illustrated in Fig.~\ref{fig:tomography}(e) and (f). Within the BZ, a single gapless point is observed, and the spin Berry curvature exhibits significant numerical values around these gapless points in Fig.~\ref{fig:tomography}(f). When evaluating the SCN with different resolution ratios and strengths of mass term, it's remarkable that the simulated SCN remains nearly unaffected by the resolution ratio of the ToF image in Fig.~\ref{fig:tomography}(g). However, the strength of mass term does influence the location of the phase transition point. Notably, the phase transition point $\Delta = \pm 3\sqrt{3} J_{NNN}$ remains entirely independent of the parameter $g$, implying that the observed deviations can be viewed as amplification of the effect on system size.

Considering both strong and weak mass term, a comparison of the simulated SCNs across various lattice sizes is presented in Fig~\ref{fig:tomography}(h). With the increasing of system size, the phase transition point approaches the theoretical point. For the effect of mass $g$, across most regions except for the vicinity of $B_z = 0$, $\xi$ predominantly assumes values close to either $0$ or $\pi$. In scenarios involving weak mass term, changes in $\xi$ tend to be abrupt, leading to the need for more and more precise samplings. Consequently, the determination of the accurate phase transition point becomes challenging, resulting in a notable discrepancy between numerical and theoretical outcomes, as evident in Fig.~\ref{fig:tomography}(h).

\section{Discussion and conclusion}
\label{sec:6}

In order to establish a connection with experiments involving cold atoms, we consider $^{6}\mathrm{Li}$ atoms with parameters as reported in \cite{PhysRevLett.118.240403, jotzu2014experimental}. For the lattice configuration, we set the depths to approximately $J_{AB}/2\pi = 5$ kHz, $J_{AA}/2\pi = 500$ Hz, and $J_{BB}/2\pi \ll J_{AA}$. The lattice constant is determined by the wavelength $\lambda = 1064$ nm of the laser beams used to create the optical lattice. The sinusoidal modulation of the lattice position along the $x$ and $y$ directions is characterized by a shaking amplitude of $0.1\lambda$. To capture the dynamics, the modulation frequency $\omega_1/2\pi$ can be configured to exceed $100$ kHz.

In experimental setups, several parameters can introduce perturbations to the ToF image. However, the trapping frequencies of the global weak underlying harmonic confinement are notably smaller than $J_{AB}$ and can thus be safely disregarded within theoretical discussions. In Fig.~\ref{fig:model 2 result}, observing the non-trivial topological phase requires a $\Delta$ range of approximately $0.06 J_{AB}$ using feasible experimental parameters. Correspondingly, this frequency value corresponds to around $300$ Hz. The main factor that affects the energy gap is the mass term that destroys spin conservation. A larger $g$ can effectively suppress thermal fluctuations.

The introduction of a spin-dependent NNN hopping term can be achieved through a global spin-dependent potential. Meanwhile, the creation of opposing artificial magnetic fluxes for each spin component can be realized via optical potential modulation. This artificial magnetic field primarily originates from the zero-order contribution of the intrinsic NNN hopping term and proves easy to prepare and detect, particularly in systems characterized by shallow potentials.

Furthermore, the broader extended Wannier function that we propose here serves a dual purpose: it not only enhances the NNN hopping amplitude but also amplifies the strength of many-body interactions. This amplification can introduce non-negligible effects stemming from many-body terms. However, the impact of these many-body interactions can be mitigated by manipulating the scattering length of atoms through Feshbach resonance.

Additionally, the extension of our work opens avenues to explore topological phase transitions and other lattice model properties by extracting ToF images in future research endeavors. It's worth noting that all the parameters discussed here are actually independent. Altering the sublattice potential $\Delta$ simultaneously affects both NN and NNN hopping, rendering the actual phase diagram more intricate than what is depicted in Fig.~\ref{fig:model 2 result}. Importantly, all these parameters are tunable via adjustments to the optical potential.

\section{Summary}

We introduce achievable simple models demonstrating non-trivial SCN in ultracold atom systems. We derive the spin-Hamiltonian and SCNs by analyzing essential parameters extracted from the quench dynamics of ToF images. It's crucial to highlight that in the topological phase, the orientation of currents in different spin sectors becomes variable, leading to quantized spin Hall currents. Moreover, the same quenching approach can be applied to topological insulators in alternative classes or systems with second-order topology, as long as it can be realized in an ultracold atom system.

In summary, we systematically build two distinct four-band models with non-trivial topological invariants. These models enable measurable SCN in cold-atom experiments, and we propose a method to effectively probe their corresponding topological invariants. The strategies and examples provided in this study offer a foundation for realizing and detecting topological states in cold-atom systems. Additionally, these insights contribute to a deeper understanding of topological excitations and the broader concept of topological order, spanning various fields like condensed-matter physics and artificial systems.

\section*{Acknowledgment}

We thank  Y.Q. Zhu, Peng He, Zhen Zheng, and S. L. Zhu for helpful discussions. This work was supported by the National Natural Science Foundation of China (Grant No. 12074180).

\appendix

\section{Effective Floquet Hamiltonian}
\label{app:1}

\begin{widetext}

Here we explore the scenario of an optical lattice described by a tight-binding model, where each site accommodates one orbital. Additionally, an external gradient field is introduced
\begin{equation}
\begin{split}
H &= \sum_{i,j} C_{i,j}^\dag H_0 C_{i,j} + C_{i+1,j}^\dag T_{\mathbf{e}_1} C_{i,j} + C_{i,j+1}^\dag T_{\mathbf{e}_2} C_{i,j} + h.c. + C_{i,j}^\dag \boldsymbol{F}(t) \cdot \mathbf{r} C_{i,j} \cr
\end{split}
\end{equation}
where $C_{i,j} = [a_{i,j},b_{i,j}]^T$. In our experiments, the modulation consists of displacing the lattice along a periodic trajectory denoted as $\mathbf{r}(t)$. This displacement introduces an inertial force $\boldsymbol{F}(t) = - m \ddot{\mathbf{r}}_{\mathrm{lat}}(t)$ acting on the atoms. To handle this additional time-dependent term, a unitary transformation can be applied to cancel it out
\begin{equation}
\begin{split}
U(t) &= \mathrm{exp} \left[ -i \sum_{i,j} C_{i,j}^\dag m \int_{0}^{t} \mathrm{d}t~\ddot{\mathbf{r}}_{\mathrm{lat}}(t) \cdot \mathbf{r} C_{i,j} \right] = \mathrm{exp} \left[ -i \sum_{i,j} C_{i,j}^\dag m \left( \dot{\mathbf{r}}_{\mathrm{lat}}(t) - \dot{\mathbf{r}}_{\mathrm{lat}}(0) \right) \cdot \mathbf{r} C_{i,j} \right]
\end{split}
\end{equation}
The modulation needs to satisfy condition: it must assume zero values at both the beginning and end of a period. To facilitate our treatment, we introduce the quantity $\mathbf{q}_{\mathrm{lat}} = m [\dot{\mathbf{r}}_{\mathrm{lat}}(t) - \dot{\mathbf{r}}_{\mathrm{lat}}(0)]$. This formulation ensures that the gauge transformation adheres to the condition $U(0) = I$. When spin-flip terms are absent, the tight-binding Hamiltonian for the spin-up component takes the following form:
\begin{equation}
\begin{split}
H_{\uparrow} &= \sum_{i,j} \Delta_{\uparrow} c_{i,j,\uparrow}^\dag c_{i,j,\uparrow} - \Delta_{\uparrow} d_{i,j,\uparrow}^\dag d_{i,j,\uparrow} \cr
& + J_{NN} \left( e^{-i \mathbf{q}_{\mathrm{lat}} \cdot \mathbf{e}_0^{ba} } c_{i,j,\uparrow}^\dag d_{i,j,\uparrow} + e^{-i \mathbf{q}_{\mathrm{lat}} \cdot \mathbf{e}_1^{ba} } c_{i+1,j,\uparrow}^\dag d_{i,j,\uparrow} + e^{-i \mathbf{q}_{\mathrm{lat}} \cdot \mathbf{e}_2^{ba} } c_{i,j+1,\downarrow}^\dag d_{i,j,\uparrow} + h.c. \right) \cr
& + J_{NNN}^{(A)} \left( e^{-i \mathbf{q}_{\mathrm{lat}} \cdot (\mathbf{e}_2^{ba} - \mathbf{e}_0^{ba}) } c_{i+1,j,\uparrow}^\dag c_{i,j,\uparrow} + e^{-i \mathbf{q}_{\mathrm{lat}} \cdot (\mathbf{e}_1^{ba} - \mathbf{e}_2^{ba}) } c_{i-1,j+1,\uparrow}^\dag c_{i,j,\uparrow} + e^{-i \mathbf{q}_{\mathrm{lat}} \cdot (\mathbf{e}_0^{ba} - \mathbf{e}_1^{ba}) } c_{i,j-1,\uparrow}^\dag c_{i,j,\uparrow} \right) \cr
& + J_{NNN}^{(B)} \left( e^{-i \mathbf{q}_{\mathrm{lat}} \cdot (\mathbf{e}_2^{ba} - \mathbf{e}_0^{ba}) } d_{i+1,j,\uparrow}^\dag d_{i,j,\uparrow} + e^{-i \mathbf{q}_{\mathrm{lat}} \cdot (\mathbf{e}_1^{ba} - \mathbf{e}_2^{ba}) } d_{i,j+1,\uparrow}^\dag d_{i,j,\uparrow} + e^{-i \mathbf{q}_{\mathrm{lat}} \cdot (\mathbf{e}_0^{ba} - \mathbf{e}_1^{ba}) } d_{i,j-1,\uparrow}^\dag d_{i,j,\uparrow} \right) + h.c.\cr
\end{split}
\end{equation}
where $\mathbf{e}_{1}^{ba}$($\mathbf{e}_{2}^{ba}$) is distance $\mathbf{r}_{i+1,j}^a - \mathbf{r}_{i,j}^b$($\mathbf{r}_{i,j+1}^a - \mathbf{r}_{i,j}^b$), $\mathbf{e}_{0}^{ba}$ is distance $\mathbf{r}_{i,j}^a - \mathbf{r}_{i,j}^b$. The unitary operator form $t_0$ to $t_0 + T$ can be written as
\begin{equation}
\begin{split}
U(t_0 + nT,t_0) &= \mathcal{T} \mathrm{exp}\left[ -i \int_{t_0}^{t_0 + nT} \mathrm{d}t^\prime~H(t^\prime) \right] = \mathrm{exp}\left[-i H_F(t_0)n T\right]
\end{split}
\end{equation}
and the effective Floquet Hamiltonian in zero and first order are
\begin{subequations}
\begin{eqnarray}
H_F^{(0)}(t_0) &=& \frac{1}{T}\int_{t_0}^{t_0 + T} \mathrm{d} t~H(t) \\
H_F^{(1)}(t_0) &=& \frac{1}{T}\frac{1}{2!i}\int_{t_0}^{t_0 + T} \mathrm{d} t_1 \int_{t_0}^{t_1} \mathrm{d}t_2~[H(t_1),H(t_2)]
\end{eqnarray}
\end{subequations}
Following the calculations outlined in the main text, the effective hopping terms undergo renormalization, obtaining a phase factor denoted as $\theta_{ij}$ in Eq.\ref{eq:10}. The resulting Floquet Bloch Hamiltonian can be conceptually regarded as a composite of two distinct and independent components. For spin-up part
\begin{subequations}
\begin{eqnarray}
B_{x} &=& \frac{J_{AB}}{T} \sum_{i} \mathrm{Re} \left[ T_1 \mathcal{J}_0(\alpha_i) e^{ -i \left( \mathbf{k} \cdot \mathbf{e}_{i} + \alpha_i \right)} + T_2 e^{ -i \mathbf{k} \cdot \mathbf{e}_{i} } \right] \\
B_{y} &=& \frac{J_{AB}}{T} \sum_{i} \mathrm{Im} \left[ T_1 \mathcal{J}_0(\alpha_i) e^{ -i \left( \mathbf{k} \cdot \mathbf{e}_{i} + \alpha_i \right)} + T_2 e^{ -i \mathbf{k} \cdot \mathbf{e}_{i} } \right]
\end{eqnarray}
\end{subequations}
and
\begin{eqnarray}
B_{z} &=& s_z \left[ \Delta + \underbrace{\frac{J_{AA} - J_{BB} }{T} \sum_{j = (i + 1)~\text{mod}~3} \left[ T_1 \mathcal{J}_0(\beta_i) \cos{ \left( \mathbf{k} \cdot (\mathbf{e}_{i} - \mathbf{e}_{i+1}) + \beta_i \right)} + T_2 \cos{ \mathbf{k} \cdot (\mathbf{e}_{i} - \mathbf{e}_{i+1}) } \right]}_{\lambda_-} \right]
\end{eqnarray}
where $\alpha_i = F_0 \hat{\mathbf{n}} \cdot \mathbf{e}_{i}/\omega_1$, $\beta_i = F_0 \hat{\mathbf{n}} \cdot (\mathbf{e}_{i} - \mathbf{e}_{(i + 1)~\text{mod}~3})/\omega_1$ and $\mathbf{e}_{0} = -a \hat{\mathbf{x}},~\mathbf{e}_{1} = \frac{a}{2} \hat{\mathbf{x}} - \frac{\sqrt{3}a}{2} \hat{\mathbf{y}},~\mathbf{e}_{2} = \frac{a}{2} \hat{\mathbf{x}} + \frac{\sqrt{3}a}{2} \hat{\mathbf{y}}$. The difference between spin up and down part is the value of NNN hopping: $J_{AA}^{\uparrow} - J_{BB}^{\uparrow} = J_{BB}^{\downarrow} - J_{AA}^{\downarrow} $. This give rise a Hamiltonian
\begin{equation}
H = \left(\begin{matrix}
\lambda_+ \mathbf{I} + \mathbf{d}_{+}\cdot \boldsymbol{\sigma} & 0 \\
0 & \lambda_+ \mathbf{I} + \mathbf{d}_{-} \cdot \boldsymbol{\sigma} \\
\end{matrix}\right)
\end{equation}
where $\mathbf{d}_{\pm} = \left[B_x, B_y, \pm(\Delta + \lambda_-) \right]$.

\section{Floquet approximation}

\label{app:2}

We have formulated an effective Floquet model as discussed earlier. However, our analysis is confined to the zero-order Floquet approximation. We extend our examination to the effective Floquet Hamiltonian for first order following the approach outlined in \cite{PhysRevX.4.031027}. Generally speaking, a Floquet gauge which has centrosymmetry can lead exact zero 1st term. However, there is no centrosymmetry in such a driving sequences even change the Floquet gauge. By computing the maximum first-order term, we can identify the parameter conditions that validate the zero-order approximation. It is noteworthy that the term $J_{AB}^2$ significantly influences the diagonal elements and serves as the primary contributor to the maximum first-order Floquet Hamiltonian. The magnitude of the maximum first-order term can be expressed as follows:
\begin{equation}
\begin{split}
1st~\mathrm{order} & = J_{AB}^2 \frac{1}{T} \frac{1}{2!i} \int_{0}^{T} d t_1 \exp{\left[ -i \mathbf{q}_{\mathrm{lat}}(t_1) \cdot \mathbf{e}_{i} \right]} \cr
&\times \int_{0}^{t_1} dt_2 \exp{\left[ i \mathbf{q}_{\mathrm{lat}}(t_2) \cdot \mathbf{e}_{j} \right]} \cr
& \le J_{AB}^2 \frac{1}{T} \frac{1}{2!i} \left[ \frac{T_1^2}{2} + T_1(T-T_1) + \frac{1}{2}(T - T_1)^2\right] \cr
& = \frac{\pi J_{AB}^2}{2\omega} \cr
\end{split}
\end{equation}
where all Bessel function $\mathcal{J}_n(z) \le 1$ and driving frequency $\omega_1$ should be sufficiently large to satisfy $\mathcal{J}_0(z) \gg \mathcal{J}_n(z)/\omega_1$. Actually, due to the contribution of $\mathcal{J}_0(\alpha_i)$ and the phase $\alpha_i$, the norm of the first-order term is even smaller than $\pi J_{AB}^2 /2\omega$. Notably, the diagonal term is chiefly influenced by $|J_{AA} - J_{BB}|$, which in our consideration is $0.1 J_{AB}$. Consequently, to ensure that $|J_{AA} - J_{AB}| > \max\left\{1\text{st}~\text{order}\right\}$.

\section{Time-of-Flight interference image Calculation}
\label{app:3}

The correlation function in real space can be calculated by
\begin{equation}
\begin{split}
\bra{G} a_{\textbf{r},\sigma}^\dag a_{\textbf{r}^\prime,\sigma} \ket{G} &= \sum_{\textbf{s},\beta,\textbf{s}^\prime,\beta^\prime} U^{(a \to c)}_{\textbf{s}\beta, \textbf{r}\sigma} (U^{(a \to c)}_{\textbf{s}^\prime\beta^\prime, \textbf{r}^\prime\sigma} )^\ast \bra{G} c_{\textbf{s},\beta}^\dag c_{\textbf{s}^\prime,\beta^\prime} \ket{G} + U^{(a \to d)}_{\textbf{s}\beta, \textbf{r}\alpha} (U^{(a \to d)}_{\textbf{s}^\prime\beta^\prime, \textbf{r}^\prime\alpha^\prime})^\ast \bra{G} d_{\textbf{s},\beta}^\dag d_{\textbf{s}^\prime,\beta^\prime} \ket{G} \cr
\bra{G} b_{\textbf{r},\sigma}^\dag b_{\textbf{r}^\prime,\sigma} \ket{G} &= \sum_{\textbf{s},\beta,\textbf{s}^\prime,\beta^\prime} U^{(b \to c)}_{\textbf{s}\beta, \textbf{r}\sigma} (U^{(b \to c)}_{\textbf{s}^\prime\beta^\prime, \textbf{r}^\prime\sigma} )^\ast \bra{G} c_{\textbf{s},\beta}^\dag c_{\textbf{s}^\prime,\beta^\prime} \ket{G} + U^{(b \to d)}_{\textbf{s}\beta, \textbf{r}\sigma} (U^{(b \to d)}_{\textbf{s}^\prime\beta^\prime, \textbf{r}^\prime\sigma})^\ast \bra{G} d_{\textbf{s},\beta}^\dag d_{\textbf{s}^\prime,\beta^\prime} \ket{G} \cr
\bra{G} a_{\textbf{r},\sigma}^\dag b_{\textbf{r}^\prime,\sigma} \ket{G} &= \sum_{\textbf{s},\beta,\textbf{s}^\prime,\beta^\prime} U^{(a \to c)}_{\textbf{s}\beta, \textbf{r}\sigma} (U^{(b \to c)}_{\textbf{s}^\prime\beta^\prime, \textbf{r}^\prime\sigma} )^\ast \bra{G} c_{\textbf{s},\beta}^\dag c_{\textbf{s}^\prime,\beta^\prime} \ket{G} + U^{(a \to d)}_{\textbf{s}\beta, \textbf{r}\sigma} (U^{(b \to d)}_{\textbf{s}^\prime\beta^\prime, \textbf{r}^\prime\sigma})^\ast \bra{G} d_{\textbf{s},\beta}^\dag d_{\textbf{s}^\prime,\beta^\prime} \ket{G} \cr
\bra{G} b_{\textbf{r},\sigma}^\dag a_{\textbf{r}^\prime,\sigma} \ket{G} &= \left( \bra{G} a_{\textbf{r},\sigma}^\dag b_{\textbf{r}^\prime,\sigma} \ket{G} \right)^\ast
\end{split}
\end{equation}
and in real space, ground state $\ket{G}$ satisfies
\begin{equation}
\begin{split}
\bra{G} c_{\textbf{s},\beta}^\dag c_{\textbf{s}^\prime,\beta^\prime} \ket{G} &= \bra{G} d_{\textbf{s},\beta}^\dag d_{\textbf{s}^\prime,\beta^\prime} \ket{G} = \delta_{\textbf{s},\textbf{s}^\prime} \delta_{\beta,\beta^\prime} \theta(E_F - \mathcal{E}) \cr
\bra{G} c_{\textbf{s},\beta}^\dag d_{\textbf{s}^\prime,\beta^\prime} \ket{G} &= \bra{G} d_{\textbf{s},\beta}^\dag c_{\textbf{s}^\prime,\beta^\prime} \ket{G} = 0
\end{split}
\end{equation}
where $\theta(E_F - \mathcal{E})$ is the Heaviside function which describes the Fermi-Dirac distribution at zero temperature. Thus, the momentum density of a finite-size model can be calculated by
\begin{equation}
\begin{split}
\bra{G} a_{\textbf{r},\alpha}^\dag a_{\textbf{r}^\prime,\alpha^\prime} \ket{G} &= \sum_{\textbf{s},\beta} U^{(a \to c)}_{\textbf{s}\beta, \textbf{r}\sigma} (U^{(a \to c)}_{\textbf{s}\beta, \textbf{r}^\prime\sigma} )^\ast \cr
\bra{G} b_{\textbf{r},\alpha}^\dag b_{\textbf{r}^\prime,\alpha^\prime} \ket{G} &= \sum_{\textbf{s},\beta} U^{(b \to c)}_{\textbf{s}\beta, \textbf{r}\sigma} (U^{(b \to d)}_{\textbf{s}\beta, \textbf{r}^\prime\sigma} )^\ast \cr
\bra{G} a_{\textbf{r},\sigma}^\dag b_{\textbf{r}^\prime,\sigma} \ket{G} &= \sum_{\textbf{s},\beta} U^{(a \to c)}_{\textbf{s}\beta, \textbf{r}\sigma} (U^{(b \to c)}_{\textbf{s}\beta, \textbf{r}^\prime\sigma} )^\ast \cr
\bra{G} b_{\textbf{r},\sigma}^\dag a_{\textbf{r}^\prime,\sigma} \ket{G} &= \left( \bra{G} a_{\textbf{r}^\prime,\sigma}^\dag b_{\textbf{r},\sigma} \ket{G} \right)^\ast
\end{split}
\end{equation}

\section{Tomography procedure}
\label{app:4}

In the context of the two-fold globally degenerate system described by Eq.\ref{eq: model 2} and Eq.\ref{eq: model 2 mass}, and considering the ground state as $\ket{G} = (\prod_{\mathbf{k} \in \text{BZ}} c^\dag_{-,2}) (\prod_{\mathbf{k} \in \text{BZ}} c^\dag_{-,1}) \ket{0}$, it suffices to measure $\braket{\Gamma_1}$, $\braket{\Gamma_{12}}$, $\braket{\Gamma_{13}}$, and $\braket{\Gamma_{15}}$ in order to complete the entire tomography process. The theoretical values for the ToF image under periodic boundary conditions are as follows:
\begin{equation}
\begin{split}
\braket{\Gamma_1} &= - 2 \cos{\varphi} \sin{\theta},~\braket{\Gamma_{12}} = - 2 \sin{\varphi} \sin{\theta} \cr
\braket{\Gamma_{13}} &= - 2 \sin{\xi} \cos{\theta},~\braket{\Gamma_{15}} = - 2 \cos{\xi} \cos{\theta} \cr
\end{split}
\end{equation}
It's important to note that all other components of the Gamma matrices are zero. The total density distribution of the ToF image in momentum space can be calculated by
\begin{equation}
\begin{split}
n(\mathbf{k}) &= f(\mathbf{k}) \braket{\Gamma_0 + \Gamma_1 - \Gamma_{23} + \Gamma_{45}} \cr
& = f(\mathbf{k}) \left( 2 - 2\cos{\varphi} \sin{\theta} \right)
\end{split}
\end{equation}
Here, we contemplate a quench that transforms the initial ground state $\ket{G}$ into a new state $\ket{\widetilde{G}} = \exp{(-i \Gamma t/2)} \ket{G}$. In the absence of a quench, we have $n(k) = \braket{\Gamma_0} + \braket{\Gamma_1}$. Considering all possible quench scenarios that can provide relevant information, we present them as follows:
\begin{equation}
\begin{split}
\Gamma_2: n(\mathbf{k},t) &= \braket{\Gamma_0} + \cos{t} \braket{\Gamma_1} + \sin{t} \braket{\Gamma_{12}} \cr
\Gamma_3: n(\mathbf{k},t) &= \braket{\Gamma_0} + \cos{t} \braket{\Gamma_1} + \sin{t} \braket{\Gamma_{13}} \cr
\Gamma_5: n(\mathbf{k},t) &= \braket{\Gamma_0} + \cos{t} \braket{\Gamma_1} + \sin{t} \braket{\Gamma_{15}} \cr
\Gamma_{12}: n(\mathbf{k},t) &= \braket{\Gamma_0} + \cos{t} \braket{\Gamma_1} - \sin{t} \braket{\Gamma_{13}} \cr
\Gamma_{13}: n(\mathbf{k},t) &= \braket{\Gamma_0} + \cos{t} \braket{\Gamma_1} + \sin{t} \braket{\Gamma_{12}} \cr
\Gamma_{14}: n(\mathbf{k},t) &= \braket{\Gamma_0} + \cos{t} \braket{\Gamma_1} + \sin{t} \braket{\Gamma_{15}} \cr
\end{split}
\end{equation}
By setting $t = 0,\pi/2,\pi$, all components can be exacted from
\begin{equation}
\frac{ n(\pi/2)}{n(0) + n(\pi)} = \frac{\braket{\Gamma_0} + \braket{\Gamma}}{2\braket{\Gamma_0}} = \frac{1}{2}\left(1 + \frac{\braket{\Gamma}}{\braket{\Gamma_0}}\right)
\end{equation}
The contribution of the broad envelope can be evaluated as background by calculating the sum at $t = 0$ and $t = \pi$.

\section{Experimental setup of optical lattice and Quench dynamics}
\label{app:5}

In order to realize different types of quench smoothly, the optical lattice can be formed by stacked hexagonal optical potential. To achieve a $\Gamma_2$ type quench, one part is formed by the setting in model 2 in Section \ref{sec:4} and another part can be achieved by three detuned standing-wave laser fields that are rotated by angles of $\pi/3$ with respect to each other:
\begin{equation}
\begin{split}
\mathcal{E}_j(\mathbf{r},t) &= E_0 e^{-i\omega^\prime t} \sin{(\mathbf{k}^\prime_i \cdot \mathbf{r} + \phi_i )} \mathbf{e}_z \cr
\end{split}
\end{equation}
and $V_s (\mathbf{r}) = - \alpha_s |\sum_{j=1}^{3} \mathcal{E}_j(\mathbf{r})|^2 $. By appropriately selecting values for the phases $\phi_i$, we can manipulate the sublattice detuning parameter $\Delta$. Specifically, for the case of a $\Gamma_2$ quench, the objective is to enhance the strength of $E_0$ to effectively suppress nearest-neighbor (NN) hopping interactions. Simultaneously, a non-zero value of $\phi_i$ is introduced, which introduces sublattice energy offsets. This introduction of an energy difference between different sublattices leads to the accumulation of a relative phase shift: $ a^\dag_{i,j} \to e^{-i\varphi} a^\dag_{i,j}$ and $b^\dag_{i,j} \to e^{i\varphi} b^\dag_{i,j}$. Consequently, this phase alteration leads to changes in the observed pattern within the ToF image.

Just as we assume when introducing the mass term, if we can achieve control over spin at the single-site level, we could manipulate the phase difference and achieve quenching in $\Gamma_{13}$ and $\Gamma_{14}$.

In the case of $\Gamma_{12}$, we require an additional dimerized optical lattice along one direction. This lattice can be achieved by introducing laser imbalance
\begin{equation}
\begin{split}
\mathcal{E}_1(\mathbf{r},t) &= \beta E_0 e^{i(\mathbf{k}_1 \cdot \mathbf{r} - \phi_1) - i \omega_L t} \mathbf{e}_x \cr
\mathcal{E}_2(\mathbf{r},t) &= -E_0 e^{i(\mathbf{k}_2 \cdot \mathbf{r} - \phi_2) - i \omega_L t} \left( \frac{1}{2} \mathbf{e}_x - \frac{\sqrt{3}}{2}\mathbf{e}_y \right) \cr
\mathcal{E}_3(\mathbf{r},t) &= -E_0 e^{i(\mathbf{k}_3 \cdot \mathbf{r} - \phi_3) - i \omega_L t} \left( \frac{1}{2} \mathbf{e}_x + \frac{\sqrt{3}}{2}\mathbf{e}_y \right) \cr
\end{split}
\end{equation}
By choosing the appropriate $\beta$, we can inhibit hopping along two directions and maintain the inter-sublattice hopping term. Further periodic driving can control the complex tunneling amplitude between sublattices. The tight-binding quench Hamiltonian with effective magnetic field we realized is
\begin{equation}
\begin{split}
H = \left( \sum_{i,j} J_{AB} e^{i\phi_{AB}} a_{i,j}^\dag b_{i,j} + h.c. \right) + \Delta \sum_{i,j} n_{i,j} s_z \sigma_z
\end{split}
\end{equation}
and its bloch Hamiltonian is
\begin{equation}
H(\mathbf{k}) = \Delta \Gamma_{15} + J_{AB} \cos{(\mathbf{k} \cdot \mathbf{e}_{ab} - \phi_{AB} )} \Gamma_{1} + J_{AB} \sin{(\mathbf{k} \cdot \mathbf{e}_{ab} - \phi_{AB} )} \Gamma_{12}
\end{equation}
where $\mathbf{e}_{ab} = \mathbf{r}_a - \mathbf{r}_b$. The momentum-dependent Hamiltonian gives rise to an evolution operator $U(\mathbf{k},t) = \mathrm{exp}[-i H(\mathbf{k})t]$:
\begin{equation}
\begin{split}
U(\mathbf{k},t) &= \cos{(\widetilde{J} t)} \Gamma_0 - i \frac{B}{2\widetilde{J}} \sin{(\widetilde{J} t)} \Gamma_{15} \cr
& - i \frac{J_{AB}}{2 \widetilde{J}} \cos{(\mathbf{k} \cdot \mathbf{e}_{ab} - \phi_{AB} )} \sin{(\widetilde{J} t)} \Gamma_{1} \cr
& - i \frac{J_{AB}}{2 \widetilde{J}} \sin{(\mathbf{k} \cdot \mathbf{e}_{ab} - \phi_{AB} )} \sin{(\widetilde{J} t)} \Gamma_{12} \cr
\end{split}
\end{equation}
where $\widetilde{J} = \sqrt{J^2_{AB} + \Delta^2}/2 $ that exhibits a momentum $\mathbf{k}$-dependent quench. For fixed $\mathbf{k}$ results, the $\mathbf{k}$-independent result $\braket{\Gamma_{12}}$ can be obtained by setting $(\mathbf{k} \cdot \mathbf{e}_{ab} - \phi_{AB}) = \pi/2$ and the theoretical ToF image is
\begin{equation}
\begin{split}
n_{ToF}(t) = \braket{\Gamma_0} + \cos{(4\widetilde{J} t)} \braket{\Gamma_{1}} + \frac{J_{AB}}{2\widetilde{J}}\sin{(4 \widetilde{J} t)} \braket{\Gamma_{13}}
\end{split}
\end{equation}
and the $\braket{\Gamma_{13}}$ part can be extracted by setting $\widetilde{J} t = \pi/8$. At each step, we can retain only the data from one column with momentum $\mathbf{k}_{\perp}$ perpendicular to $\mathbf{e}_{ab}$ to obtain the ToF image.

\end{widetext}

\nocite{*}
\bibliography{REV}

\begin{thebibliography}{91}%
\makeatletter
\providecommand \@ifxundefined [1]{%
 \@ifx{#1\undefined}
}%
\providecommand \@ifnum [1]{%
 \ifnum #1\expandafter \@firstoftwo
 \else \expandafter \@secondoftwo
 \fi
}%
\providecommand \@ifx [1]{%
 \ifx #1\expandafter \@firstoftwo
 \else \expandafter \@secondoftwo
 \fi
}%
\providecommand \natexlab [1]{#1}%
\providecommand \enquote  [1]{``#1''}%
\providecommand \bibnamefont  [1]{#1}%
\providecommand \bibfnamefont [1]{#1}%
\providecommand \citenamefont [1]{#1}%
\providecommand \href@noop [0]{\@secondoftwo}%
\providecommand \href [0]{\begingroup \@sanitize@url \@href}%
\providecommand \@href[1]{\@@startlink{#1}\@@href}%
\providecommand \@@href[1]{\endgroup#1\@@endlink}%
\providecommand \@sanitize@url [0]{\catcode `\\12\catcode `\$12\catcode `\&12\catcode `\#12\catcode `\^12\catcode `\_12\catcode `\%12\relax}%
\providecommand \@@startlink[1]{}%
\providecommand \@@endlink[0]{}%
\providecommand \url  [0]{\begingroup\@sanitize@url \@url }%
\providecommand \@url [1]{\endgroup\@href {#1}{\urlprefix }}%
\providecommand \urlprefix  [0]{URL }%
\providecommand \Eprint [0]{\href }%
\providecommand \doibase [0]{http://dx.doi.org/}%
\providecommand \selectlanguage [0]{\@gobble}%
\providecommand \bibinfo  [0]{\@secondoftwo}%
\providecommand \bibfield  [0]{\@secondoftwo}%
\providecommand \translation [1]{[#1]}%
\providecommand \BibitemOpen [0]{}%
\providecommand \bibitemStop [0]{}%
\providecommand \bibitemNoStop [0]{.\EOS\space}%
\providecommand \EOS [0]{\spacefactor3000\relax}%
\providecommand \BibitemShut  [1]{\csname bibitem#1\endcsname}%
\let\auto@bib@innerbib\@empty
\bibitem [{\citenamefont {Hasan}\ and\ \citenamefont {Kane}(2010)}]{RevModPhys.82.3045}%
  \BibitemOpen
  \bibfield  {author} {\bibinfo {author} {\bibfnamefont {M.~Z.}\ \bibnamefont {Hasan}}\ and\ \bibinfo {author} {\bibfnamefont {C.~L.}\ \bibnamefont {Kane}},\ }\bibfield  {title} {\enquote {\bibinfo {title} {Colloquium: Topological insulators},}\ }\href {\doibase 10.1103/RevModPhys.82.3045} {\bibfield  {journal} {\bibinfo  {journal} {Rev. Mod. Phys.}\ }\textbf {\bibinfo {volume} {82}},\ \bibinfo {pages} {3045--3067} (\bibinfo {year} {2010})}\BibitemShut {NoStop}%
\bibitem [{\citenamefont {Qi}\ and\ \citenamefont {Zhang}(2011)}]{RevModPhys.83.1057}%
  \BibitemOpen
  \bibfield  {author} {\bibinfo {author} {\bibfnamefont {Xiao-Liang}\ \bibnamefont {Qi}}\ and\ \bibinfo {author} {\bibfnamefont {Shou-Cheng}\ \bibnamefont {Zhang}},\ }\bibfield  {title} {\enquote {\bibinfo {title} {Topological insulators and superconductors},}\ }\href {\doibase 10.1103/RevModPhys.83.1057} {\bibfield  {journal} {\bibinfo  {journal} {Rev. Mod. Phys.}\ }\textbf {\bibinfo {volume} {83}},\ \bibinfo {pages} {1057--1110} (\bibinfo {year} {2011})}\BibitemShut {NoStop}%
\bibitem [{\citenamefont {Bernevig}\ and\ \citenamefont {Zhang}(2006)}]{PhysRevLett.96.106802}%
  \BibitemOpen
  \bibfield  {author} {\bibinfo {author} {\bibfnamefont {B.~Andrei}\ \bibnamefont {Bernevig}}\ and\ \bibinfo {author} {\bibfnamefont {Shou-Cheng}\ \bibnamefont {Zhang}},\ }\bibfield  {title} {\enquote {\bibinfo {title} {Quantum spin hall effect},}\ }\href {\doibase 10.1103/PhysRevLett.96.106802} {\bibfield  {journal} {\bibinfo  {journal} {Phys. Rev. Lett.}\ }\textbf {\bibinfo {volume} {96}},\ \bibinfo {pages} {106802} (\bibinfo {year} {2006})}\BibitemShut {NoStop}%
\bibitem [{\citenamefont {Weimer}\ \emph {et~al.}(2010)\citenamefont {Weimer}, \citenamefont {M{\"u}ller}, \citenamefont {Lesanovsky}, \citenamefont {Zoller},\ and\ \citenamefont {B{\"u}chler}}]{Weimer2010}%
  \BibitemOpen
  \bibfield  {author} {\bibinfo {author} {\bibfnamefont {Hendrik}\ \bibnamefont {Weimer}}, \bibinfo {author} {\bibfnamefont {Markus}\ \bibnamefont {M{\"u}ller}}, \bibinfo {author} {\bibfnamefont {Igor}\ \bibnamefont {Lesanovsky}}, \bibinfo {author} {\bibfnamefont {Peter}\ \bibnamefont {Zoller}}, \ and\ \bibinfo {author} {\bibfnamefont {Hans~Peter}\ \bibnamefont {B{\"u}chler}},\ }\bibfield  {title} {\enquote {\bibinfo {title} {A rydberg quantum simulator},}\ }\href {\doibase 10.1038/nphys1614} {\bibfield  {journal} {\bibinfo  {journal} {Nature Physics}\ }\textbf {\bibinfo {volume} {6}},\ \bibinfo {pages} {382--388} (\bibinfo {year} {2010})}\BibitemShut {NoStop}%
\bibitem [{\citenamefont {Barreiro}\ \emph {et~al.}(2011)\citenamefont {Barreiro}, \citenamefont {M{\"u}ller}, \citenamefont {Schindler}, \citenamefont {Nigg}, \citenamefont {Monz}, \citenamefont {Chwalla}, \citenamefont {Hennrich}, \citenamefont {Roos}, \citenamefont {Zoller},\ and\ \citenamefont {Blatt}}]{barreiro2011open}%
  \BibitemOpen
  \bibfield  {author} {\bibinfo {author} {\bibfnamefont {Julio~T}\ \bibnamefont {Barreiro}}, \bibinfo {author} {\bibfnamefont {Markus}\ \bibnamefont {M{\"u}ller}}, \bibinfo {author} {\bibfnamefont {Philipp}\ \bibnamefont {Schindler}}, \bibinfo {author} {\bibfnamefont {Daniel}\ \bibnamefont {Nigg}}, \bibinfo {author} {\bibfnamefont {Thomas}\ \bibnamefont {Monz}}, \bibinfo {author} {\bibfnamefont {Michael}\ \bibnamefont {Chwalla}}, \bibinfo {author} {\bibfnamefont {Markus}\ \bibnamefont {Hennrich}}, \bibinfo {author} {\bibfnamefont {Christian~F}\ \bibnamefont {Roos}}, \bibinfo {author} {\bibfnamefont {Peter}\ \bibnamefont {Zoller}}, \ and\ \bibinfo {author} {\bibfnamefont {Rainer}\ \bibnamefont {Blatt}},\ }\bibfield  {title} {\enquote {\bibinfo {title} {An open-system quantum simulator with trapped ions},}\ }\href {\doibase 10.1038/nature09801} {\bibfield  {journal} {\bibinfo  {journal} {Nature}\ }\textbf {\bibinfo {volume} {470}},\ \bibinfo {pages} {486--491} (\bibinfo {year} {2011})}\BibitemShut {NoStop}%
\bibitem [{\citenamefont {Zhu}\ \emph {et~al.}(2006{\natexlab{a}})\citenamefont {Zhu}, \citenamefont {Monroe},\ and\ \citenamefont {Duan}}]{PRL2006.97}%
  \BibitemOpen
  \bibfield  {author} {\bibinfo {author} {\bibfnamefont {Shi-Liang}\ \bibnamefont {Zhu}}, \bibinfo {author} {\bibfnamefont {C.}~\bibnamefont {Monroe}}, \ and\ \bibinfo {author} {\bibfnamefont {L.-M.}\ \bibnamefont {Duan}},\ }\bibfield  {title} {\enquote {\bibinfo {title} {Trapped ion quantum computation with transverse phonon modes},}\ }\href {\doibase 10.1103/physrevlett.97.050505} {\bibfield  {journal} {\bibinfo  {journal} {Phys. Rev. Lett.}\ }\textbf {\bibinfo {volume} {97}},\ \bibinfo {pages} {050505} (\bibinfo {year} {2006}{\natexlab{a}})}\BibitemShut {NoStop}%
\bibitem [{\citenamefont {Nakamura}\ \emph {et~al.}(1999)\citenamefont {Nakamura}, \citenamefont {Pashkin},\ and\ \citenamefont {Tsai}}]{nakamura1999coherent}%
  \BibitemOpen
  \bibfield  {author} {\bibinfo {author} {\bibfnamefont {Yasunobu}\ \bibnamefont {Nakamura}}, \bibinfo {author} {\bibfnamefont {Yu~A}\ \bibnamefont {Pashkin}}, \ and\ \bibinfo {author} {\bibfnamefont {JS}~\bibnamefont {Tsai}},\ }\bibfield  {title} {\enquote {\bibinfo {title} {Coherent control of macroscopic quantum states in a single-cooper-pair box},}\ }\href {https://www.nature.com/articles/19718} {\bibfield  {journal} {\bibinfo  {journal} {nature}\ }\textbf {\bibinfo {volume} {398}},\ \bibinfo {pages} {786--788} (\bibinfo {year} {1999})}\BibitemShut {NoStop}%
\bibitem [{\citenamefont {Kitagawa}\ \emph {et~al.}(2012)\citenamefont {Kitagawa}, \citenamefont {Broome}, \citenamefont {Fedrizzi}, \citenamefont {Rudner}, \citenamefont {Berg}, \citenamefont {Kassal}, \citenamefont {Aspuru-Guzik}, \citenamefont {Demler},\ and\ \citenamefont {White}}]{kitagawa2012observation}%
  \BibitemOpen
  \bibfield  {author} {\bibinfo {author} {\bibfnamefont {Takuya}\ \bibnamefont {Kitagawa}}, \bibinfo {author} {\bibfnamefont {Matthew~A}\ \bibnamefont {Broome}}, \bibinfo {author} {\bibfnamefont {Alessandro}\ \bibnamefont {Fedrizzi}}, \bibinfo {author} {\bibfnamefont {Mark~S}\ \bibnamefont {Rudner}}, \bibinfo {author} {\bibfnamefont {Erez}\ \bibnamefont {Berg}}, \bibinfo {author} {\bibfnamefont {Ivan}\ \bibnamefont {Kassal}}, \bibinfo {author} {\bibfnamefont {Al{\'a}n}\ \bibnamefont {Aspuru-Guzik}}, \bibinfo {author} {\bibfnamefont {Eugene}\ \bibnamefont {Demler}}, \ and\ \bibinfo {author} {\bibfnamefont {Andrew~G}\ \bibnamefont {White}},\ }\bibfield  {title} {\enquote {\bibinfo {title} {Observation of topologically protected bound states in photonic quantum walks},}\ }\href {\doibase 10.1038/ncomms1872} {\bibfield  {journal} {\bibinfo  {journal} {Nature communications}\ }\textbf {\bibinfo {volume} {3}},\ \bibinfo {pages} {1--7} (\bibinfo {year} {2012})}\BibitemShut {NoStop}%
\bibitem [{\citenamefont {Bloch}\ \emph {et~al.}(2008)\citenamefont {Bloch}, \citenamefont {Dalibard},\ and\ \citenamefont {Zwerger}}]{RevModPhys.80.885}%
  \BibitemOpen
  \bibfield  {author} {\bibinfo {author} {\bibfnamefont {Immanuel}\ \bibnamefont {Bloch}}, \bibinfo {author} {\bibfnamefont {Jean}\ \bibnamefont {Dalibard}}, \ and\ \bibinfo {author} {\bibfnamefont {Wilhelm}\ \bibnamefont {Zwerger}},\ }\bibfield  {title} {\enquote {\bibinfo {title} {Many-body physics with ultracold gases},}\ }\href {\doibase 10.1103/RevModPhys.80.885} {\bibfield  {journal} {\bibinfo  {journal} {Rev. Mod. Phys.}\ }\textbf {\bibinfo {volume} {80}},\ \bibinfo {pages} {885--964} (\bibinfo {year} {2008})}\BibitemShut {NoStop}%
\bibitem [{\citenamefont {Lewenstein}\ \emph {et~al.}(2007)\citenamefont {Lewenstein}, \citenamefont {Sanpera}, \citenamefont {Ahufinger}, \citenamefont {Damski}, \citenamefont {Sen},\ and\ \citenamefont {Sen}}]{lewenstein2007ultracold}%
  \BibitemOpen
  \bibfield  {author} {\bibinfo {author} {\bibfnamefont {Maciej}\ \bibnamefont {Lewenstein}}, \bibinfo {author} {\bibfnamefont {Anna}\ \bibnamefont {Sanpera}}, \bibinfo {author} {\bibfnamefont {Veronica}\ \bibnamefont {Ahufinger}}, \bibinfo {author} {\bibfnamefont {Bogdan}\ \bibnamefont {Damski}}, \bibinfo {author} {\bibfnamefont {Aditi}\ \bibnamefont {Sen}}, \ and\ \bibinfo {author} {\bibfnamefont {Ujjwal}\ \bibnamefont {Sen}},\ }\bibfield  {title} {\enquote {\bibinfo {title} {Ultracold atomic gases in optical lattices: mimicking condensed matter physics and beyond},}\ }\href {\doibase 10.1080/00018730701223200} {\bibfield  {journal} {\bibinfo  {journal} {Advances in Physics}\ }\textbf {\bibinfo {volume} {56}},\ \bibinfo {pages} {243--379} (\bibinfo {year} {2007})}\BibitemShut {NoStop}%
\bibitem [{\citenamefont {Zhang}\ \emph {et~al.}(2018)\citenamefont {Zhang}, \citenamefont {Zhu}, \citenamefont {Zhao}, \citenamefont {Yan},\ and\ \citenamefont {Zhu}}]{APDWZhang2018}%
  \BibitemOpen
  \bibfield  {author} {\bibinfo {author} {\bibfnamefont {D.-W.}\ \bibnamefont {Zhang}}, \bibinfo {author} {\bibfnamefont {Y.-Q.}\ \bibnamefont {Zhu}}, \bibinfo {author} {\bibfnamefont {Y.~X.}\ \bibnamefont {Zhao}}, \bibinfo {author} {\bibfnamefont {Hui}\ \bibnamefont {Yan}}, \ and\ \bibinfo {author} {\bibfnamefont {S.-L.}\ \bibnamefont {Zhu}},\ }\bibfield  {title} {\enquote {\bibinfo {title} {Topological quantum matter with cold atoms},}\ }\href {\doibase 10.1080/00018732.2019.1594094} {\bibfield  {journal} {\bibinfo  {journal} {Advances in Physics}\ }\textbf {\bibinfo {volume} {67}},\ \bibinfo {pages} {253} (\bibinfo {year} {2018})}\BibitemShut {NoStop}%
\bibitem [{\citenamefont {Fetter}(2009)}]{RevModPhys.81.647}%
  \BibitemOpen
  \bibfield  {author} {\bibinfo {author} {\bibfnamefont {Alexander~L.}\ \bibnamefont {Fetter}},\ }\bibfield  {title} {\enquote {\bibinfo {title} {Rotating trapped bose-einstein condensates},}\ }\href {\doibase 10.1103/RevModPhys.81.647} {\bibfield  {journal} {\bibinfo  {journal} {Rev. Mod. Phys.}\ }\textbf {\bibinfo {volume} {81}},\ \bibinfo {pages} {647--691} (\bibinfo {year} {2009})}\BibitemShut {NoStop}%
\bibitem [{\citenamefont {S\o{}rensen}\ \emph {et~al.}(2005)\citenamefont {S\o{}rensen}, \citenamefont {Demler},\ and\ \citenamefont {Lukin}}]{PhysRevLett.94.086803}%
  \BibitemOpen
  \bibfield  {author} {\bibinfo {author} {\bibfnamefont {Anders~S.}\ \bibnamefont {S\o{}rensen}}, \bibinfo {author} {\bibfnamefont {Eugene}\ \bibnamefont {Demler}}, \ and\ \bibinfo {author} {\bibfnamefont {Mikhail~D.}\ \bibnamefont {Lukin}},\ }\bibfield  {title} {\enquote {\bibinfo {title} {Fractional quantum hall states of atoms in optical lattices},}\ }\href {\doibase 10.1103/PhysRevLett.94.086803} {\bibfield  {journal} {\bibinfo  {journal} {Phys. Rev. Lett.}\ }\textbf {\bibinfo {volume} {94}},\ \bibinfo {pages} {086803} (\bibinfo {year} {2005})}\BibitemShut {NoStop}%
\bibitem [{\citenamefont {Lim}\ \emph {et~al.}(2008)\citenamefont {Lim}, \citenamefont {Smith},\ and\ \citenamefont {Hemmerich}}]{PhysRevLett.100.130402}%
  \BibitemOpen
  \bibfield  {author} {\bibinfo {author} {\bibfnamefont {Lih-King}\ \bibnamefont {Lim}}, \bibinfo {author} {\bibfnamefont {C.~Morais}\ \bibnamefont {Smith}}, \ and\ \bibinfo {author} {\bibfnamefont {Andreas}\ \bibnamefont {Hemmerich}},\ }\bibfield  {title} {\enquote {\bibinfo {title} {Staggered-vortex superfluid of ultracold bosons in an optical lattice},}\ }\href {\doibase 10.1103/PhysRevLett.100.130402} {\bibfield  {journal} {\bibinfo  {journal} {Phys. Rev. Lett.}\ }\textbf {\bibinfo {volume} {100}},\ \bibinfo {pages} {130402} (\bibinfo {year} {2008})}\BibitemShut {NoStop}%
\bibitem [{\citenamefont {Lim}\ \emph {et~al.}(2010)\citenamefont {Lim}, \citenamefont {Lazarides}, \citenamefont {Hemmerich},\ and\ \citenamefont {Morais~Smith}}]{PhysRevA.82.013616}%
  \BibitemOpen
  \bibfield  {author} {\bibinfo {author} {\bibfnamefont {Lih-King}\ \bibnamefont {Lim}}, \bibinfo {author} {\bibfnamefont {Achilleas}\ \bibnamefont {Lazarides}}, \bibinfo {author} {\bibfnamefont {Andreas}\ \bibnamefont {Hemmerich}}, \ and\ \bibinfo {author} {\bibfnamefont {C.}~\bibnamefont {Morais~Smith}},\ }\bibfield  {title} {\enquote {\bibinfo {title} {Competing pairing states for ultracold fermions in optical lattices with an artificial staggered magnetic field},}\ }\href {\doibase 10.1103/PhysRevA.82.013616} {\bibfield  {journal} {\bibinfo  {journal} {Phys. Rev. A}\ }\textbf {\bibinfo {volume} {82}},\ \bibinfo {pages} {013616} (\bibinfo {year} {2010})}\BibitemShut {NoStop}%
\bibitem [{\citenamefont {Kitagawa}\ \emph {et~al.}(2010)\citenamefont {Kitagawa}, \citenamefont {Berg}, \citenamefont {Rudner},\ and\ \citenamefont {Demler}}]{PhysRevB.82.235114}%
  \BibitemOpen
  \bibfield  {author} {\bibinfo {author} {\bibfnamefont {Takuya}\ \bibnamefont {Kitagawa}}, \bibinfo {author} {\bibfnamefont {Erez}\ \bibnamefont {Berg}}, \bibinfo {author} {\bibfnamefont {Mark}\ \bibnamefont {Rudner}}, \ and\ \bibinfo {author} {\bibfnamefont {Eugene}\ \bibnamefont {Demler}},\ }\bibfield  {title} {\enquote {\bibinfo {title} {Topological characterization of periodically driven quantum systems},}\ }\href {\doibase 10.1103/PhysRevB.82.235114} {\bibfield  {journal} {\bibinfo  {journal} {Phys. Rev. B}\ }\textbf {\bibinfo {volume} {82}},\ \bibinfo {pages} {235114} (\bibinfo {year} {2010})}\BibitemShut {NoStop}%
\bibitem [{\citenamefont {Mei}\ \emph {et~al.}(2012)\citenamefont {Mei}, \citenamefont {Zhu}, \citenamefont {Zhang}, \citenamefont {Oh},\ and\ \citenamefont {Goldman}}]{Mei2012}%
  \BibitemOpen
  \bibfield  {author} {\bibinfo {author} {\bibfnamefont {Feng}\ \bibnamefont {Mei}}, \bibinfo {author} {\bibfnamefont {S.-L.}\ \bibnamefont {Zhu}}, \bibinfo {author} {\bibfnamefont {Zhi-Ming}\ \bibnamefont {Zhang}}, \bibinfo {author} {\bibfnamefont {C.~H.}\ \bibnamefont {Oh}}, \ and\ \bibinfo {author} {\bibfnamefont {N.}~\bibnamefont {Goldman}},\ }\bibfield  {title} {\enquote {\bibinfo {title} {Simulating z$_2$ topological insulators with cold atoms in a one-dimensional optical lattice},}\ }\href {\doibase 10.1103/physreva.85.013638} {\bibfield  {journal} {\bibinfo  {journal} {Phy. Rev. A}\ }\textbf {\bibinfo {volume} {85}},\ \bibinfo {pages} {013638} (\bibinfo {year} {2012})}\BibitemShut {NoStop}%
\bibitem [{\citenamefont {Dalibard}\ \emph {et~al.}(2011)\citenamefont {Dalibard}, \citenamefont {Gerbier}, \citenamefont {Juzeli\ifmmode~\bar{u}\else \={u}\fi{}nas},\ and\ \citenamefont {\"Ohberg}}]{RevModPhys.83.1523}%
  \BibitemOpen
  \bibfield  {author} {\bibinfo {author} {\bibfnamefont {Jean}\ \bibnamefont {Dalibard}}, \bibinfo {author} {\bibfnamefont {Fabrice}\ \bibnamefont {Gerbier}}, \bibinfo {author} {\bibfnamefont {Gediminas}\ \bibnamefont {Juzeli\ifmmode~\bar{u}\else \={u}\fi{}nas}}, \ and\ \bibinfo {author} {\bibfnamefont {Patrik}\ \bibnamefont {\"Ohberg}},\ }\bibfield  {title} {\enquote {\bibinfo {title} {Colloquium: Artificial gauge potentials for neutral atoms},}\ }\href {\doibase 10.1103/RevModPhys.83.1523} {\bibfield  {journal} {\bibinfo  {journal} {Rev. Mod. Phys.}\ }\textbf {\bibinfo {volume} {83}},\ \bibinfo {pages} {1523--1543} (\bibinfo {year} {2011})}\BibitemShut {NoStop}%
\bibitem [{\citenamefont {Lin}\ \emph {et~al.}(2009{\natexlab{a}})\citenamefont {Lin}, \citenamefont {Compton}, \citenamefont {Jim{\'e}nez-Garc{\'\i}a}, \citenamefont {Porto},\ and\ \citenamefont {Spielman}}]{lin2009synthetic}%
  \BibitemOpen
  \bibfield  {author} {\bibinfo {author} {\bibfnamefont {Y-J}\ \bibnamefont {Lin}}, \bibinfo {author} {\bibfnamefont {Rob~L}\ \bibnamefont {Compton}}, \bibinfo {author} {\bibfnamefont {Karina}\ \bibnamefont {Jim{\'e}nez-Garc{\'\i}a}}, \bibinfo {author} {\bibfnamefont {James~V}\ \bibnamefont {Porto}}, \ and\ \bibinfo {author} {\bibfnamefont {Ian~B}\ \bibnamefont {Spielman}},\ }\bibfield  {title} {\enquote {\bibinfo {title} {Synthetic magnetic fields for ultracold neutral atoms},}\ }\href {\doibase 10.1038/nature08609} {\bibfield  {journal} {\bibinfo  {journal} {Nature}\ }\textbf {\bibinfo {volume} {462}},\ \bibinfo {pages} {628--632} (\bibinfo {year} {2009}{\natexlab{a}})}\BibitemShut {NoStop}%
\bibitem [{\citenamefont {Lin}\ \emph {et~al.}(2011)\citenamefont {Lin}, \citenamefont {Jim{\'e}nez-Garc{\'\i}a},\ and\ \citenamefont {Spielman}}]{lin2011spin}%
  \BibitemOpen
  \bibfield  {author} {\bibinfo {author} {\bibfnamefont {Y-J}\ \bibnamefont {Lin}}, \bibinfo {author} {\bibfnamefont {K}~\bibnamefont {Jim{\'e}nez-Garc{\'\i}a}}, \ and\ \bibinfo {author} {\bibfnamefont {Ian~B}\ \bibnamefont {Spielman}},\ }\bibfield  {title} {\enquote {\bibinfo {title} {Spin--orbit-coupled bose--einstein condensates},}\ }\href {\doibase 10.1038/nature09887} {\bibfield  {journal} {\bibinfo  {journal} {Nature}\ }\textbf {\bibinfo {volume} {471}},\ \bibinfo {pages} {83--86} (\bibinfo {year} {2011})}\BibitemShut {NoStop}%
\bibitem [{\citenamefont {Zhu}\ \emph {et~al.}(2011)\citenamefont {Zhu}, \citenamefont {Shao}, \citenamefont {Wang},\ and\ \citenamefont {Duan}}]{SLZhu2011}%
  \BibitemOpen
  \bibfield  {author} {\bibinfo {author} {\bibfnamefont {S.-L.}\ \bibnamefont {Zhu}}, \bibinfo {author} {\bibfnamefont {L.-B.}\ \bibnamefont {Shao}}, \bibinfo {author} {\bibfnamefont {Z.~D.}\ \bibnamefont {Wang}}, \ and\ \bibinfo {author} {\bibfnamefont {L.-M.}\ \bibnamefont {Duan}},\ }\bibfield  {title} {\enquote {\bibinfo {title} {Probing non-abelian statistics of majorana fermions in ultracold atomic superfluid},}\ }\href {\doibase 10.1103/physrevlett.106.100404} {\bibfield  {journal} {\bibinfo  {journal} {Phy. Rev. Lett.}\ }\textbf {\bibinfo {volume} {106}},\ \bibinfo {pages} {100404} (\bibinfo {year} {2011})}\BibitemShut {NoStop}%
\bibitem [{\citenamefont {Lin}\ \emph {et~al.}(2009{\natexlab{b}})\citenamefont {Lin}, \citenamefont {Compton}, \citenamefont {Perry}, \citenamefont {Phillips}, \citenamefont {Porto},\ and\ \citenamefont {Spielman}}]{PhysRevLett.102.130401}%
  \BibitemOpen
  \bibfield  {author} {\bibinfo {author} {\bibfnamefont {Y.-J.}\ \bibnamefont {Lin}}, \bibinfo {author} {\bibfnamefont {R.~L.}\ \bibnamefont {Compton}}, \bibinfo {author} {\bibfnamefont {A.~R.}\ \bibnamefont {Perry}}, \bibinfo {author} {\bibfnamefont {W.~D.}\ \bibnamefont {Phillips}}, \bibinfo {author} {\bibfnamefont {J.~V.}\ \bibnamefont {Porto}}, \ and\ \bibinfo {author} {\bibfnamefont {I.~B.}\ \bibnamefont {Spielman}},\ }\bibfield  {title} {\enquote {\bibinfo {title} {Bose-einstein condensate in a uniform light-induced vector potential},}\ }\href {\doibase 10.1103/PhysRevLett.102.130401} {\bibfield  {journal} {\bibinfo  {journal} {Phys. Rev. Lett.}\ }\textbf {\bibinfo {volume} {102}},\ \bibinfo {pages} {130401} (\bibinfo {year} {2009}{\natexlab{b}})}\BibitemShut {NoStop}%
\bibitem [{\citenamefont {Aidelsburger}\ \emph {et~al.}(2011)\citenamefont {Aidelsburger}, \citenamefont {Atala}, \citenamefont {Nascimb\`ene}, \citenamefont {Trotzky}, \citenamefont {Chen},\ and\ \citenamefont {Bloch}}]{PhysRevLett.107.255301}%
  \BibitemOpen
  \bibfield  {author} {\bibinfo {author} {\bibfnamefont {M.}~\bibnamefont {Aidelsburger}}, \bibinfo {author} {\bibfnamefont {M.}~\bibnamefont {Atala}}, \bibinfo {author} {\bibfnamefont {S.}~\bibnamefont {Nascimb\`ene}}, \bibinfo {author} {\bibfnamefont {S.}~\bibnamefont {Trotzky}}, \bibinfo {author} {\bibfnamefont {Y.-A.}\ \bibnamefont {Chen}}, \ and\ \bibinfo {author} {\bibfnamefont {I.}~\bibnamefont {Bloch}},\ }\bibfield  {title} {\enquote {\bibinfo {title} {Experimental realization of strong effective magnetic fields in an optical lattice},}\ }\href {\doibase 10.1103/PhysRevLett.107.255301} {\bibfield  {journal} {\bibinfo  {journal} {Phys. Rev. Lett.}\ }\textbf {\bibinfo {volume} {107}},\ \bibinfo {pages} {255301} (\bibinfo {year} {2011})}\BibitemShut {NoStop}%
\bibitem [{\citenamefont {Osterloh}\ \emph {et~al.}(2005)\citenamefont {Osterloh}, \citenamefont {Baig}, \citenamefont {Santos}, \citenamefont {Zoller},\ and\ \citenamefont {Lewenstein}}]{PhysRevLett.95.010403}%
  \BibitemOpen
  \bibfield  {author} {\bibinfo {author} {\bibfnamefont {K.}~\bibnamefont {Osterloh}}, \bibinfo {author} {\bibfnamefont {M.}~\bibnamefont {Baig}}, \bibinfo {author} {\bibfnamefont {L.}~\bibnamefont {Santos}}, \bibinfo {author} {\bibfnamefont {P.}~\bibnamefont {Zoller}}, \ and\ \bibinfo {author} {\bibfnamefont {M.}~\bibnamefont {Lewenstein}},\ }\bibfield  {title} {\enquote {\bibinfo {title} {Cold atoms in non-abelian gauge potentials: From the hofstadter "moth" to lattice gauge theory},}\ }\href {\doibase 10.1103/PhysRevLett.95.010403} {\bibfield  {journal} {\bibinfo  {journal} {Phys. Rev. Lett.}\ }\textbf {\bibinfo {volume} {95}},\ \bibinfo {pages} {010403} (\bibinfo {year} {2005})}\BibitemShut {NoStop}%
\bibitem [{\citenamefont {Li}\ \emph {et~al.}(2022)\citenamefont {Li}, \citenamefont {Zou}, \citenamefont {Du}, \citenamefont {Lv}, \citenamefont {Huang}, \citenamefont {Liang}, \citenamefont {Zhang}, \citenamefont {Yan}, \citenamefont {Zhang},\ and\ \citenamefont {Zhu}}]{JZLi2022}%
  \BibitemOpen
  \bibfield  {author} {\bibinfo {author} {\bibfnamefont {J.-Z.}\ \bibnamefont {Li}}, \bibinfo {author} {\bibfnamefont {C.-J.}\ \bibnamefont {Zou}}, \bibinfo {author} {\bibfnamefont {Y.-X.}\ \bibnamefont {Du}}, \bibinfo {author} {\bibfnamefont {Q.-X.}\ \bibnamefont {Lv}}, \bibinfo {author} {\bibfnamefont {W.}~\bibnamefont {Huang}}, \bibinfo {author} {\bibfnamefont {Z.-T.}\ \bibnamefont {Liang}}, \bibinfo {author} {\bibfnamefont {D.-W.}\ \bibnamefont {Zhang}}, \bibinfo {author} {\bibfnamefont {H.}~\bibnamefont {Yan}}, \bibinfo {author} {\bibfnamefont {S.}~\bibnamefont {Zhang}}, \ and\ \bibinfo {author} {\bibfnamefont {S.-L.}\ \bibnamefont {Zhu}},\ }\bibfield  {title} {\enquote {\bibinfo {title} {Synthetic topological vacua of yang-mills fields in bose-einstein condensates},}\ }\href {\doibase 10.1103/physrevlett.129.220402} {\bibfield  {journal} {\bibinfo  {journal} {Phys. Rev. Lett.}\ }\textbf {\bibinfo {volume} {129}},\ \bibinfo {pages} {220402} (\bibinfo {year} {2022})}\BibitemShut {NoStop}%
\bibitem [{\citenamefont {Jaksch}\ and\ \citenamefont {Zoller}(2003)}]{jaksch2003creation}%
  \BibitemOpen
  \bibfield  {author} {\bibinfo {author} {\bibfnamefont {D}~\bibnamefont {Jaksch}}\ and\ \bibinfo {author} {\bibfnamefont {P}~\bibnamefont {Zoller}},\ }\bibfield  {title} {\enquote {\bibinfo {title} {Creation of effective magnetic fields in optical lattices: the hofstadter butterfly for cold neutral atoms},}\ }\href {\doibase 10.1088/1367-2630/5/1/356} {\bibfield  {journal} {\bibinfo  {journal} {New Journal of Physics}\ }\textbf {\bibinfo {volume} {5}},\ \bibinfo {pages} {56--56} (\bibinfo {year} {2003})}\BibitemShut {NoStop}%
\bibitem [{\citenamefont {Gerbier}\ and\ \citenamefont {Dalibard}(2010)}]{gerbier2010gauge}%
  \BibitemOpen
  \bibfield  {author} {\bibinfo {author} {\bibfnamefont {Fabrice}\ \bibnamefont {Gerbier}}\ and\ \bibinfo {author} {\bibfnamefont {Jean}\ \bibnamefont {Dalibard}},\ }\bibfield  {title} {\enquote {\bibinfo {title} {Gauge fields for ultracold atoms in optical superlattices},}\ }\href {\doibase 10.1088/1367-2630/12/3/033007} {\bibfield  {journal} {\bibinfo  {journal} {New Journal of Physics}\ }\textbf {\bibinfo {volume} {12}},\ \bibinfo {pages} {033007} (\bibinfo {year} {2010})}\BibitemShut {NoStop}%
\bibitem [{\citenamefont {Goldman}\ \emph {et~al.}(2009)\citenamefont {Goldman}, \citenamefont {Kubasiak}, \citenamefont {Bermudez}, \citenamefont {Gaspard}, \citenamefont {Lewenstein},\ and\ \citenamefont {Martin-Delgado}}]{PhysRevLett.103.035301}%
  \BibitemOpen
  \bibfield  {author} {\bibinfo {author} {\bibfnamefont {N.}~\bibnamefont {Goldman}}, \bibinfo {author} {\bibfnamefont {A.}~\bibnamefont {Kubasiak}}, \bibinfo {author} {\bibfnamefont {A.}~\bibnamefont {Bermudez}}, \bibinfo {author} {\bibfnamefont {P.}~\bibnamefont {Gaspard}}, \bibinfo {author} {\bibfnamefont {M.}~\bibnamefont {Lewenstein}}, \ and\ \bibinfo {author} {\bibfnamefont {M.~A.}\ \bibnamefont {Martin-Delgado}},\ }\bibfield  {title} {\enquote {\bibinfo {title} {Non-abelian optical lattices: Anomalous quantum hall effect and dirac fermions},}\ }\href {\doibase 10.1103/PhysRevLett.103.035301} {\bibfield  {journal} {\bibinfo  {journal} {Phys. Rev. Lett.}\ }\textbf {\bibinfo {volume} {103}},\ \bibinfo {pages} {035301} (\bibinfo {year} {2009})}\BibitemShut {NoStop}%
\bibitem [{\citenamefont {Zhu}\ \emph {et~al.}(2006{\natexlab{b}})\citenamefont {Zhu}, \citenamefont {Fu}, \citenamefont {Wu}, \citenamefont {Zhang},\ and\ \citenamefont {Duan}}]{PhysRevLett.97.240401}%
  \BibitemOpen
  \bibfield  {author} {\bibinfo {author} {\bibfnamefont {Shi-Liang}\ \bibnamefont {Zhu}}, \bibinfo {author} {\bibfnamefont {Hao}\ \bibnamefont {Fu}}, \bibinfo {author} {\bibfnamefont {C.-J.}\ \bibnamefont {Wu}}, \bibinfo {author} {\bibfnamefont {S.-C.}\ \bibnamefont {Zhang}}, \ and\ \bibinfo {author} {\bibfnamefont {L.-M.}\ \bibnamefont {Duan}},\ }\bibfield  {title} {\enquote {\bibinfo {title} {Spin hall effects for cold atoms in a light-induced gauge potential},}\ }\href {\doibase 10.1103/PhysRevLett.97.240401} {\bibfield  {journal} {\bibinfo  {journal} {Phys. Rev. Lett.}\ }\textbf {\bibinfo {volume} {97}},\ \bibinfo {pages} {240401} (\bibinfo {year} {2006}{\natexlab{b}})}\BibitemShut {NoStop}%
\bibitem [{\citenamefont {Liang}\ \emph {et~al.}(2023)\citenamefont {Liang}, \citenamefont {Wei}, \citenamefont {Zhang}, \citenamefont {Wang}, \citenamefont {Zhang}, \citenamefont {Wang}, \citenamefont {Qi}, \citenamefont {Liu},\ and\ \citenamefont {Zhang}}]{PhysRevResearch.5.L012006}%
  \BibitemOpen
  \bibfield  {author} {\bibinfo {author} {\bibfnamefont {Ming-Cheng}\ \bibnamefont {Liang}}, \bibinfo {author} {\bibfnamefont {Yu-Dong}\ \bibnamefont {Wei}}, \bibinfo {author} {\bibfnamefont {Long}\ \bibnamefont {Zhang}}, \bibinfo {author} {\bibfnamefont {Xu-Jie}\ \bibnamefont {Wang}}, \bibinfo {author} {\bibfnamefont {Han}\ \bibnamefont {Zhang}}, \bibinfo {author} {\bibfnamefont {Wen-Wei}\ \bibnamefont {Wang}}, \bibinfo {author} {\bibfnamefont {Wei}\ \bibnamefont {Qi}}, \bibinfo {author} {\bibfnamefont {Xiong-Jun}\ \bibnamefont {Liu}}, \ and\ \bibinfo {author} {\bibfnamefont {Xibo}\ \bibnamefont {Zhang}},\ }\bibfield  {title} {\enquote {\bibinfo {title} {Realization of qi-wu-zhang model in spin-orbit-coupled ultracold fermions},}\ }\href {\doibase 10.1103/PhysRevResearch.5.L012006} {\bibfield  {journal} {\bibinfo  {journal} {Phys. Rev. Res.}\ }\textbf {\bibinfo {volume} {5}},\ \bibinfo {pages} {L012006} (\bibinfo {year} {2023})}\BibitemShut {NoStop}%
\bibitem [{\citenamefont {{Goldman, N.}}\ and\ \citenamefont {{Gaspard, P.}}(2007)}]{goldman2007quantum}%
  \BibitemOpen
  \bibfield  {author} {\bibinfo {author} {\bibnamefont {{Goldman, N.}}}\ and\ \bibinfo {author} {\bibnamefont {{Gaspard, P.}}},\ }\bibfield  {title} {\enquote {\bibinfo {title} {Quantum hall-like effect for cold atoms in non-abelian gauge potentials},}\ }\href {\doibase 10.1209/0295-5075/78/60001} {\bibfield  {journal} {\bibinfo  {journal} {EPL}\ }\textbf {\bibinfo {volume} {78}},\ \bibinfo {pages} {60001} (\bibinfo {year} {2007})}\BibitemShut {NoStop}%
\bibitem [{\citenamefont {Shao}\ \emph {et~al.}(2008)\citenamefont {Shao}, \citenamefont {Zhu}, \citenamefont {Sheng}, \citenamefont {Xing},\ and\ \citenamefont {Wang}}]{PhysRevLett.101.246810}%
  \BibitemOpen
  \bibfield  {author} {\bibinfo {author} {\bibfnamefont {L.~B.}\ \bibnamefont {Shao}}, \bibinfo {author} {\bibfnamefont {Shi-Liang}\ \bibnamefont {Zhu}}, \bibinfo {author} {\bibfnamefont {L.}~\bibnamefont {Sheng}}, \bibinfo {author} {\bibfnamefont {D.~Y.}\ \bibnamefont {Xing}}, \ and\ \bibinfo {author} {\bibfnamefont {Z.~D.}\ \bibnamefont {Wang}},\ }\bibfield  {title} {\enquote {\bibinfo {title} {Realizing and detecting the quantum hall effect without landau levels by using ultracold atoms},}\ }\href {\doibase 10.1103/PhysRevLett.101.246810} {\bibfield  {journal} {\bibinfo  {journal} {Phys. Rev. Lett.}\ }\textbf {\bibinfo {volume} {101}},\ \bibinfo {pages} {246810} (\bibinfo {year} {2008})}\BibitemShut {NoStop}%
\bibitem [{\citenamefont {Wu}(2008)}]{PhysRevLett.101.186807}%
  \BibitemOpen
  \bibfield  {author} {\bibinfo {author} {\bibfnamefont {Congjun}\ \bibnamefont {Wu}},\ }\bibfield  {title} {\enquote {\bibinfo {title} {Orbital analogue of the quantum anomalous hall effect in $p$-band systems},}\ }\href {\doibase 10.1103/PhysRevLett.101.186807} {\bibfield  {journal} {\bibinfo  {journal} {Phys. Rev. Lett.}\ }\textbf {\bibinfo {volume} {101}},\ \bibinfo {pages} {186807} (\bibinfo {year} {2008})}\BibitemShut {NoStop}%
\bibitem [{\citenamefont {Stanescu}\ \emph {et~al.}(2009)\citenamefont {Stanescu}, \citenamefont {Galitski}, \citenamefont {Vaishnav}, \citenamefont {Clark},\ and\ \citenamefont {Das~Sarma}}]{PhysRevA.79.053639}%
  \BibitemOpen
  \bibfield  {author} {\bibinfo {author} {\bibfnamefont {Tudor~D.}\ \bibnamefont {Stanescu}}, \bibinfo {author} {\bibfnamefont {Victor}\ \bibnamefont {Galitski}}, \bibinfo {author} {\bibfnamefont {J.~Y.}\ \bibnamefont {Vaishnav}}, \bibinfo {author} {\bibfnamefont {Charles~W.}\ \bibnamefont {Clark}}, \ and\ \bibinfo {author} {\bibfnamefont {S.}~\bibnamefont {Das~Sarma}},\ }\bibfield  {title} {\enquote {\bibinfo {title} {Topological insulators and metals in atomic optical lattices},}\ }\href {\doibase 10.1103/PhysRevA.79.053639} {\bibfield  {journal} {\bibinfo  {journal} {Phys. Rev. A}\ }\textbf {\bibinfo {volume} {79}},\ \bibinfo {pages} {053639} (\bibinfo {year} {2009})}\BibitemShut {NoStop}%
\bibitem [{\citenamefont {Alba}\ \emph {et~al.}(2011)\citenamefont {Alba}, \citenamefont {Fernandez-Gonzalvo}, \citenamefont {Mur-Petit}, \citenamefont {Pachos},\ and\ \citenamefont {Garcia-Ripoll}}]{PhysRevLett.107.235301}%
  \BibitemOpen
  \bibfield  {author} {\bibinfo {author} {\bibfnamefont {E.}~\bibnamefont {Alba}}, \bibinfo {author} {\bibfnamefont {X.}~\bibnamefont {Fernandez-Gonzalvo}}, \bibinfo {author} {\bibfnamefont {J.}~\bibnamefont {Mur-Petit}}, \bibinfo {author} {\bibfnamefont {J.~K.}\ \bibnamefont {Pachos}}, \ and\ \bibinfo {author} {\bibfnamefont {J.~J.}\ \bibnamefont {Garcia-Ripoll}},\ }\bibfield  {title} {\enquote {\bibinfo {title} {Seeing topological order in time-of-flight measurements},}\ }\href {\doibase 10.1103/PhysRevLett.107.235301} {\bibfield  {journal} {\bibinfo  {journal} {Phys. Rev. Lett.}\ }\textbf {\bibinfo {volume} {107}},\ \bibinfo {pages} {235301} (\bibinfo {year} {2011})}\BibitemShut {NoStop}%
\bibitem [{\citenamefont {B\'eri}\ and\ \citenamefont {Cooper}(2011)}]{PhysRevLett.107.145301}%
  \BibitemOpen
  \bibfield  {author} {\bibinfo {author} {\bibfnamefont {B.}~\bibnamefont {B\'eri}}\ and\ \bibinfo {author} {\bibfnamefont {N.~R.}\ \bibnamefont {Cooper}},\ }\bibfield  {title} {\enquote {\bibinfo {title} {${\mathbb{z}}_{2}$ topological insulators in ultracold atomic gases},}\ }\href {\doibase 10.1103/PhysRevLett.107.145301} {\bibfield  {journal} {\bibinfo  {journal} {Phys. Rev. Lett.}\ }\textbf {\bibinfo {volume} {107}},\ \bibinfo {pages} {145301} (\bibinfo {year} {2011})}\BibitemShut {NoStop}%
\bibitem [{\citenamefont {Goldman}\ \emph {et~al.}(2010)\citenamefont {Goldman}, \citenamefont {Satija}, \citenamefont {Nikolic}, \citenamefont {Bermudez}, \citenamefont {Martin-Delgado}, \citenamefont {Lewenstein},\ and\ \citenamefont {Spielman}}]{PhysRevLett.105.255302}%
  \BibitemOpen
  \bibfield  {author} {\bibinfo {author} {\bibfnamefont {N.}~\bibnamefont {Goldman}}, \bibinfo {author} {\bibfnamefont {I.}~\bibnamefont {Satija}}, \bibinfo {author} {\bibfnamefont {P.}~\bibnamefont {Nikolic}}, \bibinfo {author} {\bibfnamefont {A.}~\bibnamefont {Bermudez}}, \bibinfo {author} {\bibfnamefont {M.~A.}\ \bibnamefont {Martin-Delgado}}, \bibinfo {author} {\bibfnamefont {M.}~\bibnamefont {Lewenstein}}, \ and\ \bibinfo {author} {\bibfnamefont {I.~B.}\ \bibnamefont {Spielman}},\ }\bibfield  {title} {\enquote {\bibinfo {title} {Realistic time-reversal invariant topological insulators with neutral atoms},}\ }\href {\doibase 10.1103/PhysRevLett.105.255302} {\bibfield  {journal} {\bibinfo  {journal} {Phys. Rev. Lett.}\ }\textbf {\bibinfo {volume} {105}},\ \bibinfo {pages} {255302} (\bibinfo {year} {2010})}\BibitemShut {NoStop}%
\bibitem [{\citenamefont {Atala}\ \emph {et~al.}(2013)\citenamefont {Atala}, \citenamefont {Aidelsburger}, \citenamefont {Barreiro}, \citenamefont {Abanin}, \citenamefont {Kitagawa}, \citenamefont {Demler},\ and\ \citenamefont {Bloch}}]{atala2013direct}%
  \BibitemOpen
  \bibfield  {author} {\bibinfo {author} {\bibfnamefont {Marcos}\ \bibnamefont {Atala}}, \bibinfo {author} {\bibfnamefont {Monika}\ \bibnamefont {Aidelsburger}}, \bibinfo {author} {\bibfnamefont {Julio~T}\ \bibnamefont {Barreiro}}, \bibinfo {author} {\bibfnamefont {Dmitry}\ \bibnamefont {Abanin}}, \bibinfo {author} {\bibfnamefont {Takuya}\ \bibnamefont {Kitagawa}}, \bibinfo {author} {\bibfnamefont {Eugene}\ \bibnamefont {Demler}}, \ and\ \bibinfo {author} {\bibfnamefont {Immanuel}\ \bibnamefont {Bloch}},\ }\bibfield  {title} {\enquote {\bibinfo {title} {Direct measurement of the zak phase in topological bloch bands},}\ }\href {\doibase 10.1038/nphys2790} {\bibfield  {journal} {\bibinfo  {journal} {Nature Physics}\ }\textbf {\bibinfo {volume} {9}},\ \bibinfo {pages} {795--800} (\bibinfo {year} {2013})}\BibitemShut {NoStop}%
\bibitem [{\citenamefont {Duca}\ \emph {et~al.}(2015)\citenamefont {Duca}, \citenamefont {Li}, \citenamefont {Reitter}, \citenamefont {Bloch}, \citenamefont {Schleier-Smith},\ and\ \citenamefont {Schneider}}]{duca2015aharonov}%
  \BibitemOpen
  \bibfield  {author} {\bibinfo {author} {\bibfnamefont {Lucia}\ \bibnamefont {Duca}}, \bibinfo {author} {\bibfnamefont {Tracy}\ \bibnamefont {Li}}, \bibinfo {author} {\bibfnamefont {Martin}\ \bibnamefont {Reitter}}, \bibinfo {author} {\bibfnamefont {Immanuel}\ \bibnamefont {Bloch}}, \bibinfo {author} {\bibfnamefont {Monika}\ \bibnamefont {Schleier-Smith}}, \ and\ \bibinfo {author} {\bibfnamefont {Ulrich}\ \bibnamefont {Schneider}},\ }\bibfield  {title} {\enquote {\bibinfo {title} {An aharonov-bohm interferometer for determining bloch band topology},}\ }\href {\doibase 10.1126/science.1259052} {\bibfield  {journal} {\bibinfo  {journal} {Science}\ }\textbf {\bibinfo {volume} {347}},\ \bibinfo {pages} {288--292} (\bibinfo {year} {2015})}\BibitemShut {NoStop}%
\bibitem [{\citenamefont {Grusdt}\ \emph {et~al.}(2016)\citenamefont {Grusdt}, \citenamefont {Yao}, \citenamefont {Abanin}, \citenamefont {Fleischhauer},\ and\ \citenamefont {Demler}}]{grusdt2016interferometric}%
  \BibitemOpen
  \bibfield  {author} {\bibinfo {author} {\bibfnamefont {Fabian}\ \bibnamefont {Grusdt}}, \bibinfo {author} {\bibfnamefont {Norman~Y}\ \bibnamefont {Yao}}, \bibinfo {author} {\bibfnamefont {D}~\bibnamefont {Abanin}}, \bibinfo {author} {\bibfnamefont {Michael}\ \bibnamefont {Fleischhauer}}, \ and\ \bibinfo {author} {\bibfnamefont {E}~\bibnamefont {Demler}},\ }\bibfield  {title} {\enquote {\bibinfo {title} {Interferometric measurements of many-body topological invariants using mobile impurities},}\ }\href {\doibase 10.1038/ncomms11994} {\bibfield  {journal} {\bibinfo  {journal} {Nature communications}\ }\textbf {\bibinfo {volume} {7}},\ \bibinfo {pages} {1--9} (\bibinfo {year} {2016})}\BibitemShut {NoStop}%
\bibitem [{\citenamefont {Abanin}\ \emph {et~al.}(2013)\citenamefont {Abanin}, \citenamefont {Kitagawa}, \citenamefont {Bloch},\ and\ \citenamefont {Demler}}]{PhysRevLett.110.165304}%
  \BibitemOpen
  \bibfield  {author} {\bibinfo {author} {\bibfnamefont {Dmitry~A.}\ \bibnamefont {Abanin}}, \bibinfo {author} {\bibfnamefont {Takuya}\ \bibnamefont {Kitagawa}}, \bibinfo {author} {\bibfnamefont {Immanuel}\ \bibnamefont {Bloch}}, \ and\ \bibinfo {author} {\bibfnamefont {Eugene}\ \bibnamefont {Demler}},\ }\bibfield  {title} {\enquote {\bibinfo {title} {Interferometric approach to measuring band topology in 2d optical lattices},}\ }\href {\doibase 10.1103/PhysRevLett.110.165304} {\bibfield  {journal} {\bibinfo  {journal} {Phys. Rev. Lett.}\ }\textbf {\bibinfo {volume} {110}},\ \bibinfo {pages} {165304} (\bibinfo {year} {2013})}\BibitemShut {NoStop}%
\bibitem [{\citenamefont {Grusdt}\ \emph {et~al.}(2014)\citenamefont {Grusdt}, \citenamefont {Abanin},\ and\ \citenamefont {Demler}}]{PhysRevA.89.043621}%
  \BibitemOpen
  \bibfield  {author} {\bibinfo {author} {\bibfnamefont {F.}~\bibnamefont {Grusdt}}, \bibinfo {author} {\bibfnamefont {D.}~\bibnamefont {Abanin}}, \ and\ \bibinfo {author} {\bibfnamefont {E.}~\bibnamefont {Demler}},\ }\bibfield  {title} {\enquote {\bibinfo {title} {Measuring ${\mathbb{z}}_{2}$ topological invariants in optical lattices using interferometry},}\ }\href {\doibase 10.1103/PhysRevA.89.043621} {\bibfield  {journal} {\bibinfo  {journal} {Phys. Rev. A}\ }\textbf {\bibinfo {volume} {89}},\ \bibinfo {pages} {043621} (\bibinfo {year} {2014})}\BibitemShut {NoStop}%
\bibitem [{\citenamefont {Price}\ and\ \citenamefont {Cooper}(2012)}]{PhysRevA.85.033620}%
  \BibitemOpen
  \bibfield  {author} {\bibinfo {author} {\bibfnamefont {H.~M.}\ \bibnamefont {Price}}\ and\ \bibinfo {author} {\bibfnamefont {N.~R.}\ \bibnamefont {Cooper}},\ }\bibfield  {title} {\enquote {\bibinfo {title} {Mapping the berry curvature from semiclassical dynamics in optical lattices},}\ }\href {\doibase 10.1103/PhysRevA.85.033620} {\bibfield  {journal} {\bibinfo  {journal} {Phys. Rev. A}\ }\textbf {\bibinfo {volume} {85}},\ \bibinfo {pages} {033620} (\bibinfo {year} {2012})}\BibitemShut {NoStop}%
\bibitem [{\citenamefont {Dauphin}\ and\ \citenamefont {Goldman}(2013)}]{PhysRevLett.111.135302}%
  \BibitemOpen
  \bibfield  {author} {\bibinfo {author} {\bibfnamefont {Alexandre}\ \bibnamefont {Dauphin}}\ and\ \bibinfo {author} {\bibfnamefont {Nathan}\ \bibnamefont {Goldman}},\ }\bibfield  {title} {\enquote {\bibinfo {title} {Extracting the chern number from the dynamics of a fermi gas: Implementing a quantum hall bar for cold atoms},}\ }\href {\doibase 10.1103/PhysRevLett.111.135302} {\bibfield  {journal} {\bibinfo  {journal} {Phys. Rev. Lett.}\ }\textbf {\bibinfo {volume} {111}},\ \bibinfo {pages} {135302} (\bibinfo {year} {2013})}\BibitemShut {NoStop}%
\bibitem [{\citenamefont {Jotzu}\ \emph {et~al.}(2014)\citenamefont {Jotzu}, \citenamefont {Messer}, \citenamefont {Desbuquois}, \citenamefont {Lebrat}, \citenamefont {Uehlinger}, \citenamefont {Greif},\ and\ \citenamefont {Esslinger}}]{jotzu2014experimental}%
  \BibitemOpen
  \bibfield  {author} {\bibinfo {author} {\bibfnamefont {Gregor}\ \bibnamefont {Jotzu}}, \bibinfo {author} {\bibfnamefont {Michael}\ \bibnamefont {Messer}}, \bibinfo {author} {\bibfnamefont {R{\'e}mi}\ \bibnamefont {Desbuquois}}, \bibinfo {author} {\bibfnamefont {Martin}\ \bibnamefont {Lebrat}}, \bibinfo {author} {\bibfnamefont {Thomas}\ \bibnamefont {Uehlinger}}, \bibinfo {author} {\bibfnamefont {Daniel}\ \bibnamefont {Greif}}, \ and\ \bibinfo {author} {\bibfnamefont {Tilman}\ \bibnamefont {Esslinger}},\ }\bibfield  {title} {\enquote {\bibinfo {title} {Experimental realization of the topological haldane model with ultracold fermions},}\ }\href {\doibase 10.1038/nature13915} {\bibfield  {journal} {\bibinfo  {journal} {Nature}\ }\textbf {\bibinfo {volume} {515}},\ \bibinfo {pages} {237--240} (\bibinfo {year} {2014})}\BibitemShut {NoStop}%
\bibitem [{\citenamefont {Aidelsburger}\ \emph {et~al.}(2015)\citenamefont {Aidelsburger}, \citenamefont {Lohse}, \citenamefont {Schweizer}, \citenamefont {Atala}, \citenamefont {Barreiro}, \citenamefont {Nascimb{\`e}ne}, \citenamefont {Cooper}, \citenamefont {Bloch},\ and\ \citenamefont {Goldman}}]{aidelsburger2015measuring}%
  \BibitemOpen
  \bibfield  {author} {\bibinfo {author} {\bibfnamefont {Monika}\ \bibnamefont {Aidelsburger}}, \bibinfo {author} {\bibfnamefont {Michael}\ \bibnamefont {Lohse}}, \bibinfo {author} {\bibfnamefont {Christian}\ \bibnamefont {Schweizer}}, \bibinfo {author} {\bibfnamefont {Marcos}\ \bibnamefont {Atala}}, \bibinfo {author} {\bibfnamefont {Julio~T}\ \bibnamefont {Barreiro}}, \bibinfo {author} {\bibfnamefont {Sylvain}\ \bibnamefont {Nascimb{\`e}ne}}, \bibinfo {author} {\bibfnamefont {NR}~\bibnamefont {Cooper}}, \bibinfo {author} {\bibfnamefont {Immanuel}\ \bibnamefont {Bloch}}, \ and\ \bibinfo {author} {\bibfnamefont {Nathan}\ \bibnamefont {Goldman}},\ }\bibfield  {title} {\enquote {\bibinfo {title} {Measuring the chern number of hofstadter bands with ultracold bosonic atoms},}\ }\href {\doibase 10.1038/nphys3171} {\bibfield  {journal} {\bibinfo  {journal} {Nature Physics}\ }\textbf {\bibinfo {volume} {11}},\ \bibinfo {pages} {162--166} (\bibinfo {year} {2015})}\BibitemShut {NoStop}%
\bibitem [{\citenamefont {Deng}\ \emph {et~al.}(2014)\citenamefont {Deng}, \citenamefont {Wang},\ and\ \citenamefont {Duan}}]{PhysRevA.90.041601}%
  \BibitemOpen
  \bibfield  {author} {\bibinfo {author} {\bibfnamefont {Dong-Ling}\ \bibnamefont {Deng}}, \bibinfo {author} {\bibfnamefont {Sheng-Tao}\ \bibnamefont {Wang}}, \ and\ \bibinfo {author} {\bibfnamefont {L.-M.}\ \bibnamefont {Duan}},\ }\bibfield  {title} {\enquote {\bibinfo {title} {Direct probe of topological order for cold atoms},}\ }\href {\doibase 10.1103/PhysRevA.90.041601} {\bibfield  {journal} {\bibinfo  {journal} {Phys. Rev. A}\ }\textbf {\bibinfo {volume} {90}},\ \bibinfo {pages} {041601} (\bibinfo {year} {2014})}\BibitemShut {NoStop}%
\bibitem [{\citenamefont {Hauke}\ \emph {et~al.}(2014)\citenamefont {Hauke}, \citenamefont {Lewenstein},\ and\ \citenamefont {Eckardt}}]{PhysRevLett.113.045303}%
  \BibitemOpen
  \bibfield  {author} {\bibinfo {author} {\bibfnamefont {Philipp}\ \bibnamefont {Hauke}}, \bibinfo {author} {\bibfnamefont {Maciej}\ \bibnamefont {Lewenstein}}, \ and\ \bibinfo {author} {\bibfnamefont {Andr\'e}\ \bibnamefont {Eckardt}},\ }\bibfield  {title} {\enquote {\bibinfo {title} {Tomography of band insulators from quench dynamics},}\ }\href {\doibase 10.1103/PhysRevLett.113.045303} {\bibfield  {journal} {\bibinfo  {journal} {Phys. Rev. Lett.}\ }\textbf {\bibinfo {volume} {113}},\ \bibinfo {pages} {045303} (\bibinfo {year} {2014})}\BibitemShut {NoStop}%
\bibitem [{\citenamefont {Fläschner}\ \emph {et~al.}(2016)\citenamefont {Fläschner}, \citenamefont {Rem}, \citenamefont {Tarnowski}, \citenamefont {Vogel}, \citenamefont {Lühmann}, \citenamefont {Sengstock},\ and\ \citenamefont {Weitenberg}}]{doi:10.1126/science.aad4568}%
  \BibitemOpen
  \bibfield  {author} {\bibinfo {author} {\bibfnamefont {N.}~\bibnamefont {Fläschner}}, \bibinfo {author} {\bibfnamefont {B.~S.}\ \bibnamefont {Rem}}, \bibinfo {author} {\bibfnamefont {M.}~\bibnamefont {Tarnowski}}, \bibinfo {author} {\bibfnamefont {D.}~\bibnamefont {Vogel}}, \bibinfo {author} {\bibfnamefont {D.-S.}\ \bibnamefont {Lühmann}}, \bibinfo {author} {\bibfnamefont {K.}~\bibnamefont {Sengstock}}, \ and\ \bibinfo {author} {\bibfnamefont {C.}~\bibnamefont {Weitenberg}},\ }\bibfield  {title} {\enquote {\bibinfo {title} {Experimental reconstruction of the berry curvature in a floquet bloch band},}\ }\href {\doibase 10.1126/science.aad4568} {\bibfield  {journal} {\bibinfo  {journal} {Science}\ }\textbf {\bibinfo {volume} {352}},\ \bibinfo {pages} {1091--1094} (\bibinfo {year} {2016})}\BibitemShut {NoStop}%
\bibitem [{\citenamefont {Yi}\ \emph {et~al.}(2023)\citenamefont {Yi}, \citenamefont {Yu}, \citenamefont {Yuan}, \citenamefont {Jiao}, \citenamefont {Yang}, \citenamefont {Jiang}, \citenamefont {Zhang}, \citenamefont {Chen},\ and\ \citenamefont {Pan}}]{PhysRevResearch.5.L032016}%
  \BibitemOpen
  \bibfield  {author} {\bibinfo {author} {\bibfnamefont {Chang-Rui}\ \bibnamefont {Yi}}, \bibinfo {author} {\bibfnamefont {Jinlong}\ \bibnamefont {Yu}}, \bibinfo {author} {\bibfnamefont {Huan}\ \bibnamefont {Yuan}}, \bibinfo {author} {\bibfnamefont {Rui-Heng}\ \bibnamefont {Jiao}}, \bibinfo {author} {\bibfnamefont {Yu-Meng}\ \bibnamefont {Yang}}, \bibinfo {author} {\bibfnamefont {Xiao}\ \bibnamefont {Jiang}}, \bibinfo {author} {\bibfnamefont {Jin-Yi}\ \bibnamefont {Zhang}}, \bibinfo {author} {\bibfnamefont {Shuai}\ \bibnamefont {Chen}}, \ and\ \bibinfo {author} {\bibfnamefont {Jian-Wei}\ \bibnamefont {Pan}},\ }\bibfield  {title} {\enquote {\bibinfo {title} {Extracting the quantum geometric tensor of an optical raman lattice by bloch-state tomography},}\ }\href {\doibase 10.1103/PhysRevResearch.5.L032016} {\bibfield  {journal} {\bibinfo  {journal} {Phys. Rev. Res.}\ }\textbf {\bibinfo {volume} {5}},\ \bibinfo {pages} {L032016} (\bibinfo {year} {2023})}\BibitemShut {NoStop}%
\bibitem [{\citenamefont {Li}\ \emph {et~al.}(2016)\citenamefont {Li}, \citenamefont {Duca}, \citenamefont {Reitter}, \citenamefont {Grusdt}, \citenamefont {Demler}, \citenamefont {Endres}, \citenamefont {Schleier-Smith}, \citenamefont {Bloch},\ and\ \citenamefont {Schneider}}]{doi:10.1126/science.aad5812}%
  \BibitemOpen
  \bibfield  {author} {\bibinfo {author} {\bibfnamefont {Tracy}\ \bibnamefont {Li}}, \bibinfo {author} {\bibfnamefont {Lucia}\ \bibnamefont {Duca}}, \bibinfo {author} {\bibfnamefont {Martin}\ \bibnamefont {Reitter}}, \bibinfo {author} {\bibfnamefont {Fabian}\ \bibnamefont {Grusdt}}, \bibinfo {author} {\bibfnamefont {Eugene}\ \bibnamefont {Demler}}, \bibinfo {author} {\bibfnamefont {Manuel}\ \bibnamefont {Endres}}, \bibinfo {author} {\bibfnamefont {Monika}\ \bibnamefont {Schleier-Smith}}, \bibinfo {author} {\bibfnamefont {Immanuel}\ \bibnamefont {Bloch}}, \ and\ \bibinfo {author} {\bibfnamefont {Ulrich}\ \bibnamefont {Schneider}},\ }\bibfield  {title} {\enquote {\bibinfo {title} {Bloch state tomography using wilson lines},}\ }\href {\doibase 10.1126/science.aad5812} {\bibfield  {journal} {\bibinfo  {journal} {Science}\ }\textbf {\bibinfo {volume} {352}},\ \bibinfo {pages} {1094--1097} (\bibinfo {year} {2016})}\BibitemShut {NoStop}%
\bibitem [{\citenamefont {Tarnowski}\ \emph {et~al.}(2017)\citenamefont {Tarnowski}, \citenamefont {Nuske}, \citenamefont {Fl\"aschner}, \citenamefont {Rem}, \citenamefont {Vogel}, \citenamefont {Freystatzky}, \citenamefont {Sengstock}, \citenamefont {Mathey},\ and\ \citenamefont {Weitenberg}}]{PhysRevLett.118.240403}%
  \BibitemOpen
  \bibfield  {author} {\bibinfo {author} {\bibfnamefont {Matthias}\ \bibnamefont {Tarnowski}}, \bibinfo {author} {\bibfnamefont {Marlon}\ \bibnamefont {Nuske}}, \bibinfo {author} {\bibfnamefont {Nick}\ \bibnamefont {Fl\"aschner}}, \bibinfo {author} {\bibfnamefont {Benno}\ \bibnamefont {Rem}}, \bibinfo {author} {\bibfnamefont {Dominik}\ \bibnamefont {Vogel}}, \bibinfo {author} {\bibfnamefont {Lukas}\ \bibnamefont {Freystatzky}}, \bibinfo {author} {\bibfnamefont {Klaus}\ \bibnamefont {Sengstock}}, \bibinfo {author} {\bibfnamefont {Ludwig}\ \bibnamefont {Mathey}}, \ and\ \bibinfo {author} {\bibfnamefont {Christof}\ \bibnamefont {Weitenberg}},\ }\bibfield  {title} {\enquote {\bibinfo {title} {Observation of topological bloch-state defects and their merging transition},}\ }\href {\doibase 10.1103/PhysRevLett.118.240403} {\bibfield  {journal} {\bibinfo  {journal} {Phys. Rev. Lett.}\ }\textbf {\bibinfo {volume} {118}},\ \bibinfo {pages} {240403} (\bibinfo {year} {2017})}\BibitemShut {NoStop}%
\bibitem [{\citenamefont {Streda}(1982)}]{Streda_1982}%
  \BibitemOpen
  \bibfield  {author} {\bibinfo {author} {\bibfnamefont {P}~\bibnamefont {Streda}},\ }\bibfield  {title} {\enquote {\bibinfo {title} {Theory of quantised hall conductivity in two dimensions},}\ }\href {\doibase 10.1088/0022-3719/15/22/005} {\bibfield  {journal} {\bibinfo  {journal} {Journal of Physics C: Solid State Physics}\ }\textbf {\bibinfo {volume} {15}},\ \bibinfo {pages} {L717--L721} (\bibinfo {year} {1982})}\BibitemShut {NoStop}%
\bibitem [{\citenamefont {Umucal\ifmmode \imath \else~\i \fi{}lar}\ \emph {et~al.}(2008)\citenamefont {Umucal\ifmmode \imath \else~\i \fi{}lar}, \citenamefont {Zhai},\ and\ \citenamefont {Oktel}}]{PhysRevLett.100.070402}%
  \BibitemOpen
  \bibfield  {author} {\bibinfo {author} {\bibfnamefont {R.~O.}\ \bibnamefont {Umucal\ifmmode \imath \else~\i \fi{}lar}}, \bibinfo {author} {\bibfnamefont {Hui}\ \bibnamefont {Zhai}}, \ and\ \bibinfo {author} {\bibfnamefont {M.~\"O.}\ \bibnamefont {Oktel}},\ }\bibfield  {title} {\enquote {\bibinfo {title} {Trapped fermi gases in rotating optical lattices: Realization and detection of the topological hofstadter insulator},}\ }\href {\doibase 10.1103/PhysRevLett.100.070402} {\bibfield  {journal} {\bibinfo  {journal} {Phys. Rev. Lett.}\ }\textbf {\bibinfo {volume} {100}},\ \bibinfo {pages} {070402} (\bibinfo {year} {2008})}\BibitemShut {NoStop}%
\bibitem [{\citenamefont {Zhu}\ \emph {et~al.}(2013)\citenamefont {Zhu}, \citenamefont {Wang}, \citenamefont {Chan},\ and\ \citenamefont {Duan}}]{SLZhu2013}%
  \BibitemOpen
  \bibfield  {author} {\bibinfo {author} {\bibfnamefont {Shi-Liang}\ \bibnamefont {Zhu}}, \bibinfo {author} {\bibfnamefont {Z.-D.}\ \bibnamefont {Wang}}, \bibinfo {author} {\bibfnamefont {Y.-H.}\ \bibnamefont {Chan}}, \ and\ \bibinfo {author} {\bibfnamefont {L.-M.}\ \bibnamefont {Duan}},\ }\bibfield  {title} {\enquote {\bibinfo {title} {Topological bose-mott insulators in a one-dimensional optical superlattice},}\ }\href {\doibase 10.1103/physrevlett.110.075303} {\bibfield  {journal} {\bibinfo  {journal} {Phys. Rev. Lett.}\ }\textbf {\bibinfo {volume} {110}},\ \bibinfo {pages} {075303} (\bibinfo {year} {2013})}\BibitemShut {NoStop}%
\bibitem [{\citenamefont {Zhao}\ \emph {et~al.}(2011)\citenamefont {Zhao}, \citenamefont {Bray-Ali}, \citenamefont {Williams}, \citenamefont {Spielman},\ and\ \citenamefont {Satija}}]{PhysRevA.84.063629}%
  \BibitemOpen
  \bibfield  {author} {\bibinfo {author} {\bibfnamefont {Erhai}\ \bibnamefont {Zhao}}, \bibinfo {author} {\bibfnamefont {Noah}\ \bibnamefont {Bray-Ali}}, \bibinfo {author} {\bibfnamefont {Carl~J.}\ \bibnamefont {Williams}}, \bibinfo {author} {\bibfnamefont {I.~B.}\ \bibnamefont {Spielman}}, \ and\ \bibinfo {author} {\bibfnamefont {Indubala~I.}\ \bibnamefont {Satija}},\ }\bibfield  {title} {\enquote {\bibinfo {title} {Chern numbers hiding in time-of-flight images},}\ }\href {\doibase 10.1103/PhysRevA.84.063629} {\bibfield  {journal} {\bibinfo  {journal} {Phys. Rev. A}\ }\textbf {\bibinfo {volume} {84}},\ \bibinfo {pages} {063629} (\bibinfo {year} {2011})}\BibitemShut {NoStop}%
\bibitem [{\citenamefont {Li}\ \emph {et~al.}(2008)\citenamefont {Li}, \citenamefont {Shao}, \citenamefont {Sheng},\ and\ \citenamefont {Xing}}]{PhysRevA.78.053617}%
  \BibitemOpen
  \bibfield  {author} {\bibinfo {author} {\bibfnamefont {Fuxiang}\ \bibnamefont {Li}}, \bibinfo {author} {\bibfnamefont {L.~B.}\ \bibnamefont {Shao}}, \bibinfo {author} {\bibfnamefont {L.}~\bibnamefont {Sheng}}, \ and\ \bibinfo {author} {\bibfnamefont {D.~Y.}\ \bibnamefont {Xing}},\ }\bibfield  {title} {\enquote {\bibinfo {title} {Simulation of the quantum hall effect in a staggered modulated magnetic field with ultracold atoms},}\ }\href {\doibase 10.1103/PhysRevA.78.053617} {\bibfield  {journal} {\bibinfo  {journal} {Phys. Rev. A}\ }\textbf {\bibinfo {volume} {78}},\ \bibinfo {pages} {053617} (\bibinfo {year} {2008})}\BibitemShut {NoStop}%
\bibitem [{\citenamefont {Lang}\ \emph {et~al.}(2012)\citenamefont {Lang}, \citenamefont {Cai},\ and\ \citenamefont {Chen}}]{PhysRevLett.108.220401}%
  \BibitemOpen
  \bibfield  {author} {\bibinfo {author} {\bibfnamefont {Li-Jun}\ \bibnamefont {Lang}}, \bibinfo {author} {\bibfnamefont {Xiaoming}\ \bibnamefont {Cai}}, \ and\ \bibinfo {author} {\bibfnamefont {Shu}\ \bibnamefont {Chen}},\ }\bibfield  {title} {\enquote {\bibinfo {title} {Edge states and topological phases in one-dimensional optical superlattices},}\ }\href {\doibase 10.1103/PhysRevLett.108.220401} {\bibfield  {journal} {\bibinfo  {journal} {Phys. Rev. Lett.}\ }\textbf {\bibinfo {volume} {108}},\ \bibinfo {pages} {220401} (\bibinfo {year} {2012})}\BibitemShut {NoStop}%
\bibitem [{\citenamefont {Zhang}\ \emph {et~al.}(2020)\citenamefont {Zhang}, \citenamefont {Tang}, \citenamefont {Lang}, \citenamefont {Yan},\ and\ \citenamefont {Zhu}}]{SCDWZhang2020}%
  \BibitemOpen
  \bibfield  {author} {\bibinfo {author} {\bibfnamefont {Dan-Wei}\ \bibnamefont {Zhang}}, \bibinfo {author} {\bibfnamefont {Ling-Zhi}\ \bibnamefont {Tang}}, \bibinfo {author} {\bibfnamefont {Li-Jun}\ \bibnamefont {Lang}}, \bibinfo {author} {\bibfnamefont {Hui}\ \bibnamefont {Yan}}, \ and\ \bibinfo {author} {\bibfnamefont {Shi-Liang}\ \bibnamefont {Zhu}},\ }\bibfield  {title} {\enquote {\bibinfo {title} {Non-hermitian topological anderson insulators},}\ }\href {\doibase 10.1007/s11433-020-1521-9} {\bibfield  {journal} {\bibinfo  {journal} {Sci. China Phys. Mech. Astron.}\ }\textbf {\bibinfo {volume} {63}},\ \bibinfo {pages} {267062} (\bibinfo {year} {2020})}\BibitemShut {NoStop}%
\bibitem [{\citenamefont {Liu}\ \emph {et~al.}(2010)\citenamefont {Liu}, \citenamefont {Liu}, \citenamefont {Wu},\ and\ \citenamefont {Sinova}}]{PhysRevA.81.033622}%
  \BibitemOpen
  \bibfield  {author} {\bibinfo {author} {\bibfnamefont {Xiong-Jun}\ \bibnamefont {Liu}}, \bibinfo {author} {\bibfnamefont {Xin}\ \bibnamefont {Liu}}, \bibinfo {author} {\bibfnamefont {Congjun}\ \bibnamefont {Wu}}, \ and\ \bibinfo {author} {\bibfnamefont {Jairo}\ \bibnamefont {Sinova}},\ }\bibfield  {title} {\enquote {\bibinfo {title} {Quantum anomalous hall effect with cold atoms trapped in a square lattice},}\ }\href {\doibase 10.1103/PhysRevA.81.033622} {\bibfield  {journal} {\bibinfo  {journal} {Phys. Rev. A}\ }\textbf {\bibinfo {volume} {81}},\ \bibinfo {pages} {033622} (\bibinfo {year} {2010})}\BibitemShut {NoStop}%
\bibitem [{\citenamefont {Goldman}\ \emph {et~al.}(2012)\citenamefont {Goldman}, \citenamefont {Beugnon},\ and\ \citenamefont {Gerbier}}]{PhysRevLett.108.255303}%
  \BibitemOpen
  \bibfield  {author} {\bibinfo {author} {\bibfnamefont {Nathan}\ \bibnamefont {Goldman}}, \bibinfo {author} {\bibfnamefont {J\'er\^ome}\ \bibnamefont {Beugnon}}, \ and\ \bibinfo {author} {\bibfnamefont {Fabrice}\ \bibnamefont {Gerbier}},\ }\bibfield  {title} {\enquote {\bibinfo {title} {Detecting chiral edge states in the hofstadter optical lattice},}\ }\href {\doibase 10.1103/PhysRevLett.108.255303} {\bibfield  {journal} {\bibinfo  {journal} {Phys. Rev. Lett.}\ }\textbf {\bibinfo {volume} {108}},\ \bibinfo {pages} {255303} (\bibinfo {year} {2012})}\BibitemShut {NoStop}%
\bibitem [{\citenamefont {Stanescu}\ \emph {et~al.}(2010)\citenamefont {Stanescu}, \citenamefont {Galitski},\ and\ \citenamefont {Das~Sarma}}]{PhysRevA.82.013608}%
  \BibitemOpen
  \bibfield  {author} {\bibinfo {author} {\bibfnamefont {Tudor~D.}\ \bibnamefont {Stanescu}}, \bibinfo {author} {\bibfnamefont {Victor}\ \bibnamefont {Galitski}}, \ and\ \bibinfo {author} {\bibfnamefont {S.}~\bibnamefont {Das~Sarma}},\ }\bibfield  {title} {\enquote {\bibinfo {title} {Topological states in two-dimensional optical lattices},}\ }\href {\doibase 10.1103/PhysRevA.82.013608} {\bibfield  {journal} {\bibinfo  {journal} {Phys. Rev. A}\ }\textbf {\bibinfo {volume} {82}},\ \bibinfo {pages} {013608} (\bibinfo {year} {2010})}\BibitemShut {NoStop}%
\bibitem [{\citenamefont {Buchhold}\ \emph {et~al.}(2012)\citenamefont {Buchhold}, \citenamefont {Cocks},\ and\ \citenamefont {Hofstetter}}]{PhysRevA.85.063614}%
  \BibitemOpen
  \bibfield  {author} {\bibinfo {author} {\bibfnamefont {Michael}\ \bibnamefont {Buchhold}}, \bibinfo {author} {\bibfnamefont {Daniel}\ \bibnamefont {Cocks}}, \ and\ \bibinfo {author} {\bibfnamefont {Walter}\ \bibnamefont {Hofstetter}},\ }\bibfield  {title} {\enquote {\bibinfo {title} {Effects of smooth boundaries on topological edge modes in optical lattices},}\ }\href {\doibase 10.1103/PhysRevA.85.063614} {\bibfield  {journal} {\bibinfo  {journal} {Phys. Rev. A}\ }\textbf {\bibinfo {volume} {85}},\ \bibinfo {pages} {063614} (\bibinfo {year} {2012})}\BibitemShut {NoStop}%
\bibitem [{\citenamefont {Killi}\ and\ \citenamefont {Paramekanti}(2012)}]{PhysRevA.85.061606}%
  \BibitemOpen
  \bibfield  {author} {\bibinfo {author} {\bibfnamefont {Matthew}\ \bibnamefont {Killi}}\ and\ \bibinfo {author} {\bibfnamefont {Arun}\ \bibnamefont {Paramekanti}},\ }\bibfield  {title} {\enquote {\bibinfo {title} {Use of quantum quenches to probe the equilibrium current patterns of ultracold atoms in an optical lattice},}\ }\href {\doibase 10.1103/PhysRevA.85.061606} {\bibfield  {journal} {\bibinfo  {journal} {Phys. Rev. A}\ }\textbf {\bibinfo {volume} {85}},\ \bibinfo {pages} {061606} (\bibinfo {year} {2012})}\BibitemShut {NoStop}%
\bibitem [{\citenamefont {Reichl}\ and\ \citenamefont {Mueller}(2014)}]{PhysRevA.89.063628}%
  \BibitemOpen
  \bibfield  {author} {\bibinfo {author} {\bibfnamefont {Matthew~D.}\ \bibnamefont {Reichl}}\ and\ \bibinfo {author} {\bibfnamefont {Erich~J.}\ \bibnamefont {Mueller}},\ }\bibfield  {title} {\enquote {\bibinfo {title} {Floquet edge states with ultracold atoms},}\ }\href {\doibase 10.1103/PhysRevA.89.063628} {\bibfield  {journal} {\bibinfo  {journal} {Phys. Rev. A}\ }\textbf {\bibinfo {volume} {89}},\ \bibinfo {pages} {063628} (\bibinfo {year} {2014})}\BibitemShut {NoStop}%
\bibitem [{\citenamefont {Zhu}\ \emph {et~al.}(2007)\citenamefont {Zhu}, \citenamefont {Wang},\ and\ \citenamefont {Duan}}]{Zhu2007}%
  \BibitemOpen
  \bibfield  {author} {\bibinfo {author} {\bibfnamefont {S.-L.}\ \bibnamefont {Zhu}}, \bibinfo {author} {\bibfnamefont {B.}~\bibnamefont {Wang}}, \ and\ \bibinfo {author} {\bibfnamefont {L.-M.}\ \bibnamefont {Duan}},\ }\bibfield  {title} {\enquote {\bibinfo {title} {Simulation and detection of dirac fermions with cold atoms in an optical lattice},}\ }\href {\doibase 10.1103/physrevlett.98.260402} {\bibfield  {journal} {\bibinfo  {journal} {Phys. Rev. Lett.}\ }\textbf {\bibinfo {volume} {98}},\ \bibinfo {pages} {260402} (\bibinfo {year} {2007})}\BibitemShut {NoStop}%
\bibitem [{\citenamefont {Kennedy}\ \emph {et~al.}(2013)\citenamefont {Kennedy}, \citenamefont {Siviloglou}, \citenamefont {Miyake}, \citenamefont {Burton},\ and\ \citenamefont {Ketterle}}]{PhysRevLett.111.225301}%
  \BibitemOpen
  \bibfield  {author} {\bibinfo {author} {\bibfnamefont {Colin~J.}\ \bibnamefont {Kennedy}}, \bibinfo {author} {\bibfnamefont {Georgios~A.}\ \bibnamefont {Siviloglou}}, \bibinfo {author} {\bibfnamefont {Hirokazu}\ \bibnamefont {Miyake}}, \bibinfo {author} {\bibfnamefont {William~Cody}\ \bibnamefont {Burton}}, \ and\ \bibinfo {author} {\bibfnamefont {Wolfgang}\ \bibnamefont {Ketterle}},\ }\bibfield  {title} {\enquote {\bibinfo {title} {Spin-orbit coupling and quantum spin hall effect for neutral atoms without spin flips},}\ }\href {\doibase 10.1103/PhysRevLett.111.225301} {\bibfield  {journal} {\bibinfo  {journal} {Phys. Rev. Lett.}\ }\textbf {\bibinfo {volume} {111}},\ \bibinfo {pages} {225301} (\bibinfo {year} {2013})}\BibitemShut {NoStop}%
\bibitem [{\citenamefont {Aidelsburger}\ \emph {et~al.}(2013)\citenamefont {Aidelsburger}, \citenamefont {Atala}, \citenamefont {Lohse}, \citenamefont {Barreiro}, \citenamefont {Paredes},\ and\ \citenamefont {Bloch}}]{PhysRevLett.111.185301}%
  \BibitemOpen
  \bibfield  {author} {\bibinfo {author} {\bibfnamefont {M.}~\bibnamefont {Aidelsburger}}, \bibinfo {author} {\bibfnamefont {M.}~\bibnamefont {Atala}}, \bibinfo {author} {\bibfnamefont {M.}~\bibnamefont {Lohse}}, \bibinfo {author} {\bibfnamefont {J.~T.}\ \bibnamefont {Barreiro}}, \bibinfo {author} {\bibfnamefont {B.}~\bibnamefont {Paredes}}, \ and\ \bibinfo {author} {\bibfnamefont {I.}~\bibnamefont {Bloch}},\ }\bibfield  {title} {\enquote {\bibinfo {title} {Realization of the hofstadter hamiltonian with ultracold atoms in optical lattices},}\ }\href {\doibase 10.1103/PhysRevLett.111.185301} {\bibfield  {journal} {\bibinfo  {journal} {Phys. Rev. Lett.}\ }\textbf {\bibinfo {volume} {111}},\ \bibinfo {pages} {185301} (\bibinfo {year} {2013})}\BibitemShut {NoStop}%
\bibitem [{\citenamefont {Kane}\ and\ \citenamefont {Mele}(2005)}]{PhysRevLett.95.146802}%
  \BibitemOpen
  \bibfield  {author} {\bibinfo {author} {\bibfnamefont {C.~L.}\ \bibnamefont {Kane}}\ and\ \bibinfo {author} {\bibfnamefont {E.~J.}\ \bibnamefont {Mele}},\ }\bibfield  {title} {\enquote {\bibinfo {title} {${Z}_{2}$ topological order and the quantum spin hall effect},}\ }\href {\doibase 10.1103/PhysRevLett.95.146802} {\bibfield  {journal} {\bibinfo  {journal} {Phys. Rev. Lett.}\ }\textbf {\bibinfo {volume} {95}},\ \bibinfo {pages} {146802} (\bibinfo {year} {2005})}\BibitemShut {NoStop}%
\bibitem [{\citenamefont {Sheng}\ \emph {et~al.}(2006)\citenamefont {Sheng}, \citenamefont {Weng}, \citenamefont {Sheng},\ and\ \citenamefont {Haldane}}]{PhysRevLett.97.036808}%
  \BibitemOpen
  \bibfield  {author} {\bibinfo {author} {\bibfnamefont {D.~N.}\ \bibnamefont {Sheng}}, \bibinfo {author} {\bibfnamefont {Z.~Y.}\ \bibnamefont {Weng}}, \bibinfo {author} {\bibfnamefont {L.}~\bibnamefont {Sheng}}, \ and\ \bibinfo {author} {\bibfnamefont {F.~D.~M.}\ \bibnamefont {Haldane}},\ }\bibfield  {title} {\enquote {\bibinfo {title} {Quantum spin-hall effect and topologically invariant chern numbers},}\ }\href {\doibase 10.1103/PhysRevLett.97.036808} {\bibfield  {journal} {\bibinfo  {journal} {Phys. Rev. Lett.}\ }\textbf {\bibinfo {volume} {97}},\ \bibinfo {pages} {036808} (\bibinfo {year} {2006})}\BibitemShut {NoStop}%
\bibitem [{\citenamefont {Li}\ \emph {et~al.}(2010)\citenamefont {Li}, \citenamefont {Sheng}, \citenamefont {Sheng},\ and\ \citenamefont {Xing}}]{PhysRevB.82.165104}%
  \BibitemOpen
  \bibfield  {author} {\bibinfo {author} {\bibfnamefont {Huichao}\ \bibnamefont {Li}}, \bibinfo {author} {\bibfnamefont {L.}~\bibnamefont {Sheng}}, \bibinfo {author} {\bibfnamefont {D.~N.}\ \bibnamefont {Sheng}}, \ and\ \bibinfo {author} {\bibfnamefont {D.~Y.}\ \bibnamefont {Xing}},\ }\bibfield  {title} {\enquote {\bibinfo {title} {Chern number of thin films of the topological insulator ${\text{bi}}_{2}{\text{se}}_{3}$},}\ }\href {\doibase 10.1103/PhysRevB.82.165104} {\bibfield  {journal} {\bibinfo  {journal} {Phys. Rev. B}\ }\textbf {\bibinfo {volume} {82}},\ \bibinfo {pages} {165104} (\bibinfo {year} {2010})}\BibitemShut {NoStop}%
\bibitem [{\citenamefont {Haldane}(2004)}]{PhysRevLett.93.206602}%
  \BibitemOpen
  \bibfield  {author} {\bibinfo {author} {\bibfnamefont {F.~D.~M.}\ \bibnamefont {Haldane}},\ }\bibfield  {title} {\enquote {\bibinfo {title} {Berry curvature on the fermi surface: Anomalous hall effect as a topological fermi-liquid property},}\ }\href {\doibase 10.1103/PhysRevLett.93.206602} {\bibfield  {journal} {\bibinfo  {journal} {Phys. Rev. Lett.}\ }\textbf {\bibinfo {volume} {93}},\ \bibinfo {pages} {206602} (\bibinfo {year} {2004})}\BibitemShut {NoStop}%
\bibitem [{\citenamefont {Prodan}(2009)}]{PhysRevB.80.125327}%
  \BibitemOpen
  \bibfield  {author} {\bibinfo {author} {\bibfnamefont {Emil}\ \bibnamefont {Prodan}},\ }\bibfield  {title} {\enquote {\bibinfo {title} {Robustness of the spin-chern number},}\ }\href {\doibase 10.1103/PhysRevB.80.125327} {\bibfield  {journal} {\bibinfo  {journal} {Phys. Rev. B}\ }\textbf {\bibinfo {volume} {80}},\ \bibinfo {pages} {125327} (\bibinfo {year} {2009})}\BibitemShut {NoStop}%
\bibitem [{\citenamefont {Lv}\ \emph {et~al.}(2021)\citenamefont {Lv}, \citenamefont {Du}, \citenamefont {Liang}, \citenamefont {Liu}, \citenamefont {Liang}, \citenamefont {Chen}, \citenamefont {Zhou}, \citenamefont {Zhang}, \citenamefont {Zhang}, \citenamefont {Ai}, \citenamefont {Yan},\ and\ \citenamefont {Zhu}}]{PhysRevLett.127.136802}%
  \BibitemOpen
  \bibfield  {author} {\bibinfo {author} {\bibfnamefont {Qing-Xian}\ \bibnamefont {Lv}}, \bibinfo {author} {\bibfnamefont {Yan-Xiong}\ \bibnamefont {Du}}, \bibinfo {author} {\bibfnamefont {Zhen-Tao}\ \bibnamefont {Liang}}, \bibinfo {author} {\bibfnamefont {Hong-Zhi}\ \bibnamefont {Liu}}, \bibinfo {author} {\bibfnamefont {Jia-Hao}\ \bibnamefont {Liang}}, \bibinfo {author} {\bibfnamefont {Lin-Qing}\ \bibnamefont {Chen}}, \bibinfo {author} {\bibfnamefont {Li-Ming}\ \bibnamefont {Zhou}}, \bibinfo {author} {\bibfnamefont {Shan-Chao}\ \bibnamefont {Zhang}}, \bibinfo {author} {\bibfnamefont {Dan-Wei}\ \bibnamefont {Zhang}}, \bibinfo {author} {\bibfnamefont {Bao-Quan}\ \bibnamefont {Ai}}, \bibinfo {author} {\bibfnamefont {Hui}\ \bibnamefont {Yan}}, \ and\ \bibinfo {author} {\bibfnamefont {Shi-Liang}\ \bibnamefont {Zhu}},\ }\bibfield  {title} {\enquote {\bibinfo {title} {Measurement of spin chern numbers in quantum simulated topological insulators},}\ }\href {\doibase 10.1103/PhysRevLett.127.136802} {\bibfield
  {journal} {\bibinfo  {journal} {Phys. Rev. Lett.}\ }\textbf {\bibinfo {volume} {127}},\ \bibinfo {pages} {136802} (\bibinfo {year} {2021})}\BibitemShut {NoStop}%
\bibitem [{\citenamefont {Yang}\ \emph {et~al.}(2011)\citenamefont {Yang}, \citenamefont {Xu}, \citenamefont {Sheng}, \citenamefont {Wang}, \citenamefont {Xing},\ and\ \citenamefont {Sheng}}]{PhysRevLett.107.066602}%
  \BibitemOpen
  \bibfield  {author} {\bibinfo {author} {\bibfnamefont {Yunyou}\ \bibnamefont {Yang}}, \bibinfo {author} {\bibfnamefont {Zhong}\ \bibnamefont {Xu}}, \bibinfo {author} {\bibfnamefont {L.}~\bibnamefont {Sheng}}, \bibinfo {author} {\bibfnamefont {Baigeng}\ \bibnamefont {Wang}}, \bibinfo {author} {\bibfnamefont {D.~Y.}\ \bibnamefont {Xing}}, \ and\ \bibinfo {author} {\bibfnamefont {D.~N.}\ \bibnamefont {Sheng}},\ }\bibfield  {title} {\enquote {\bibinfo {title} {Time-reversal-symmetry-broken quantum spin hall effect},}\ }\href {\doibase 10.1103/PhysRevLett.107.066602} {\bibfield  {journal} {\bibinfo  {journal} {Phys. Rev. Lett.}\ }\textbf {\bibinfo {volume} {107}},\ \bibinfo {pages} {066602} (\bibinfo {year} {2011})}\BibitemShut {NoStop}%
\bibitem [{\citenamefont {Zhang}\ \emph {et~al.}(2016)\citenamefont {Zhang}, \citenamefont {Zhao}, \citenamefont {Liu}, \citenamefont {Xue}, \citenamefont {Zhu},\ and\ \citenamefont {Wang}}]{DWZhang2016}%
  \BibitemOpen
  \bibfield  {author} {\bibinfo {author} {\bibfnamefont {Dan-Wei}\ \bibnamefont {Zhang}}, \bibinfo {author} {\bibfnamefont {Y.~X.}\ \bibnamefont {Zhao}}, \bibinfo {author} {\bibfnamefont {Rui-Bin}\ \bibnamefont {Liu}}, \bibinfo {author} {\bibfnamefont {Zheng-Yuan}\ \bibnamefont {Xue}}, \bibinfo {author} {\bibfnamefont {Shi-Liang}\ \bibnamefont {Zhu}}, \ and\ \bibinfo {author} {\bibfnamefont {Z.~D.}\ \bibnamefont {Wang}},\ }\bibfield  {title} {\enquote {\bibinfo {title} {Quantum simulation of exotic pt-invariant topological nodal loop bands with ultracold atoms in an optical lattice},}\ }\href {\doibase 10.1103/physreva.93.043617} {\bibfield  {journal} {\bibinfo  {journal} {Phy. Rev. A}\ }\textbf {\bibinfo {volume} {93}},\ \bibinfo {pages} {043617} (\bibinfo {year} {2016})}\BibitemShut {NoStop}%
\bibitem [{\citenamefont {Soltan-Panahi}\ \emph {et~al.}(2011)\citenamefont {Soltan-Panahi}, \citenamefont {Struck}, \citenamefont {Hauke}, \citenamefont {Bick}, \citenamefont {Plenkers}, \citenamefont {Meineke}, \citenamefont {Becker}, \citenamefont {Windpassinger}, \citenamefont {Lewenstein},\ and\ \citenamefont {Sengstock}}]{soltan2011multi}%
  \BibitemOpen
  \bibfield  {author} {\bibinfo {author} {\bibfnamefont {Parvis}\ \bibnamefont {Soltan-Panahi}}, \bibinfo {author} {\bibfnamefont {Julian}\ \bibnamefont {Struck}}, \bibinfo {author} {\bibfnamefont {Philipp}\ \bibnamefont {Hauke}}, \bibinfo {author} {\bibfnamefont {Andreas}\ \bibnamefont {Bick}}, \bibinfo {author} {\bibfnamefont {Wiebke}\ \bibnamefont {Plenkers}}, \bibinfo {author} {\bibfnamefont {Georg}\ \bibnamefont {Meineke}}, \bibinfo {author} {\bibfnamefont {Christoph}\ \bibnamefont {Becker}}, \bibinfo {author} {\bibfnamefont {Patrick}\ \bibnamefont {Windpassinger}}, \bibinfo {author} {\bibfnamefont {Maciej}\ \bibnamefont {Lewenstein}}, \ and\ \bibinfo {author} {\bibfnamefont {Klaus}\ \bibnamefont {Sengstock}},\ }\bibfield  {title} {\enquote {\bibinfo {title} {Multi-component quantum gases in spin-dependent hexagonal lattices},}\ }\href {\doibase 10.1038/nphys1916} {\bibfield  {journal} {\bibinfo  {journal} {Nature Physics}\ }\textbf {\bibinfo {volume} {7}},\ \bibinfo {pages} {434--440} (\bibinfo {year}
  {2011})}\BibitemShut {NoStop}%
\bibitem [{\citenamefont {Lee}\ \emph {et~al.}(2009)\citenamefont {Lee}, \citenamefont {Gr\'emaud}, \citenamefont {Han}, \citenamefont {Englert},\ and\ \citenamefont {Miniatura}}]{PhysRevA.80.043411}%
  \BibitemOpen
  \bibfield  {author} {\bibinfo {author} {\bibfnamefont {Kean~Loon}\ \bibnamefont {Lee}}, \bibinfo {author} {\bibfnamefont {Beno\^{\i}t}\ \bibnamefont {Gr\'emaud}}, \bibinfo {author} {\bibfnamefont {Rui}\ \bibnamefont {Han}}, \bibinfo {author} {\bibfnamefont {Berthold-Georg}\ \bibnamefont {Englert}}, \ and\ \bibinfo {author} {\bibfnamefont {Christian}\ \bibnamefont {Miniatura}},\ }\bibfield  {title} {\enquote {\bibinfo {title} {Ultracold fermions in a graphene-type optical lattice},}\ }\href {\doibase 10.1103/PhysRevA.80.043411} {\bibfield  {journal} {\bibinfo  {journal} {Phys. Rev. A}\ }\textbf {\bibinfo {volume} {80}},\ \bibinfo {pages} {043411} (\bibinfo {year} {2009})}\BibitemShut {NoStop}%
\bibitem [{\citenamefont {Struck}\ \emph {et~al.}(2012)\citenamefont {Struck}, \citenamefont {\"Olschl\"ager}, \citenamefont {Weinberg}, \citenamefont {Hauke}, \citenamefont {Simonet}, \citenamefont {Eckardt}, \citenamefont {Lewenstein}, \citenamefont {Sengstock},\ and\ \citenamefont {Windpassinger}}]{PhysRevLett.108.225304}%
  \BibitemOpen
  \bibfield  {author} {\bibinfo {author} {\bibfnamefont {J.}~\bibnamefont {Struck}}, \bibinfo {author} {\bibfnamefont {C.}~\bibnamefont {\"Olschl\"ager}}, \bibinfo {author} {\bibfnamefont {M.}~\bibnamefont {Weinberg}}, \bibinfo {author} {\bibfnamefont {P.}~\bibnamefont {Hauke}}, \bibinfo {author} {\bibfnamefont {J.}~\bibnamefont {Simonet}}, \bibinfo {author} {\bibfnamefont {A.}~\bibnamefont {Eckardt}}, \bibinfo {author} {\bibfnamefont {M.}~\bibnamefont {Lewenstein}}, \bibinfo {author} {\bibfnamefont {K.}~\bibnamefont {Sengstock}}, \ and\ \bibinfo {author} {\bibfnamefont {P.}~\bibnamefont {Windpassinger}},\ }\bibfield  {title} {\enquote {\bibinfo {title} {Tunable gauge potential for neutral and spinless particles in driven optical lattices},}\ }\href {\doibase 10.1103/PhysRevLett.108.225304} {\bibfield  {journal} {\bibinfo  {journal} {Phys. Rev. Lett.}\ }\textbf {\bibinfo {volume} {108}},\ \bibinfo {pages} {225304} (\bibinfo {year} {2012})}\BibitemShut {NoStop}%
\bibitem [{\citenamefont {Hauke}\ \emph {et~al.}(2012)\citenamefont {Hauke}, \citenamefont {Tieleman}, \citenamefont {Celi}, \citenamefont {\"Olschl\"ager}, \citenamefont {Simonet}, \citenamefont {Struck}, \citenamefont {Weinberg}, \citenamefont {Windpassinger}, \citenamefont {Sengstock}, \citenamefont {Lewenstein},\ and\ \citenamefont {Eckardt}}]{PhysRevLett.109.145301}%
  \BibitemOpen
  \bibfield  {author} {\bibinfo {author} {\bibfnamefont {Philipp}\ \bibnamefont {Hauke}}, \bibinfo {author} {\bibfnamefont {Olivier}\ \bibnamefont {Tieleman}}, \bibinfo {author} {\bibfnamefont {Alessio}\ \bibnamefont {Celi}}, \bibinfo {author} {\bibfnamefont {Christoph}\ \bibnamefont {\"Olschl\"ager}}, \bibinfo {author} {\bibfnamefont {Juliette}\ \bibnamefont {Simonet}}, \bibinfo {author} {\bibfnamefont {Julian}\ \bibnamefont {Struck}}, \bibinfo {author} {\bibfnamefont {Malte}\ \bibnamefont {Weinberg}}, \bibinfo {author} {\bibfnamefont {Patrick}\ \bibnamefont {Windpassinger}}, \bibinfo {author} {\bibfnamefont {Klaus}\ \bibnamefont {Sengstock}}, \bibinfo {author} {\bibfnamefont {Maciej}\ \bibnamefont {Lewenstein}}, \ and\ \bibinfo {author} {\bibfnamefont {Andr\'e}\ \bibnamefont {Eckardt}},\ }\bibfield  {title} {\enquote {\bibinfo {title} {Non-abelian gauge fields and topological insulators in shaken optical lattices},}\ }\href {\doibase 10.1103/PhysRevLett.109.145301} {\bibfield  {journal} {\bibinfo  {journal}
  {Phys. Rev. Lett.}\ }\textbf {\bibinfo {volume} {109}},\ \bibinfo {pages} {145301} (\bibinfo {year} {2012})}\BibitemShut {NoStop}%
\bibitem [{\citenamefont {Zheng}\ and\ \citenamefont {Zhai}(2014)}]{PhysRevA.89.061603}%
  \BibitemOpen
  \bibfield  {author} {\bibinfo {author} {\bibfnamefont {Wei}\ \bibnamefont {Zheng}}\ and\ \bibinfo {author} {\bibfnamefont {Hui}\ \bibnamefont {Zhai}},\ }\bibfield  {title} {\enquote {\bibinfo {title} {Floquet topological states in shaking optical lattices},}\ }\href {\doibase 10.1103/PhysRevA.89.061603} {\bibfield  {journal} {\bibinfo  {journal} {Phys. Rev. A}\ }\textbf {\bibinfo {volume} {89}},\ \bibinfo {pages} {061603} (\bibinfo {year} {2014})}\BibitemShut {NoStop}%
\bibitem [{\citenamefont {Gro{\ss}mann}\ and\ \citenamefont {Hänggi}(1992)}]{Gro_mann_1992}%
  \BibitemOpen
  \bibfield  {author} {\bibinfo {author} {\bibfnamefont {F}~\bibnamefont {Gro{\ss}mann}}\ and\ \bibinfo {author} {\bibfnamefont {P}~\bibnamefont {Hänggi}},\ }\href {\doibase 10.1209/0295-5075/18/7/001} {\ \textbf {\bibinfo {volume} {18}},\ \bibinfo {pages} {571--576} (\bibinfo {year} {1992})}\BibitemShut {NoStop}%
\bibitem [{\citenamefont {Dunlap}\ and\ \citenamefont {Kenkre}(1986)}]{PhysRevB.34.3625}%
  \BibitemOpen
  \bibfield  {author} {\bibinfo {author} {\bibfnamefont {D.~H.}\ \bibnamefont {Dunlap}}\ and\ \bibinfo {author} {\bibfnamefont {V.~M.}\ \bibnamefont {Kenkre}},\ }\bibfield  {title} {\enquote {\bibinfo {title} {Dynamic localization of a charged particle moving under the influence of an electric field},}\ }\href {\doibase 10.1103/PhysRevB.34.3625} {\bibfield  {journal} {\bibinfo  {journal} {Phys. Rev. B}\ }\textbf {\bibinfo {volume} {34}},\ \bibinfo {pages} {3625--3633} (\bibinfo {year} {1986})}\BibitemShut {NoStop}%
\bibitem [{\citenamefont {Holthaus}(1992)}]{PhysRevLett.69.351}%
  \BibitemOpen
  \bibfield  {author} {\bibinfo {author} {\bibfnamefont {Martin}\ \bibnamefont {Holthaus}},\ }\bibfield  {title} {\enquote {\bibinfo {title} {Collapse of minibands in far-infrared irradiated superlattices},}\ }\href {\doibase 10.1103/PhysRevLett.69.351} {\bibfield  {journal} {\bibinfo  {journal} {Phys. Rev. Lett.}\ }\textbf {\bibinfo {volume} {69}},\ \bibinfo {pages} {351--354} (\bibinfo {year} {1992})}\BibitemShut {NoStop}%
\bibitem [{\citenamefont {Eckardt}\ \emph {et~al.}(2005)\citenamefont {Eckardt}, \citenamefont {Weiss},\ and\ \citenamefont {Holthaus}}]{PhysRevLett.95.260404}%
  \BibitemOpen
  \bibfield  {author} {\bibinfo {author} {\bibfnamefont {Andr\'e}\ \bibnamefont {Eckardt}}, \bibinfo {author} {\bibfnamefont {Christoph}\ \bibnamefont {Weiss}}, \ and\ \bibinfo {author} {\bibfnamefont {Martin}\ \bibnamefont {Holthaus}},\ }\bibfield  {title} {\enquote {\bibinfo {title} {Superfluid-insulator transition in a periodically driven optical lattice},}\ }\href {\doibase 10.1103/PhysRevLett.95.260404} {\bibfield  {journal} {\bibinfo  {journal} {Phys. Rev. Lett.}\ }\textbf {\bibinfo {volume} {95}},\ \bibinfo {pages} {260404} (\bibinfo {year} {2005})}\BibitemShut {NoStop}%
\bibitem [{\citenamefont {Kierig}\ \emph {et~al.}(2008)\citenamefont {Kierig}, \citenamefont {Schnorrberger}, \citenamefont {Schietinger}, \citenamefont {Tomkovic},\ and\ \citenamefont {Oberthaler}}]{PhysRevLett.100.190405}%
  \BibitemOpen
  \bibfield  {author} {\bibinfo {author} {\bibfnamefont {E.}~\bibnamefont {Kierig}}, \bibinfo {author} {\bibfnamefont {U.}~\bibnamefont {Schnorrberger}}, \bibinfo {author} {\bibfnamefont {A.}~\bibnamefont {Schietinger}}, \bibinfo {author} {\bibfnamefont {J.}~\bibnamefont {Tomkovic}}, \ and\ \bibinfo {author} {\bibfnamefont {M.~K.}\ \bibnamefont {Oberthaler}},\ }\bibfield  {title} {\enquote {\bibinfo {title} {Single-particle tunneling in strongly driven double-well potentials},}\ }\href {\doibase 10.1103/PhysRevLett.100.190405} {\bibfield  {journal} {\bibinfo  {journal} {Phys. Rev. Lett.}\ }\textbf {\bibinfo {volume} {100}},\ \bibinfo {pages} {190405} (\bibinfo {year} {2008})}\BibitemShut {NoStop}%
\bibitem [{\citenamefont {Zenesini}\ \emph {et~al.}(2009)\citenamefont {Zenesini}, \citenamefont {Lignier}, \citenamefont {Ciampini}, \citenamefont {Morsch},\ and\ \citenamefont {Arimondo}}]{PhysRevLett.102.100403}%
  \BibitemOpen
  \bibfield  {author} {\bibinfo {author} {\bibfnamefont {Alessandro}\ \bibnamefont {Zenesini}}, \bibinfo {author} {\bibfnamefont {Hans}\ \bibnamefont {Lignier}}, \bibinfo {author} {\bibfnamefont {Donatella}\ \bibnamefont {Ciampini}}, \bibinfo {author} {\bibfnamefont {Oliver}\ \bibnamefont {Morsch}}, \ and\ \bibinfo {author} {\bibfnamefont {Ennio}\ \bibnamefont {Arimondo}},\ }\bibfield  {title} {\enquote {\bibinfo {title} {Coherent control of dressed matter waves},}\ }\href {\doibase 10.1103/PhysRevLett.102.100403} {\bibfield  {journal} {\bibinfo  {journal} {Phys. Rev. Lett.}\ }\textbf {\bibinfo {volume} {102}},\ \bibinfo {pages} {100403} (\bibinfo {year} {2009})}\BibitemShut {NoStop}%
\bibitem [{\citenamefont {Struck}\ \emph {et~al.}(2011)\citenamefont {Struck}, \citenamefont {Ölschläger}, \citenamefont {Targat}, \citenamefont {Soltan-Panahi}, \citenamefont {Eckardt}, \citenamefont {Lewenstein}, \citenamefont {Windpassinger},\ and\ \citenamefont {Sengstock}}]{science.1207239}%
  \BibitemOpen
  \bibfield  {author} {\bibinfo {author} {\bibfnamefont {J.}~\bibnamefont {Struck}}, \bibinfo {author} {\bibfnamefont {C.}~\bibnamefont {Ölschläger}}, \bibinfo {author} {\bibfnamefont {R.~Le}\ \bibnamefont {Targat}}, \bibinfo {author} {\bibfnamefont {P.}~\bibnamefont {Soltan-Panahi}}, \bibinfo {author} {\bibfnamefont {A.}~\bibnamefont {Eckardt}}, \bibinfo {author} {\bibfnamefont {M.}~\bibnamefont {Lewenstein}}, \bibinfo {author} {\bibfnamefont {P.}~\bibnamefont {Windpassinger}}, \ and\ \bibinfo {author} {\bibfnamefont {K.}~\bibnamefont {Sengstock}},\ }\bibfield  {title} {\enquote {\bibinfo {title} {Quantum simulation of frustrated classical magnetism in triangular optical lattices},}\ }\href {\doibase 10.1126/science.1207239} {\bibfield  {journal} {\bibinfo  {journal} {Science}\ }\textbf {\bibinfo {volume} {333}},\ \bibinfo {pages} {996--999} (\bibinfo {year} {2011})}\BibitemShut {NoStop}%
\bibitem [{\citenamefont {Sacha}\ \emph {et~al.}(2012)\citenamefont {Sacha}, \citenamefont {Targo\ifmmode~\acute{n}\else \'{n}\fi{}ska},\ and\ \citenamefont {Zakrzewski}}]{PhysRevA.85.053613}%
  \BibitemOpen
  \bibfield  {author} {\bibinfo {author} {\bibfnamefont {Krzysztof}\ \bibnamefont {Sacha}}, \bibinfo {author} {\bibfnamefont {Katarzyna}\ \bibnamefont {Targo\ifmmode~\acute{n}\else \'{n}\fi{}ska}}, \ and\ \bibinfo {author} {\bibfnamefont {Jakub}\ \bibnamefont {Zakrzewski}},\ }\bibfield  {title} {\enquote {\bibinfo {title} {Frustration and time-reversal symmetry breaking for fermi and bose-fermi systems},}\ }\href {\doibase 10.1103/PhysRevA.85.053613} {\bibfield  {journal} {\bibinfo  {journal} {Phys. Rev. A}\ }\textbf {\bibinfo {volume} {85}},\ \bibinfo {pages} {053613} (\bibinfo {year} {2012})}\BibitemShut {NoStop}%
\bibitem [{\citenamefont {Fukui}\ \emph {et~al.}(2005)\citenamefont {Fukui}, \citenamefont {Hatsugai},\ and\ \citenamefont {Suzuki}}]{JPSJ.74.1674}%
  \BibitemOpen
  \bibfield  {author} {\bibinfo {author} {\bibfnamefont {Takahiro}\ \bibnamefont {Fukui}}, \bibinfo {author} {\bibfnamefont {Yasuhiro}\ \bibnamefont {Hatsugai}}, \ and\ \bibinfo {author} {\bibfnamefont {Hiroshi}\ \bibnamefont {Suzuki}},\ }\bibfield  {title} {\enquote {\bibinfo {title} {Chern numbers in discretized brillouin zone: Efficient method of computing (spin) hall conductances},}\ }\href {\doibase 10.1143/JPSJ.74.1674} {\bibfield  {journal} {\bibinfo  {journal} {Journal of the Physical Society of Japan}\ }\textbf {\bibinfo {volume} {74}},\ \bibinfo {pages} {1674--1677} (\bibinfo {year} {2005})}\BibitemShut {NoStop}%
\bibitem [{\citenamefont {Goldman}\ and\ \citenamefont {Dalibard}(2014)}]{PhysRevX.4.031027}%
  \BibitemOpen
  \bibfield  {author} {\bibinfo {author} {\bibfnamefont {N.}~\bibnamefont {Goldman}}\ and\ \bibinfo {author} {\bibfnamefont {J.}~\bibnamefont {Dalibard}},\ }\bibfield  {title} {\enquote {\bibinfo {title} {Periodically driven quantum systems: Effective hamiltonians and engineered gauge fields},}\ }\href {\doibase 10.1103/PhysRevX.4.031027} {\bibfield  {journal} {\bibinfo  {journal} {Phys. Rev. X}\ }\textbf {\bibinfo {volume} {4}},\ \bibinfo {pages} {031027} (\bibinfo {year} {2014})}\BibitemShut {NoStop}%
\end{thebibliography}%

\end{document}